# On Fairness of Systemic Risk Measures

Francesca Biagini[*]     Jean-Pierre Fouque [†]     Marco Frittelli[‡]

Thilo Meyer-Brandis[§]

April 22, 2019

**Abstract**

In our previous paper, "A Unified Approach to Systemic Risk Measures via Acceptance Set" (*Mathematical Finance, 2018*), we have introduced a general class of systemic risk measures that allow for random allocations to individual banks before aggregation of their risks. In the present paper, we prove the dual representation of a particular subclass of such systemic risk measures and the existence and uniqueness of the optimal allocation related to them. We also introduce an associated utility maximization problem which has the same optimal solution as the systemic risk measure. In addition, the optimizer in the dual formulation provides a *risk allocation* which is fair from the point of view of the individual financial institutions. The case with exponential utilities which allows for explicit computation is treated in details.

**Acknowledgment:** The third author would like to thank Enea Monzio Compagnoni for very helpful discussion and relevant insights on the whole paper during the preparation of his Laurea thesis, his Laurea student Giacomo Bizzarrini, as well as his Ph.D. student Alessandro Doldi for his careful reading and decisive contribution to Section 4.4.1.

[*]Department of Mathematics, University of Munich, Theresienstraße 39, 80333 Munich, Germany. *francesca.biagini@math.lmu.de*. Secondary affiliation: Department of Mathematics, University of Oslo, Box 1053, Blindern, 0316, Oslo, Norway.

[†]Department of Statistics & Applied Probability, University of California, Santa Barbara, CA 93106-3110, *fouque@pstat.ucsb.edu*. Work supported by NSF grants DMS-1409434 and DMS-1814091.

[‡]Dipartimento di Matematica, Università degli Studi di Milano, Via Saldini 50, 20133 Milano, Italy, *marco.frittelli@unimi.it*.

[§]Department of Mathematics, University of Munich, Theresienstraße 39, 80333 Munich, Germany. *meyerbr@math.lmu.de*. Part of this research was performed while F. Biagini, M. Frittelli and T. Meyer-Brandis were visiting the University of California Santa Barbara.

# 1 Introduction

Consider a vector $\mathbf{X} = (X^1, \ldots, X^N) \in L^0(\Omega, \mathcal{F}, \mathbb{P}, \mathbb{R}^N)$ of $N$ random variables denoting a configuration of risky (financial) factors at a future time $T$ associated to a system of $N$ financial institutions/banks.

One of the first proposals in the framework of risk measures to measure the systemic risk of $\mathbf{X}$, see [16], was to consider the map

$$\rho(\mathbf{X}) := \inf\{m \in \mathbb{R} \mid \Lambda(\mathbf{X}) + m \in \mathbb{A}\}, \quad (1.1)$$

where

$$\Lambda : \mathbb{R}^N \to \mathbb{R},$$

is an aggregation rule that aggregates the $N$-dimensional risk factors into a univariate risk factor, and

$$\mathbb{A} \subseteq L^0(\Omega, \mathcal{F}, \mathbb{P}, \mathbb{R}),$$

is an acceptance set of real valued random variables. As within the framework of univariate monetary risk measures, systemic risk might again be interpreted as the minimal cash amount that secures the system when it is added to the total aggregated system loss $\Lambda(\mathbf{X})$, given that $\Lambda(\mathbf{X})$ allows for a monetary loss interpretation. Note, however, that in (1.1) systemic risk is the minimal capital added to secure the system *after aggregating individual risks*.

It might be more relevant to measure systemic risk as the minimal cash that secures the aggregated system by adding the capital into the single institutions *before aggregating their individual risks*. This way of measuring systemic risk can be expressed by

$$\rho(\mathbf{X}) := \inf\left\{\sum_{i=1}^{N} m_i \mid \mathbf{m} = (m_1, \cdots, m_N) \in \mathbb{R}^N, \Lambda(\mathbf{X} + \mathbf{m}) \in \mathbb{A}\right\}. \quad (1.2)$$

Here, the amount $m_i$ is added to the financial position $X^i$ of institution $i \in \{1, \cdots, N\}$ before the corresponding total loss $\Lambda(\mathbf{X} + \mathbf{m})$ is computed (we refer to [3], [7] and [27]).

One of the main novelty of our paper [7] was the possibility of adding to $\mathbf{X}$ not merely a vector $\mathbf{m} = (m_1, \cdots, m_N) \in \mathbb{R}^N$ of deterministic cash, but, more generally, a random vector $\mathbf{Y} \in \mathcal{C}$ for some given class $\mathcal{C}$. In particular, the main example considered in [7] and studied further in this paper, is given by choosing the aggregation function

$$\Lambda(\mathbf{x}) = \sum_{n=1}^{N} u_n(x_n)$$

for utility functions $u_n$, $n = 1, \cdots, N$, the acceptance set $\mathbb{A} = \{Z \in L^1(\Omega, \mathcal{F}, \mathbb{P}, \mathbb{R}), \mathbb{E}[Z] \geq B\}$ for a given constant $B$, and the class $\mathcal{C}$ such that

$$\mathcal{C} \subseteq \mathcal{C}_\mathbb{R} \cap \mathcal{L}, \text{ where } \mathcal{C}_\mathbb{R} := \left\{\mathbf{Y} \in L^0(\Omega, \mathcal{F}, \mathbb{P}, \mathbb{R}^N) \mid \sum_{n=1}^{N} Y^n \in \mathbb{R}\right\}, \quad (1.3)$$

where the subspace $\mathcal{L} \subseteq L^0(\Omega, \mathcal{F}, \mathbb{P}, \mathbb{R}^N)$ will be specified later. Here, the notation $\sum_{n=1}^N Y^n \in \mathbb{R}$ means that $\sum_{n=1}^N Y^n$ is equal to some deterministic constant in $\mathbb{R}$, even though each single $Y^n$, $n = 1, \cdots, N$, is a random variable. Under these assumptions the systemic risk measure considered in [7] takes the form

$$\rho(\mathbf{X}) := \inf_{\mathbf{Y} \in \mathcal{C} \subset \mathcal{C}_{\mathbb{R}}} \left\{ \sum_{n=1}^N Y^n \mid \mathbb{E}\left[\sum_{n=1}^N u_n(X^n + Y^n)\right] \geq B \right\}, \quad (1.4)$$

and can still be interpreted as the minimal total cash amount $\sum_{n=1}^N Y^n \in \mathbb{R}$ needed today to secure the system by distributing the cash at the future time $T$ among the components of the risk vector $\mathbf{X}$. However, while the total capital requirement $\sum_{n=1}^N Y^n$ is determined today, contrary to (1.2) the individual allocation $Y^i(\omega)$ to institution $i$ does not need to be decided today but in general depends on the scenario $\omega$ realized at time $T$. This total cash amount $\rho(\mathbf{X})$ is composed today through the formula $\sum_{n=1}^N \rho^n(\mathbf{X}) = \rho(\mathbf{X})$, where each $\rho^n(\mathbf{X}) \in \mathbb{R}$ is the risk allocation of each bank, as explained in Definition 1.2. Thus, one prominent example that can be modelled by considering random allocations is the default fund of a CCP that is liable for any participating institution. We will come back to this mechanism in Section 5.

By considering scenario dependent allocations we are also taking into account the possible dependencies among the banks, as the budget constraints in (1.4) will not depend only on the marginal distribution of $\mathbf{X}$, as it would happen for deterministic $Y^n$.

**Definition 1.1.** *We say that the scenario dependent allocation $\mathbf{Y}_{\mathbf{X}} = (Y_{\mathbf{X}}^n)_{n=1,\ldots N} \in \mathcal{C}$ is a systemic optimal allocation for $\rho(\mathbf{X})$, defined in (1.4), if it satisfies $\rho(\mathbf{X}) = \sum_{n=1}^N Y_{\mathbf{X}}^n$ and $\mathbb{E}\left[\sum_{n=1}^N u_n(X^n + Y_{\mathbf{X}}^n))\right] \geq B$ .*

As two of the main results of the paper

- we study in Section 3 the dual formulation of the systemic risk measure (1.4):

$$\rho(\mathbf{X}) = \max_{\mathbf{Q} \in \mathcal{D}} \left\{ \sum_{n=1}^N \mathbb{E}_{Q^n}[-X^n] - \alpha_B(\mathbf{Q}) \right\}, \quad (1.5)$$

  where $\mathbf{Q} := (Q^1, \cdots, Q^N)$, and the penalty function $\alpha_B$ and the domain $\mathcal{D}$ are specified in Section 3. In particular, we establish existence and uniqueness of the optimizer $\mathbf{Q}_{\mathbf{X}} \in \mathcal{D}$ of (1.5).

- we show in Section 4 existence and uniqueness of the systemic optimal allocation $\mathbf{Y}_{\mathbf{X}}$ for the systemic risk measure (1.4).

We now associate to the risk minimization problem (1.4) the following related utility maximization problem that will play a central role in this paper:

$$\pi(\mathbf{X}) := \sup_{\mathbf{Y} \in \mathcal{C} \subset \mathcal{C}_{\mathbb{R}}} \left\{ \mathbb{E}\left[\sum_{n=1}^N u_n(X^n + Y^n)\right] \mid \sum_{n=1}^N Y^n \leq A \right\}. \quad (1.6)$$

If we interpret $\sum_{n=1}^{N} u_n(X^n + Y^n)$ as the aggregated utility of the system after allocating $\mathbf{Y}$, then $\pi(\mathbf{X})$ can be interpreted as the maximal expected utility of the system over all random allocations $\mathbf{Y} \in \mathcal{C}$ such that the aggregated budget constraint $\sum_{n=1}^{N} Y^n \leq A$ holds for a given constant $A$. In the following, we may write $\rho(\mathbf{X}) = \rho_B(\mathbf{X})$ and $\pi(\mathbf{X}) = \pi_A(\mathbf{X})$ in order to express the dependence on the minimal level of expected utility $B \in \mathbb{R}$ and on the maximal budget level $A \in \mathbb{R}$, respectively. We will see in Section 4.1 that

$$B = \pi_A(\mathbf{X}) \text{ if and only if } A = \rho_B(\mathbf{X}),$$

and, in these cases, the two problems $\pi_A(\mathbf{X})$ and $\rho_B(\mathbf{X})$ have the same unique optimal solution $\mathbf{Y}_{\mathbf{X}}$. From this, we infer that once a level $\rho(\mathbf{X})$ of total systemic risk has been determined, then

- *the systemic optimal allocation $\mathbf{Y}_{\mathbf{X}}$ for $\rho$ maximizes the expected system utility among all random allocations of total cost less or equal to $\rho(\mathbf{X})$.*

Once the total systemic risk has been identified as $\rho(\mathbf{X})$, the second essential question is how to allocate the total risk to the individual institutions.

**Definition 1.2.** *We say that a vector $(\rho^n(\mathbf{X}))_{n=1,\dots,N} \in \mathbb{R}^N$ is a systemic risk allocation of $\rho(\mathbf{X})$ if it fulfills $\sum_{n=1}^{N} \rho^n(\mathbf{X}) = \rho(\mathbf{X})$.*

The requirement $\sum_{n=1}^{N} \rho^n(\mathbf{X}) = \rho(\mathbf{X})$ is known as the *"Full Allocation"* property, see for example [13]. In the case of deterministic allocations $\mathbf{Y} \in \mathbb{R}^N$, i.e. $\mathcal{C} = \mathbb{R}^N$, the optimal deterministic $\mathbf{Y}_{\mathbf{X}}$ represents a canonical risk allocation $\rho^n(\mathbf{X}) := Y_{\mathbf{X}}^n$. For general (random) allocations $\mathbf{Y} \in \mathcal{C} \subset \mathcal{C}_{\mathbb{R}}$, we do not have any more such canonical way to determine $\rho^n(\mathbf{X})$, however we will provide evidence that a good choice is:

$$\rho^n(\mathbf{X}) := \mathbb{E}_{Q_{\mathbf{X}}^n}[Y_{\mathbf{X}}^n] \text{ for } n = 1, \cdots, N, \tag{1.7}$$

where $\mathbf{Q}_{\mathbf{X}}$ is the optimizer of the dual problem (1.5).

To this end, suppose that a probability vector $\mathbf{Q} = (Q^1, \cdots, Q^N)$ is given for the system and consider an alternative formulation of the systemic utility maximization problem in terms of the valuation provided by $\mathbf{Q}$:

$$\pi^{\mathbf{Q}}(\mathbf{X}) = \pi_A^{\mathbf{Q}}(\mathbf{X}) := \sup_{\mathbf{Y} \in \mathcal{L}} \left\{ \mathbb{E}\left[\sum_{n=1}^{N} u_n(X^n + Y^n)\right] \mid \sum_{n=1}^{N} \mathbb{E}_{Q^n}[Y^n] \leq A \right\}. \tag{1.8}$$

Note that in (1.8) (as well as in (1.9)) the allocation $\mathbf{Y}$ belongs to a vector space $\mathcal{L}$ of random variables (introduced later) without requiring that $\mathbf{Y} \in \mathcal{C}_{\mathbb{R}}$ (that is adding up to a deterministic quantity). Thus, for $\pi^{\mathbf{Q}}(\mathbf{X})$ we maximize the expected systemic utility among all $\mathbf{Y} \in \mathcal{L}$ satisfying the budget constraint $\sum_{n=1}^{N} \mathbb{E}_{Q^n}[Y^n] \leq A$.

Similarily, we can introduce a systemic risk measure in terms of the vector of probability measures $\mathbf{Q}$ by

$$\rho^{\mathbf{Q}}(\mathbf{X}) = \rho_B^{\mathbf{Q}}(\mathbf{X}) := \inf_{\mathbf{Y} \in \mathcal{L}} \left\{ \sum_{n=1}^N \mathbb{E}_{Q^n}[Y^n] \mid \mathbb{E}\left[\sum_{n=1}^N u_n(X^n + Y^n)\right] \geq B \right\}, \quad (1.9)$$

For $\rho^{\mathbf{Q}}(\mathbf{X})$, we thus look for the minimal systemic cost $\sum_{n=1}^N \mathbb{E}_{Q^n}[Y^n]$ among all $\mathbf{Y} \in \mathcal{L}$ satisfying the acceptability (utility) constraint $\mathbb{E}\left[\sum_{n=1}^N u_n(X^n + Y^n)\right] \geq B$.

A priori, $\rho$ and $\rho^{\mathbf{Q}}$ defined in (1.4) and (1.9) are quite different objects: even if they both subsume the same systemic budget constraint, $\rho$ is defined only through the computation of the cash amount $\sum_{n=1}^N Y^n \in \mathbb{R}$, while in $\rho^{\mathbf{Q}}$ the risk is defined by calculating the value (or the cost) of the random allocations: $\sum_{n=1}^N \mathbb{E}_{Q^n}[Y^n]$. A similar comparison applies to $\pi$ and $\pi^{\mathbf{Q}}$.

*Remark* 1.3. To better understand such comparison, we make an analogy with the classical (univariate) utility maximization from terminal wealth in securities markets. Let $\mathcal{K} := \{(H.S)_T \mid H \text{ admissible}\}$, where $(H.S)_T$ is the stochastic integral, and let

$$U(x) = \sup \{\mathbb{E}[u(x + K)] \mid K \in \mathcal{K}\}$$

be the utility from the initial wealth $x \in \mathbb{R}$ and from optimally investing in the securities $S$ adopting admissible strategies $H$. In this case, there is no need to introduce a cost operator, as we are investing in replicable contingent claims having, by definition, initial value $x$. On the contrary,

$$U^Q(x) = \sup \{\mathbb{E}[u(x + K)] \mid \mathbb{E}_Q[K] \leq 0\}$$

is the optimal utility function when a probability vector $Q$ is given. A priori, the two problems are of different nature, unless one shows (see i.e. [6]) that for a particular probability measure $Q_x$ the two problems have the same optimal value and:

$$U(x) = U^{Q_x}(x) = \min_{Q \in \mathcal{M}} U^Q(x)$$

where $\mathcal{M}$ is the set of martingale measures. From the mathematical point of view, once the minimax martingale measure $Q_x$ is determined, $U^{Q_x}(x)$ is easier to solve than $U(x)$, and the solution to $U^{Q_x}(x)$ can then be used to find the solution to $U(x)$. Also for the financial application, one may use $Q_x$ to compute the *fair price* (see [21] and Remark 3.2.2 [23]) of a contingent claim $C$ by computing $\mathbb{E}_{Q_x}[C]$.

In view of the analogy in the above remark, in this paper we also prove that

(i) the optimizer $\mathbf{Q_X} = (Q_{\mathbf{X}}^1, \cdots Q_{\mathbf{X}}^N)$ of the dual problem (1.5) satisfies:

$$\rho_B(\mathbf{X}) = \rho_B^{\mathbf{Q_X}}(\mathbf{X}), \quad \pi_A(\mathbf{X}) = \pi_A^{\mathbf{Q_X}}(\mathbf{X});$$

(ii) all these four problems have the same (unique) optimal solution $\mathbf{Y_X}$, when $A := \rho_B(\mathbf{X})$;

(iii) $\mathbf{Q_X}$ provides a systemic risk allocation $(\mathbb{E}_{Q_\mathbf{X}^1}[Y_\mathbf{X}^1], \cdots, (\mathbb{E}_{Q_\mathbf{X}^N}[Y_\mathbf{X}^N])$, with

$$\sum_{n=1}^{N} \mathbb{E}_{Q_{\mathbf{X}^n}}[Y_\mathbf{X}^n] = \rho_B(\mathbf{X}), \tag{1.10}$$

(iv) and

$$\rho_B(\mathbf{X}) = \max_{\mathbf{Q} \in \mathcal{D}} \rho_B^{\mathbf{Q}}(\mathbf{X}) = \rho_B^{\mathbf{Q_X}}(\mathbf{X}),$$

where the domain $\mathcal{D}$ is defined in (3.3) and it replaces, in the analogy with utility maximization, the set of martingale measures.

Hence $\rho_B^{\mathbf{Q_X}}$ is a valid alternative to $\rho_B$ (same value and optimal solution) and this justifies its use to compute the systemic risk. In addition, (1.10) shows that the operator assigned by $\mathbb{E}_{\mathbf{Q_X}}[\cdot]$ evaluates the risk component $Y_\mathbf{X}^n$ of the optimal allocation accordingly to $\rho_B$ (not only to $\rho_B^{\mathbf{Q_X}}$) and proves that the definition in (1.7) provides indeed a systemic risk allocation for $\rho(\mathbf{X})$. In Section 5, we further elaborate on this interpretation, we study in detail the properties of the systemic risk probability vector $\mathbf{Q_X}$ and in particular we provide the formula for the marginal risk contribution:

$$\frac{d}{d\varepsilon}\rho(\mathbf{X}+\varepsilon\mathbf{V})|_{\varepsilon=0} = -\sum_{n=1}^{N} \mathbb{E}_{Q_\mathbf{X}^n}[V^n] \text{ for } \mathbf{V} \in \mathcal{L}.$$

We also discuss certain properties inferred from the above results that argue for the fairness of the systemic risk allocation.

Based on the above exposition, we structure the remaining paper as follows. In Section 2 we introduce the technical setting within Orlicz spaces and the main assumptions, and we show that our optimization problems are well posed. In Section 3 we study the dual representation (1.5) of the systemic risk measure, notably existence and uniqueness of the dual optimizer $\mathbf{Q_X}$ is proved in Proposition 3.1 and Corollary 4.13. In Section 4 we deal with existence and uniqueness of optimal solutions of the primal problems (1.4), (1.6) and (1.8), (1.9). To guarantee existence, we need to enlarge the environment and consider appropriate spaces of integrable random variables. In Section 5 we derive cash additivity and risk marginal contribution properties of the systemic risk measure $\rho(\mathbf{X})$, and fairness properties of the optimal allocations $\rho^n(\mathbf{X})$. The case with *exponential utilities* and *grouping of institutions* will be treated in details in Section 6, where additional sensitivity and monotonicity properties will be established as well.

We conclude this Section with a literature overview on systemic risk. In [20], [12] and [19] one can find empirical studies on banking networks, while interbank lending has been studied via interacting diffusions and mean field approach in several papers like [31], [29], [15], [38], [5]. Among the many contributions on systemic risk modeling, we mention the classical contagion model proposed by [26], the default model of [34], the illiquidity cascade models of [33], [37] and [40], the asset fire sale cascade model by [18] and [14], as well as the model in [46] that additionally includes cross-holdings. Further works on network modeling are [1], [44], [2], [35], [4], [24] and [25]. See also the references therein. For an exhaustive overview on the literature on systemic risk we refer the reader to the recent volumes of [36] and of [30].

## 2 The setting

We now introduce the setting and discuss some fundamental properties of our systemic risk measures. Given a probability space $(\Omega, \mathcal{F}, \mathbb{P})$, we consider the space of random vectors

$$L^0 := L^0(\mathbb{P}; \mathbb{R}^N) := \{\mathbf{X} = (X^1, \ldots, X^N) \mid X^n \in L^0(\Omega, \mathcal{F}, \mathbb{P}; \mathbb{R}),\ n = 1, \cdots, N\}.$$

The measurable space $(\Omega, \mathcal{F})$ will be fixed throughout the paper and will not appear in the notations. Unless we need to specify a different probability, we will also suppress $\mathbb{P}$ from the notations and simply write $L^0(\mathbb{R}^N)$. In addition, we will sometimes suppress $\mathbb{R}^d$, $d = 1, ..., N$, in the notation of the vector spaces, when the dimension of the random vector is clear from the context. We assume that $L^0(\mathbb{R}^N)$ is a vector lattice equipped with componentwise order relation: $\mathbf{X}_1 \geq \mathbf{X}_2$ if $X_1^i \geq X_2^i$ $\mathbb{P}$-a.s. for all $i = 1, \cdots, N$.

When $\mathbf{Q} = (Q^1, ..., Q^N)$ is a vector of probability measures on $(\Omega, \mathcal{F})$, we set $L^1(\mathbf{Q}) := \{\mathbf{X} = (X^1, \ldots, X^N) \mid X^n \in L^1(Q^n),\ n = 1, \cdots, N\}$. Unless differently stated, all inequalities between random vectors are meant to be $\mathbb{P}$-a.s. inequalities.

A vector $\mathbf{X} = (X^1, \ldots, X^N) \in L^0$ denotes a configuration of risky factors at a future time $T$ associated to a system of $N$ entities.

### 2.1 Orlicz setting

We will consider systemic risk measures defined on Orlicz spaces, see [41] for further details on Orlicz spaces. This presents several advantages. From a mathematical point of view, it is a more general setting than $L^\infty$, but at the same time it simplifies the analysis, since the topology is order continuous and there are no singular elements in the dual space. Furthermore, it has been shown in [10] that the Orlicz setting is the natural one to embed utility maximization problems, as the natural integrability condition $\mathbb{E}[u(X)] > -\infty$ is implied by $\mathbb{E}[\phi(X)] < +\infty$. Univariate convex risk measures on Orlicz spaces have been introduced and studied by [17] and [9].

Let $u : \mathbb{R} \to \mathbb{R}$ be a concave and increasing function satisfying $\lim_{x \to -\infty} \frac{u(x)}{x} = +\infty$.

Consider $\phi(x) := -u(-|x|) + u(0)$. Then $\phi : \mathbb{R} \to [0, +\infty)$ is a strict Young function, i.e., it is finite valued, even and convex on $\mathbb{R}$ with $\phi(0) = 0$ and $\lim_{x \to +\infty} \frac{\phi(x)}{x} = +\infty$. The Orlicz space $L^\phi$ and Orlicz Heart $M^\phi$ are respectively defined by

$$L^\phi := \left\{ X \in L^0(\mathbb{R}) \mid \mathbb{E}[\phi(\alpha X)] < +\infty \text{ for some } \alpha > 0 \right\}, \qquad (2.1)$$

$$M^\phi := \left\{ X \in L^0(\mathbb{R}) \mid \mathbb{E}[\phi(\alpha X)] < +\infty \text{ for all } \alpha > 0 \right\}, \qquad (2.2)$$

and they are Banach spaces when endowed with the Luxemburg norm. The topological dual of $M^\phi$ is the Orlicz space $L^{\phi^*}$, where the convex conjugate $\phi^*$ of $\phi$, defined by

$$\phi^*(y) := \sup_{x \in \mathbb{R}} \left\{ xy - \phi(x) \right\}, \; y \in \mathbb{R},$$

is also a strict Young function. Note that

$$\mathbb{E}[u(X)] > -\infty \text{ if } \mathbb{E}[\phi(X)] < +\infty. \qquad (2.3)$$

*Remark* 2.1. It is well known that $L^\infty(\mathbb{P}; \mathbb{R}) \subseteq M^\phi \subseteq L^\phi \subseteq L^1(\mathbb{P}; \mathbb{R})$. In addition, from the Fenchel inequality $xy \leq \phi(x) + \phi^*(y)$ we obtain

$$(\alpha |X|) \left( \lambda \frac{dQ}{d\mathbb{P}} \right) \leq \phi(\alpha |X|) + \phi^* \left( \lambda \frac{dQ}{d\mathbb{P}} \right)$$

for some probability measure $Q \ll \mathbb{P}$, and we immediately deduce that $\frac{dQ}{d\mathbb{P}} \in L^{\phi^*}$ implies $L^\phi \subseteq L^1(Q; \mathbb{R})$.

Given the utility functions $u_1, \cdots, u_N : \mathbb{R} \to \mathbb{R}$, satisfying the above conditions, with associated Young functions $\phi_1, \cdots, \phi_N$, we define

$$\mathcal{L} = M^\Phi := M^{\phi_1} \times \cdots \times M^{\phi_N}, \quad L^\Phi := L^{\phi_1} \times \cdots \times L^{\phi_N}. \qquad (2.4)$$

### 2.2 Assumptions and some properties of $\rho$

We consider systemic risk measures $\rho : M^\Phi \to \mathbb{R} \cup \{\infty\} \cup \{-\infty\}$ with

$$\rho(\mathbf{X}) := \inf_{\mathbf{Y} \in \mathcal{C} \subset \mathcal{C}_\mathbb{R}} \left\{ \sum_{n=1}^N Y^n \mid \mathbb{E}\left[ \sum_{n=1}^N u_n(X^n + Y^n) \right] \geq B \right\}, \qquad (2.5)$$

as in (1.4), where the notation $\mathbb{E}\left[ \sum_{n=1}^N u_n(X^n + Y^n) \right] \geq B$ also means that $\sum_{n=1}^N u_n(X^n + Y^n) \in L^1(\mathbb{P})$ and the linear space $\mathcal{C}_\mathbb{R}$ was introduced in (1.3). Note that there is no loss of generality in assuming that $u_n(0) = 0$ (simply replace $B$ with $B - \sum_{n=1}^N u_n(0)$).

The following are standing assumptions for the rest of the paper.

*Assumption* 2.2.  1. $\mathcal{C}_0 \subseteq \mathcal{C}_\mathbb{R}$ and $\mathcal{C} = \mathcal{C}_0 \cap M^\Phi$ is a convex cone satisfying $\mathbb{R}^N \subseteq \mathcal{C} \subseteq \mathcal{C}_\mathbb{R}$.

2. For all $n = 1, \cdots, N$, $u_n : \mathbb{R} \to \mathbb{R}$ is increasing, strictly concave, differentiable and satisfies the Inada conditions

$$u'_n(-\infty) := \lim_{x \to -\infty} u'_n(x) = +\infty, \quad u'_n(+\infty) := \lim_{x \to +\infty} u'_n(x) = 0.$$

3. $B < \Lambda(+\infty)$, i.e., there exists $\mathbf{M} \in \mathbb{R}^N$ such that $\sum_{n=1}^N u_n(M^n) \geq B$.

4. For all $n = 1, \cdots, N$, it holds that for any probability measure $Q \ll \mathbb{P}$

$$\mathbb{E}\left[v_n\left(\frac{dQ}{d\mathbb{P}}\right)\right] < \infty \quad \text{iff} \quad \mathbb{E}\left[v_n\left(\lambda \frac{dQ}{d\mathbb{P}}\right)\right] < \infty, \quad \forall \lambda > 0,$$

where $v_n(y) := \sup_{x \in \mathbb{R}} \{u_n(x) - xy\}$ denotes the convex conjugate of $u_n$.

Also, from the Fenchel inequality $u_n(X) \leq X \frac{dQ}{d\mathbb{P}} + v_n\left(\frac{dQ}{d\mathbb{P}}\right)$ $\mathbb{P}$ a.s., we immediately deduce that if $X \in L^1(Q)$ and $\mathbb{E}\left[v_n\left(\frac{dQ}{d\mathbb{P}}\right)\right] < \infty$ for some probability measure $Q \ll \mathbb{P}$, then $\mathbb{E}[u_n(X)] < +\infty$. We remark that some further useful properties of the convex conjugate $v_n$ are collected in Lemma A.5.

Item 4 in Assumption 2.2 is related to the Reasonable Asymptotic Elasticity condition on utility functions, which was introduced in [45]. This assumption, even though quite weak (see [8] Section 2.2), is fundamental to guarantee the existence of the optimal solution to classical utility maximization problems (see [45] and [8]). In this paper it is necessary in Section A.3 and for the results of Section 4.

*Remark* 2.3. Note that the duality results presented in Propositions 3.1 and 3.3 hold true even under the following weaker assumptions on the utility functions: For all $n = 1, ..., N$, $u_n$ is increasing, concave and $\lim_{x \to -\infty} \frac{u_n(x)}{x} = +\infty$.

The *domain* of $\rho$ is defined by $dom(\rho) := \{\mathbf{X} \in M^\Phi \mid \rho(\mathbf{X}) < +\infty\}$. The proof of the following proposition, which exploits the behavior of $u_n$ at $-\infty$, is given in Appendix A.1.

**Proposition 2.4.** *(a) For all $\mathbf{X} \in M^\Phi$ we have $\rho(\mathbf{X}) > -\infty$. The map $\rho : M^\Phi \to \mathbb{R} \cup \{+\infty\}$ defined in (2.5) is finitely valued, monotone decreasing, convex, continuous and subdifferentiable on the Orlicz Heart $M^\Phi = dom(\rho)$.*

*(b) Furthermore*

$$\rho(\mathbf{X}) = \rho^=(\mathbf{X}) := \inf\left\{\sum_{n=1}^N Y^n \mid \mathbf{Y} \in \mathcal{C}, \mathbb{E}\left[\sum_{n=1}^N u_n(X^n + Y^n)\right] = B\right\}, \quad \mathbf{X} \in dom(\rho).$$

*If there exists an optimal allocation $\mathbf{Y}_\mathbf{X} = \{Y_\mathbf{X}^n\}_n \in \mathcal{C}_0 \cap M^\Phi$ of $\rho(\mathbf{X})$, then it is unique.*

*Example* 2.5. We complete this subsection by introducing one relevant example for the set of admissible random elements, which we denote by $\mathcal{C}^{(\mathbf{n})}$.

*Definition 2.6.* For $h \in \{1, \cdots, N\}$, let $\mathbf{n} := (n_1, \cdots, n_h) \in \mathbb{N}^h$, with $n_{m-1} < n_m$ for all $m = 1, \cdots, h$, $n_0 := 0$ and $n_h := N$. We set $I_m := \{n_{m-1}+1, \cdots, n_m\}$ for each $m = 1, \cdots, h$. We introduce the following family of allocations $\mathcal{C}^{(\mathbf{n})} = \mathcal{C}_0^{(\mathbf{n})} \cap M^\Phi$, where

$$\mathcal{C}_0^{(\mathbf{n})} = \left\{ \mathbf{Y} \in L^0(\mathbb{R}^N) \mid \exists\, d = (d_1, \cdots, d_h) \in \mathbb{R}^h : \sum_{i \in I_m} Y^i = d_m \text{ for } m = 1, \cdots, h \right\} \subseteq \mathcal{C}_\mathbb{R}. \quad (2.6)$$

Definition 2.6 models a cluster $C = (C_1, \cdots, C_h)$ of financial institutions, which is a partition of $\{X^1, \cdots, X^N\}$. The constraint on $\mathbf{Y}$ is then that the components of $\mathbf{Y}$ must sum up to a real number in each element $C_i$ of the cluster, i.e., $\sum_{j \in C_i} Y^j \in \mathbb{R}$.

For a given $\mathbf{n} := (n_1, \cdots, n_h)$, the values $(d_1, \cdots, d_h)$ may change, but the number of elements in each of the $h$ groups $I_m$ is fixed by $\mathbf{n}$. It is then easily seen that $\mathcal{C}^{(\mathbf{n})}$ is a linear space containing $\mathbb{R}^N$ and closed with respect to convergence in probability. We point out that the family $\mathcal{C}^{(\mathbf{n})}$ admits two extreme cases:

(i) the strongest restriction occurs when $h = N$, i.e. we consider exactly $N$ groups, and in this case $\mathcal{C}^{(\mathbf{n})} = \mathbb{R}^N$ corresponds to the deterministic case;

(ii) on the opposite side, we have only one group $h = 1$ and $\mathcal{C}^{(\mathbf{n})} = \mathcal{C}_\mathbb{R} \cap M^\Phi$ is the largest possible class, corresponding to arbitrary random injection $\mathbf{Y} \in M^\Phi$ with the only constraint $\sum_{n=1}^N Y^n \in \mathbb{R}$.

## 3 Dual representation of $\rho$

We now investigate the dual representation of systemic risk measures of the form (2.5). When $\mathbf{Z} \in M^\Phi$ and $\xi \in L^{\Phi^*}$, we set $\mathbb{E}[\xi \mathbf{Z}] := \sum_{n=1}^N \mathbb{E}[\xi^n Z^n]$ and, for $\frac{d\mathbf{Q}}{d\mathbb{P}} \in L_+^{\Phi^*}$, $\mathbb{E}_\mathbf{Q}[\mathbf{Z}] = \sum_{n=1}^N \mathbb{E}_{Q^n}[Z^n]$. We will frequently identify the density $\frac{dQ}{d\mathbb{P}}$ with the associated probability measure $Q \ll \mathbb{P}$.

**Proposition 3.1.** *For any $\mathbf{X} \in M^\Phi$,*

$$\rho_B(\mathbf{X}) = \max_{\mathbf{Q} \in \mathcal{D}} \left\{ \sum_{n=1}^N \mathbb{E}_{Q^n}[-X^n] - \alpha_B(\mathbf{Q}) \right\}, \quad (3.1)$$

*where the penalty function is given by*

$$\alpha_B(\mathbf{Q}) := \sup_{\mathbf{Z} \in \mathcal{A}} \left\{ \sum_{n=1}^N \mathbb{E}_{Q^n}[-Z^n] \right\}, \quad (3.2)$$

*with* $\mathcal{A} := \left\{ \mathbf{Z} \in M^\Phi \mid \sum_{n=1}^N \mathbb{E}[u_n(Z^n)] \geq B \right\}$ *and*

$$\mathcal{D} := dom(\alpha_B) \cap \left\{ \frac{d\mathbf{Q}}{d\mathbb{P}} \in L_+^{\Phi^*} \mid Q^n(\Omega) = 1\ \forall n \text{ and } \sum_{n=1}^N (\mathbb{E}_{Q^n}[Y^n] - Y^n) \leq 0 \text{ for all } \mathbf{Y} \in \mathcal{C}_0 \cap M^\Phi \right\},$$
(3.3)

*where* $dom(\alpha_B) := \{\, \mathbf{Q} = (Q^1, \cdots, Q^N) \mid Q^n << P\ \forall n \text{ and } \alpha_B(\mathbf{Q}) < +\infty\}$.

(i) Suppose that for some $i, j \in \{1, \cdots, N\}$, $i \neq j$, we have $\pm(e_i 1_A - e_j 1_A) \in \mathcal{C}$ for all $A \in \mathcal{F}$. Then

$$\mathcal{D} = dom(\alpha_B) \cap \left\{ \frac{d\mathbf{Q}}{d\mathbb{P}} \in L_+^{\Phi^*} \mid Q^n(\Omega) = 1 \; \forall n, \; Q^i = Q^j \text{ and } \sum_{n=1}^{N}(\mathbb{E}_{Q^n}[Y^n] - Y^n) \leq 0 \text{ for all } \mathbf{Y} \in \mathcal{C} \right\}.$$

(ii) Suppose that $\pm(e_i 1_A - e_j 1_A) \in \mathcal{C}$ for all $i, j$ and all $A \in \mathcal{F}$. Then

$$\mathcal{D} = dom(\alpha_B) \cap \left\{ \frac{d\mathbf{Q}}{d\mathbb{P}} \in L_+^{\Phi^*} \mid Q^n(\Omega) = 1, \; Q^n = Q, \; \forall n \right\}.$$

*Proof.* The dual representation (3.1) is a consequence of Proposition 2.4, Theorem A.2 and of Propositions 3.9 and 3.11 in [32], taking into consideration that $\mathcal{C}$ is a convex cone, the dual space of the Orlicz Heart $M^\Phi$ is the Orlicz space $L^{\Phi^*}$ and $M^\Phi = dom(\rho)$. Note that from Theorem A.2 we know that the dual elements $\xi \in L_+^{\Phi^*}$ are positive but a priori not normalized. However, we obtain $\mathbb{E}[\xi^n] = 1$ by taking as $\mathbf{Y} = \pm e_j \in \mathbb{R}^N$, and using $\sum_{n=1}^{N}(\xi^n(Y^n) - Y^n) \leq 0$ for all $\mathbf{Y} \in \mathcal{C}$, so that $\xi^j(1) - 1 \leq 0$ and $\xi^j(-1) + 1 \leq 0$ imply $\xi^j(1) = 1$. This shows the form of the domain $\mathcal{D}$ in (3.3). Furthermore:

(i) Take $\mathbf{Y} := e_i 1_A - e_j 1_A \in \mathcal{C}$. From $\sum_{n=1}^{N}(Q^n(Y^n) - Y^n) \leq 0$ we obtain $Q^i(1_A) - 1_A + Q^j(-1_A) + 1_A \leq 0$, i.e., $Q^i(A) - Q^j(A) \leq 0$ and similarly taking $\mathbf{Y} := -e_i 1_A + e_j 1_A \in \mathcal{C}$, we get $Q^j(A) - Q^i(A) \leq 0$.

(ii) From (i), we obtain $Q^i = Q^j$. In addition, we get $\sum_{n=1}^{N}(\mathbb{E}_Q[Y^n] - Y^n) = \mathbb{E}_Q[\sum_{n=1}^{N} Y^n] - \sum_{n=1}^{N} Y^n) = 0$, as $\sum_{n=1}^{N} Y^n \in \mathbb{R}$.

□

Proposition 3.1 guarantees the existence of a maximizer $\mathbf{Q_X}$ to the dual problem (3.1) and that $\alpha_B(\mathbf{Q_X}) < +\infty$. Uniqueness will be proved in Corollary 4.13.

**Definition 3.2.** *Let $\mathbf{X} \in M^\Phi$. An optimal solution of the dual problem (3.1) is a vector of probability measures $\mathbf{Q_X} = (Q_\mathbf{X}^1, \cdots, Q_\mathbf{X}^N)$ verifying $\frac{d\mathbf{Q_X}}{d\mathbb{P}} \in \mathcal{D}$ and*

$$\rho_B(\mathbf{X}) = \sum_{n=1}^{N} \mathbb{E}_{Q_\mathbf{X}^n}[-X^n] - \alpha_B(\mathbf{Q_X}). \tag{3.4}$$

A vector $\mathbf{Q}$ of probability measures having density in $\mathcal{D}$ could be viewed, in the systemic $N$-dimensional one period setting, as the counterpart of the notion of ($\mathbb{P}$-absolutely continuous) martingale measures. Indeed, as $\mathbf{Y} \in \mathcal{C}_0 \subseteq \mathcal{C}_\mathbb{R}$, $\sum_{n=1}^{N} Y^n \in \mathbb{R}$ is the total amount to be allocated to the $N$ institutions and then the total cost or value $\sum_{n=1}^{N} \mathbb{E}_{Q^n}[Y^n]$ should at most be equal to $\sum_{n=1}^{N} Y^n$, for any "fair" valuation operator $\mathbb{E}_\mathbf{Q}[\cdot]$, which is the case if $\frac{d\mathbf{Q}}{d\mathbb{P}} \in \mathcal{D}$.

There exists a simple relation among $\rho_B$, $\rho_B^{\mathbf{Q}}$ and $\alpha_B(\mathbf{Q})$ defined in (2.5), (1.9), and (3.2), respectively.

**Proposition 3.3.** *We have*

$$\rho_B^{\mathbf{Q}}(\mathbf{X}) = -\sum_{n=1}^{N} \mathbb{E}_{Q^n}[X^n] - \alpha_B(\mathbf{Q}), \tag{3.5}$$

*and*

$$\rho_B(\mathbf{X}) = \max_{\frac{d\mathbf{Q}}{d\mathbb{P}} \in \mathcal{D}} \rho_B^{\mathbf{Q}}(\mathbf{X}) = \rho_B^{\mathbf{Q_X}}(\mathbf{X}), \tag{3.6}$$

*where $\mathbf{Q_X}$ is an optimal solution of the dual problem (3.1).*

*Proof.* We have

$$\begin{aligned}
-\alpha_B(\mathbf{Q}) &= \inf\left\{\sum_{n=1}^{N} \mathbb{E}_{Q^n}[Z^n] \mid \mathbf{Z} \in M^{\Phi} \text{ and } \sum_{n=1}^{N} \mathbb{E}[u_n(Z^n)] \geq B\right\} \\
&= \inf\left\{\sum_{n=1}^{N} \mathbb{E}_{Q^n}[X^n + Y^n] \mid \mathbf{Y} \in M^{\Phi} \text{ and } \sum_{n=1}^{N} \mathbb{E}[u_n(X^n + Y^n)] \geq B\right\} \\
&= \sum_{n=1}^{N} \mathbb{E}_{Q^n}[X^n] + \rho_B^{\mathbf{Q}}(\mathbf{X}),
\end{aligned}$$

which proves (3.5). Then from (3.5) and (3.4) we deduce

$$\rho_B^{\mathbf{Q_X}}(\mathbf{X}) = -\sum_{n=1}^{N} \mathbb{E}_{Q_{\mathbf{X}}^n}[X^n] - \alpha_B(\mathbf{Q_X}) = \rho_B(\mathbf{X})$$

and from (3.1) and (3.5) $\rho_B(\mathbf{X}) = \max_{\mathbf{Q} \in \mathcal{D}} \rho_B^{\mathbf{Q}}(\mathbf{X})$. $\square$

**Proposition 3.4.** *When $\alpha_B(\mathbf{Q}) < +\infty$, then the penalty function in (3.2) can be written as*

$$\alpha_B(\mathbf{Q}) := \sup_{\mathbf{Z} \in \mathcal{A}} \left\{\sum_{n=1}^{N} \mathbb{E}_{Q^n}[-Z^n]\right\} = \inf_{\lambda > 0}\left(-\frac{1}{\lambda}B + \frac{1}{\lambda}\sum_{n=1}^{N} \mathbb{E}\left[v_n\left(\lambda \frac{dQ^n}{d\mathbb{P}}\right)\right]\right), \tag{3.7}$$

*and $\mathbb{E}\left[v_n\left(\lambda \frac{dQ^n}{d\mathbb{P}}\right)\right] < \infty$ for all $n$ and all $\lambda > 0$. In addition, the infimum is attained in (3.7), i.e.,*

$$\alpha_B(\mathbf{Q}) = \sum_{n=1}^{N} \mathbb{E}\left[\frac{dQ^n}{d\mathbb{P}} v_n'\left(\lambda^* \frac{dQ^n}{d\mathbb{P}}\right)\right], \tag{3.8}$$

*where $\lambda^* > 0$ is the unique solution of the equation*[1]

$$-B + \sum_{n=1}^{N} \mathbb{E}\left[v_n\left(\lambda \frac{dQ^n}{d\mathbb{P}}\right)\right] - \lambda \sum_{n=1}^{N} \mathbb{E}\left[\frac{dQ^n}{d\mathbb{P}} v_n'\left(\lambda \frac{dQ^n}{d\mathbb{P}}\right)\right] = 0. \tag{3.9}$$

---
[1]Note that $\lambda^*$ will depend on $B$, $(u_n)_{n=1,\cdots,N}$ and $\left(\frac{dQ_n}{d\mathbb{P}}\right)_{n=1,\cdots,N}$.

*Proof.* In Appendix A.2.1. □

*Example* 3.5. Consider the grouping of Example 2.5. As $\mathcal{C}^{(\mathbf{n})}$ is a linear space containing $\mathbb{R}^N$, the dual representation (3.1) applies. In addition in each group we have $\pm(e_i 1_A - e_j 1_A) \in \mathcal{C}^{(\mathbf{n})}$ for all $i,j$ *in the same group* and for all $A \in \mathcal{F}$. Therefore, in each group the components $Q^i$, $i \in I_m$, of the dual elements are all the same, i.e., $Q^i = Q^j$, for all $i, j \in I_m$, and the representation (3.1) becomes

$$\rho_B(\mathbf{X}) = \max_{\mathbf{Q} \in \mathcal{D}} \left\{ \sum_{m=1}^{h} \sum_{k \in I_m} \left( \mathbb{E}_{Q^m}[-X^k] \right) - \alpha_B(\mathbf{Q}) \right\} = \max_{\mathbf{Q} \in \mathcal{D}} \left\{ \sum_{m=1}^{h} \mathbb{E}_{Q^m}[-\overline{X}_m] - \alpha_B(\mathbf{Q}) \right\}, \tag{3.10}$$

with

$$\mathcal{D} := dom(\alpha_B) \cap \left\{ \frac{d\mathbf{Q}}{d\mathbb{P}} \in L_+^{\Phi^*} \mid Q^i = Q^j \,\forall i,j \in I_m, Q^i(\Omega) = 1 \right\} \tag{3.11}$$

and $\overline{X}_m := \sum_{k \in I_m} X^k$. Indeed,

$$\sum_{n=1}^{N}(\mathbb{E}_{Q^n}[Y^n] - Y^n) = \sum_{m=1}^{h} \sum_{k \in I_m} (\mathbb{E}_{Q^m}[Y^k] - Y^k) = \sum_{m=1}^{h} \left( \mathbb{E}_{Q^m}\left[\sum_{k \in I_m} Y^k\right] - \sum_{k \in I_m} Y^k \right) = 0,$$

as $\sum_{k \in I_m} Y^k = d_m \in \mathbb{R}$. If we have only one single group, all components of a dual element $\mathbf{Q} \in \mathcal{D}$ are the same.

When $\mathbf{Q} = (Q^1, \cdots, Q^n)_{n=1,\cdots,N} \in \mathcal{D}$, defined in (3.11), then $(\mathbb{E}_{\mathbf{Q}_1}[Y_{\mathbf{X}}^1], ..., \mathbb{E}_{\mathbf{Q}_N}[Y_{\mathbf{X}}^N])$ is a systemic risk allocation as in Definition (1.1), i.e.,

$$\sum_{n=1}^{N} \mathbb{E}_{Q^n}[Y_{\mathbf{X}}^n] = \sum_{m=1}^{h} \sum_{k \in I_m} \mathbb{E}_{Q^m}[Y_{\mathbf{X}}^k] = \sum_{m=1}^{h} \mathbb{E}_{Q^m}\left[\sum_{k \in I_m} Y_{\mathbf{X}}^k\right] = \sum_{m=1}^{h} d_m = \rho(\mathbf{X}). \tag{3.12}$$

*Example* 3.6. Consider $u_n : \mathbb{R} \to \mathbb{R}$, $u_n(x) = -e^{-\alpha_n x}/\alpha_n$, $\alpha_n > 0$ for each $n$, and let $B < 0$. Then, $v'_n(y) = \frac{1}{\alpha_n} \ln(y)$. From the first order condition (3.9) we obtain that the minimizer is $\lambda^* = -\frac{B}{\beta}$, with $\beta := \sum_{n=1}^{N} \frac{1}{\alpha_n}$. Therefore, from (3.8) we have

$$\alpha_B(\mathbf{Q}) = \sum_{n=1}^{N} \mathbb{E}\left[\frac{dQ^n}{d\mathbb{P}} v'_n\left(\lambda^* \frac{dQ^n}{d\mathbb{P}}\right)\right] = \sum_{n=1}^{N} \frac{1}{\alpha_n} \left( H(Q^n, \mathbb{P}) + \ln\left(-\frac{B}{\beta}\right) \right), \tag{3.13}$$

where $H(Q^n, \mathbb{P}) := \mathbb{E}\left[\frac{dQ^n}{d\mathbb{P}} \ln\left(\frac{dQ^n}{d\mathbb{P}}\right)\right]$ is the relative entropy.

# 4 Existence of the optimal solutions

In this section we deal with existence and uniqueness of optimal allocations for $\rho_B(\mathbf{X})$ and the other related primal optimization problems introduced in Section 1. **Throughout this entire section**, we assume $\mathbf{X} \in M^\Phi$ and that $\mathbf{Q} = (Q^1, ..., Q^N)$ satisfies $Q^n \ll \mathbb{P}$, $\frac{d\mathbf{Q}}{d\mathbb{P}} \in L^{\Phi^*}$ and $\alpha_B(\mathbf{Q}) < +\infty$, or equivalently $\rho_B^{\mathbf{Q}}(\mathbf{X}) > -\infty$. Recall from Proposition 3.4 that this implies $\mathbb{E}\left[v_n\left(\lambda \frac{dQ^n}{dP}\right)\right] < +\infty$ for all $n$ and all $\lambda > 0$. Set

$$L^1(\mathbb{P}; \mathbf{Q}) := (L^1(\mathbb{P}; \mathbb{R}^N) \cap L^1(\mathbf{Q}; \mathbb{R}^N)) \supseteq L^{\Phi} \supseteq M^{\Phi}, \quad (4.1)$$

where the inclusions follow from Remark 2.1 and $\frac{d\mathbf{Q}}{d\mathbb{P}} \in L^{\Phi^*}$.

W.l.o.g. we may assume that $u_i(0) = 0$, $1 \leq i \leq N$ and observe that then

$$u_i(x_i) = u_i(x_i^+) + u_i(-x_i^-). \quad (4.2)$$

When the utility functions $u_n$ are of exponential type, the Orlicz Heart $M^{\Phi}$ is sufficiently large and contains the optimal allocation $\mathbf{Y_X}$ to $\rho_B(\mathbf{X})$, see Section 6. This of course also happens in the case of general utility functions on a finite probability space.

As shown in Section 4.3, in general, we cannot expect to find the optimal solution $\mathbf{Y_Q}$ for the problem $\rho_B^{\mathbf{Q}}(\mathbf{X})$ in the space $M^{\Phi}$, but only in the larger space $L^1(\mathbf{Q})$ and this motivates the introduction of several extended problems.

Let $B \in \mathbb{R}$ and define

$$\widetilde{\rho}_B^{\mathbf{Q}}(\mathbf{X}) \; := \; \inf_{\mathbf{Y} \in L^1(\mathbb{P}; \mathbf{Q})} \left\{ \sum_{n=1}^{N} \mathbb{E}_{Q^n}[Y^n] \;\Big|\; \mathbb{E}\left[\sum_{n=1}^{N} u_n(X^n + Y^n)\right] \geq B \right\},$$

$$\widehat{\rho}_B^{\mathbf{Q}}(\mathbf{X}) \; := \; \inf_{\mathbf{Y} \in L^1(\mathbf{Q})} \left\{ \sum_{n=1}^{N} \mathbb{E}_{Q^n}[Y^n] \;\Big|\; \mathbb{E}\left[\sum_{n=1}^{N} u_n(X^n + Y^n)\right] \geq B \right\},$$

$$\widetilde{\rho}_B(\mathbf{X}) \; := \; \inf_{\mathbf{Y} \in \mathcal{C}_0 \cap L^1(\mathbb{P}; \mathbf{Q_X})} \left\{ \sum_{n=1}^{N} Y^n \;\Big|\; \mathbb{E}\left[\sum_{n=1}^{N} u_n(X^n + Y^n)\right] \geq B \right\}.$$

Analogously, we define $\widetilde{\pi}_A^{\mathbf{Q}}(\mathbf{X})$, $\widehat{\pi}_A^{\mathbf{Q}}(\mathbf{X})$ and $\widetilde{\pi}_A(\mathbf{X})$ for $A \in \mathbb{R}$ by using optimization (1.8). We will show in (4.9) and (4.10), that these extensions from $M^{\Phi}$ to integrable random variables do not change the optimal values.

In order to prove the existence of the optimal allocation for $\widetilde{\rho}_B(\mathbf{X})$ we will proceed in several steps. In Theorem 4.10 we first prove the existence of the optimal solution $\widehat{\mathbf{Y}}_{\mathbf{Q}} \in L^1(\mathbf{Q})$ for $\widehat{\rho}_B^{\mathbf{Q}}(\mathbf{X})$. Then in Proposition 4.11 we show that the optimizer to $\rho_B(\mathbf{X})$ or to $\widetilde{\rho}_B(\mathbf{X})$, when it exists, coincides with $\widehat{\mathbf{Y}}_{\mathbf{Q_X}} \in L^1(\mathbf{Q_X})$. The next key step is to show the existence of $\mathbf{Y} \in L^1(\mathbb{P})$ which is, as specified in Theorem 4.14, the candidate solution to the extended problem and then to prove that $\mathbf{Y} \in L^1(\mathbf{Q_X})$. In a final step (see Theorem 4.19, Proposition 4.22 and Corollary 4.23) we prove that $\rho_B(\mathbf{X}) = \widetilde{\rho}_B(\mathbf{X})$ and that such $\mathbf{Y} \in L^1(\mathbb{P}; \mathbf{Q_X})$, hereafter denoted with $\widetilde{\mathbf{Y}}_{\mathbf{X}}$, is the optimizer of the extended problem $\widetilde{\rho}_B(\mathbf{X})$, and hence it coincides with $\widehat{\mathbf{Y}}_{\mathbf{Q_X}}$.

### 4.1 On $\rho_B(\mathbf{X})$ and $\pi_A(\mathbf{X})$.

Recall that under Assumptions 2.2, $\mathcal{C}$ is a convex cone and therefore, if $\mathbf{Y} \in \mathcal{C}$, then $\mathbf{Y} + \delta \in \mathcal{C}$ for every deterministic $\delta \in \mathbb{R}^N$. Note that $\rho_B^{\mathbf{Q}}(\mathbf{X}) < +\infty$ and $\pi_A^{\mathbf{Q}}(\mathbf{X}) > -\infty$.

**Proposition 4.1.** *(a) $B = \pi_A(\mathbf{X})$ if and only if $A = \rho_B(\mathbf{X})$.*

*(b) If $B = \widetilde{\pi}_A(\mathbf{X})$ then $A = \widetilde{\rho}_B(\mathbf{X})$.*

*(c) In case $A = \rho_B(\mathbf{X})$ and there exists an optimal solution of one of the two problems $\pi_A(\mathbf{X})$ or $\rho_B(\mathbf{X})$, then it is the unique optimal solution of both problems.*

*Proof.* (a) ($\Leftarrow$) Let $A = \rho_B(\mathbf{X})$ and suppose first that $\pi_A(\mathbf{X}) > B$. Then there must exist $\widetilde{\mathbf{Y}} \in \mathcal{C}_0 \cap M^\Phi$ such that $\sum_{n=1}^N \widetilde{Y}^n \leq A$ and $\mathbb{E}\left[\sum_{n=1}^N u_n(X^n + \widetilde{Y}^n)\right] > B$. The continuity of $u_n$ and $\mathbb{E}[u_n(Z^n)] > -\infty$ for all $\mathbf{Z} \in M^\Phi$ imply that there exists $\varepsilon > 0$ and $\widehat{\mathbf{Y}} := \widetilde{\mathbf{Y}} - \varepsilon \mathbf{1} \in \mathcal{C}_0 \cap M^\Phi$ such that $\mathbb{E}\left[\sum_{n=1}^N u_n(X^n + \widehat{Y}^n)\right] \geq B$ and $\sum_{n=1}^N \widehat{Y}^n < A$. This is in contradiction with $A = \rho_B(\mathbf{X})$.

Suppose now that $\pi_A(\mathbf{X}) < B$. Then there exists $\delta > 0$ such that

$$\mathbb{E}\left[\sum_{n=1}^N u_n(X^n + Y^n)\right] \leq B - \delta,$$

for all $\mathbf{Y} \in \mathcal{C}_0 \cap M^\Phi$ such that $\sum_{n=1}^N Y^n \leq A$. As $A = \rho_B(\mathbf{X})$, for all $\varepsilon > 0$, there exists $\mathbf{Y}_\varepsilon \in \mathcal{C}_0 \cap M^\Phi$ such that $\sum_{n=1}^N Y_\varepsilon^n \leq A + \varepsilon$ and $\mathbb{E}\left[\sum_{n=1}^N u_n(X^n + Y_\varepsilon^n)\right] \geq B$. For any $\eta \geq \varepsilon \geq \sum_{n=1}^N Y_\varepsilon^n - A$, we get $\sum_{n=1}^N (Y_\varepsilon^n - \frac{\eta}{N}) \leq A + \varepsilon - \eta \leq A$. By the continuity of $u_n$ and $\mathbb{E}[u_n(Z^n)] > -\infty$ for all $\mathbf{Z} \in M^\Phi$, we may select $\varepsilon > 0$ and $\eta \geq \varepsilon$ small enough so that $\mathbb{E}\left[\sum_{n=1}^N u_n(X^n + Y_\varepsilon^n - \frac{\eta}{N})\right] > B - \delta$. As $\widehat{\mathbf{Y}} := (Y_\varepsilon^n - \frac{\eta}{N})_n \in \mathcal{C}_0 \cap M^\Phi$, we obtain a contradiction.

($\Rightarrow$) Let $B = \pi_A(\mathbf{X})$ and suppose first that $\rho_B(\mathbf{X}) < A$. Then, there must exist $\widetilde{\mathbf{Y}} \in \mathcal{C}_0 \cap M^\Phi$ such that $\mathbb{E}\left[\sum_{n=1}^N u_n(X^n + \widetilde{Y}^n)\right] \geq B$ and $\sum_{n=1}^N \widetilde{Y}^n < A$. Then, there exists $\varepsilon > 0$ and $\widehat{\mathbf{Y}} := \widetilde{\mathbf{Y}} + \varepsilon \mathbf{1} \in \mathcal{C}_0 \cap M^\Phi$ such that $\sum_{n=1}^N \widehat{Y}^n \leq A$ and $\mathbb{E}\left[\sum_{n=1}^N u_n(X^n + \widehat{Y}^n)\right] > B$. This is in contradiction with $B = \pi_A(\mathbf{X})$.

Suppose now that $\rho_B(\mathbf{X}) > A$. Then, there exists $\delta > 0$ such that $\sum_{n=1}^N Y^n \geq A + \delta$ for all $\mathbf{Y} \in \mathcal{C}_0 \cap M^\Phi$ such that $\mathbb{E}\left[\sum_{n=1}^N u_n(X^n + Y^n)\right] \geq B$. As $B = \pi_A(\mathbf{X})$, for all $\varepsilon > 0$ there exists $\mathbf{Y}_\varepsilon \in \mathcal{C}_0 \cap M^\Phi$ such that $\mathbb{E}\left[\sum_{n=1}^N u_n(X^n + Y_\varepsilon^n)\right] > B - \varepsilon$ and $\sum_{n=1}^N Y_\varepsilon^n \leq A$. Define $\eta_\varepsilon := \inf\left\{a > 0 : \mathbb{E}\left[\sum_{n=1}^N u_n(X^n + Y_\varepsilon^n + \frac{a}{N})\right] \geq B\right\}$ and note that $\eta_\varepsilon \downarrow 0$ if $\varepsilon \downarrow 0$. Select $\varepsilon > 0$ such that $\eta_\varepsilon < \delta$. Then, for any $0 < \beta < \delta - \eta_\varepsilon$ we have

$$\mathbb{E}\left[\sum_{n=1}^N u_n(X^n + Y_\varepsilon^n + \frac{\eta_\varepsilon + \beta}{N})\right] \geq B,$$

and $\sum_{n=1}^N (Y_\varepsilon^n + \frac{\eta_\varepsilon + \beta}{N}) \leq A + \eta_\varepsilon + \beta < A + \delta$. As $(Y_\varepsilon^n + \frac{\eta_\varepsilon + \beta}{N}) \in \mathcal{C}_0 \cap M^\Phi$, we obtain a contradiction.

(b) The implication $B = \widetilde{\pi}_A(\mathbf{X}) \Rightarrow A = \widetilde{\rho}_B(\mathbf{X})$ follows exactly in the same way of the implication ($\Rightarrow$) in (a), replacing $M^\Phi$ with $L^1(\mathbb{P}; \mathbf{Q_X})$.

(c) Suppose that there exists $\mathbf{Y} \in \mathcal{C}_0 \cap M^\Phi$ that is the optimal solution of problem (1.4). As $A := \rho_B(\mathbf{X})$, then $\sum_{n=1}^N Y^n = A$ and the constraint in problem (1.6) is fulfilled for $\mathbf{Y}$. By (a), $B = \pi_A(\mathbf{X}) \geq \mathbb{E}\left[\sum_{n=1}^N u_n(X^n + Y^n)\right] \geq B$ and we deduce that $\mathbf{Y}$ is an optimal solution of problem (1.6). Suppose that there exists $\mathbf{Y} \in \mathcal{C}_0 \cap M^\Phi$ that is the optimal solution of problem (1.6) and set $B := \pi_A(\mathbf{X})$. Then $\mathbb{E}\left[\sum_{n=1}^N u_n(X^n + Y^n)\right] = B$ and the constraint in problem (1.4) is fulfilled for $\mathbf{Y}$. By (a), $A = \rho_B(\mathbf{X}) \leq \sum_{n=1}^N Y^n \leq A$ and we deduce that $\mathbf{Y}$ is an optimal solution of problem (1.4). As $\rho_B(\mathbf{X})$ admits at most one solution by Proposition 2.4, the same must be true for $\pi_A(\mathbf{X})$. $\square$

**Proposition 4.2.** (a) $B = \pi_A^{\mathbf{Q}}(\mathbf{X})$ if and only if $A = \rho_B^{\mathbf{Q}}(\mathbf{X})$. (b) If $B = \widetilde{\pi}_A^{\mathbf{Q}}(\mathbf{X})$ then $A = \widetilde{\rho}_B^{\mathbf{Q}}(\mathbf{X})$, similarly for $\widehat{\pi}_A^{\mathbf{Q}}$ and $\widehat{\rho}_B^{\mathbf{Q}}$. (c) When $A = \rho_B^{\mathbf{Q}}(\mathbf{X})$ and $B = \pi_A^{\mathbf{Q}}(\mathbf{X})$, and there exists an optimal solution of one of the two problems $\pi_A^{\mathbf{Q}}(\mathbf{X})$ or $\rho_B^{\mathbf{Q}}(\mathbf{X})$, then it is the unique optimal solution of both problems. (d) In item (c) we may replace $\pi_A^{\mathbf{Q}}, \rho_B^{\mathbf{Q}}$ with $\widetilde{\pi}_A^{\mathbf{Q}}, \widetilde{\rho}_B^{\mathbf{Q}}$ or with $\widehat{\pi}_A^{\mathbf{Q}}, \widehat{\rho}_B^{\mathbf{Q}}$.

*Proof.* Use step by step the same arguments in the proof of Proposition 4.1 replacing $\sum_{n=1}^N Y^n$ with $\sum_{n=1}^N \mathbb{E}_{Q^n}[Y^n]$. The uniqueness in Item (c) is a consequence of Remark 4.9. $\square$

When using $\mathbf{Q} = \mathbf{Q}_{\mathbf{X}}$, we already proved that $\rho_B(\mathbf{X}) = \rho_B^{\mathbf{Q}_{\mathbf{X}}}(\mathbf{X})$. Similarly:

**Corollary 4.3.** *Let* $A := \rho_B(\mathbf{X})$. *Then* $\pi_A(\mathbf{X}) = \pi_A^{\mathbf{Q}_{\mathbf{X}}}(\mathbf{X})$.

*Proof.* As $A = \rho_B(\mathbf{X}) \in \mathbb{R}$, then by Proposition 3.3 $A = \rho_B(\mathbf{X}) = \rho_B^{\mathbf{Q}_{\mathbf{X}}}(\mathbf{X})$. By Proposition 4.1 (a), respectively Proposition 4.2 (a), we deduce: $B = \pi_A(\mathbf{X})$, resp. $B = \pi_A^{\mathbf{Q}_{\mathbf{X}}}(\mathbf{X})$, hence $\pi_A(\mathbf{X}) = \pi_A^{\mathbf{Q}_{\mathbf{X}}}(\mathbf{X})$. $\square$

### 4.2 On the optimal values

The main contribution of this section is to show that the optimal values coincide, see (4.9) and (4.10), and that, see (4.12)

$$\pi_A^{\mathbf{Q}}(\mathbf{X}) = \max_{\sum_{n=1}^N a^n = A} \sum_{n=1}^N U_n(a^n), \ A \in \mathbb{R},$$

where

$$U_n(a^n) := \sup\left\{\mathbb{E}\left[u_n(X^n + W)\right] \mid W \in M^{\phi_n}, \ \mathbb{E}_{Q^n}[W] \leq a^n\right\} \quad (4.3)$$

and $\mathbf{a} \in \mathbb{R}^N$. In the sequel we write $U_n^{Q^n}(a^n)$, when we need to emphasize the dependence on $Q^n$. Note that $\mathbb{E}[u_n(X^n + W)] \leq u_n(\mathbb{E}[X^n + W]) < +\infty$ for all $X^n, W \in M^{\phi_n} \subseteq L^1(\mathbb{P}; \mathbb{R})$. The conditions $X^n, W \in M^{\phi_n}$ imply that $\mathbb{E}[u_n(X^n + W)] > -\infty$, from which it follows that $U_n(a^n) > -\infty$. As $\frac{d\mathbf{Q}}{dP} \in L^{\Phi^*}$, then $W \in M^{\phi_n}$ implies $W \in L^1(Q^n)$ and the problem (4.3) is well posed. Due to the monotonicity and concavity of $u_n$, $U_n$ is monotone increasing, concave and continuous on $\mathbb{R}$ and we may replace, in the definition of $U_n$, the inequality with the equality sign. However, in general the optimal solution to (4.3) will only exist on a larger domain, as suggested by the well known result reported in Proposition A.6. This leads to introduce the auxiliary problems:

$$\widehat{U}_n(a^n) := \sup\left\{\mathbb{E}\left[u_n(X^n + W)\right] \mid W \in L^1(Q^n),\, \mathbb{E}_{Q^n}[W] \leq a^n\right\},$$
$$\widetilde{U}_n(a^n) := \sup\left\{\mathbb{E}\left[u_n(X^n + W)\right] \mid W \in L^1(\mathbb{P}, Q^n),\, \mathbb{E}_{Q^n}[W] \leq a^n\right\}, \quad (4.4)$$

where $L^1(\mathbb{P}, Q^n)$ is defined as in (4.1). The following proposition is the multi-dimensional version of well known utility maximization problems. Its proof is based on the extended Namioka-Klee Theorem and is deferred to the Apppendix A.4.

**Proposition 4.4.** *We have that*

$$U_n(a^n) = \widetilde{U}_n(a^n) = \widehat{U}_n(a^n) < +\infty. \quad (4.5)$$

*If $U_n(a^n) < u_n(+\infty)$ then*

$$U^n : \mathbb{R} \to \mathbb{R} \text{ is differentiable, } U_n(-\infty) = -\infty, U'_n > 0, U'_n(-\infty) = +\infty, U'_n(+\infty) = 0 \quad (4.6)$$

*and*

$$U_n(a^n) = \inf_{\lambda > 0}\left\{\lambda\left(\mathbb{E}_{Q^n}[X^n] + a^n\right) + \mathbb{E}\left[v_n\left(\lambda \frac{dQ^n}{dP}\right)\right]\right\}. \quad (4.7)$$

We now show that the optimal values are the same.

**Lemma 4.5.** *Let $A := \rho_B^{\mathbf{Q}}(\mathbf{X})$ and $\pi_A^{\mathbf{Q}}(\mathbf{X}) < +\infty$. Then*

$$\pi_A^{\mathbf{Q}}(\mathbf{X}) = \sup_{\mathbf{Y} \in M^\Phi}\left\{\mathbb{E}\left[\sum_{n=1}^N u_n(X^n + Y^n)\right] \mid \sum_{n=1}^N \mathbb{E}_{Q^n}[Y^n] = A\right\} := \pi_A^{\mathbf{Q},=}(\mathbf{X}) \quad (4.8)$$

*and*

$$\pi_A^{\mathbf{Q}}(\mathbf{X}) = \sup_{\sum_{n=1}^N a^n = A} \sum_{n=1}^N U_n(a^n) = \widetilde{\pi}_A^{\mathbf{Q}}(\mathbf{X}) = \widehat{\pi}_A^{\mathbf{Q}}(\mathbf{X}), \quad (4.9)$$
$$\rho_B^{\mathbf{Q}}(\mathbf{X}) = \widetilde{\rho}_B^{\mathbf{Q}}(\mathbf{X}) = \widehat{\rho}_B^{\mathbf{Q}}(\mathbf{X}). \quad (4.10)$$

*Proof.* Clearly, $+\infty > \pi_A^{\mathbf{Q}}(\mathbf{X}) \geq \pi_A^{\mathbf{Q},=}(\mathbf{X})$. By contradiction suppose that $\pi_A^{\mathbf{Q}}(\mathbf{X}) > \pi_A^{\mathbf{Q},=}(\mathbf{X})$ and take $\varepsilon > 0$ such that $\pi_A^{\mathbf{Q}}(\mathbf{X}) - \varepsilon > \pi_A^{\mathbf{Q},=}(\mathbf{X})$. By definition of $\pi_A^{\mathbf{Q}}(\mathbf{X})$ there exists $\mathbf{Y} \in M^\Phi$ satisfying $\sum_{n=1}^N \mathbb{E}_{Q^n}[Y^n] < A$ and $\mathbb{E}\left[\sum_{n=1}^N u_n(X^n + Y^n)\right] > \pi_A^{\mathbf{Q}}(\mathbf{X}) - \varepsilon$. Take $\widetilde{Y}^n = Y^n + \delta$, $\delta \in \mathbb{R}_+$, such that $\sum_{n=1}^N \mathbb{E}_{Q^n}\left[\widetilde{Y}^n\right] = A$. Then $\pi_A^{\mathbf{Q},=}(\mathbf{X}) \geq$ $\mathbb{E}\left[\sum_{n=1}^{N} u_n(X^n + \widetilde{Y}^n)\right] \geq \mathbb{E}\left[\sum_{n=1}^{N} u_n(X^n + Y^n)\right] > \pi_A^{\mathbf{Q}}(\mathbf{X}) - \varepsilon > \pi_A^{\mathbf{Q},=}(\mathbf{X})$, a contradiction. Hence (4.8) holds true. Note that

$$M^{\Phi} = \left\{ \mathbf{Y} = \mathbf{a} + \mathbf{Z} \mid \mathbf{a} \in \mathbb{R}^N \text{ and } \mathbf{Z} \in M^{\Phi} \text{ such that } \mathbb{E}_{Q^n}[Z^n] = 0 \text{ for each } n \right\}.$$

Indeed, just take $\mathbf{Y} \in M^{\Phi}$ and let $a^n := \mathbb{E}_{Q^n}[Y^n] \in \mathbb{R}$ and $Z^n := Y^n - a^n \in M^{\phi_n}$. Then

$$\begin{aligned}
\pi_A^{\mathbf{Q}}(\mathbf{X}) &= \sup_{\mathbf{Y} \in M^{\Phi}} \left\{ \mathbb{E}\left[\sum_{n=1}^{N} u_n(X^n + Y^n)\right] \mid \sum_{n=1}^{N} \mathbb{E}_{Q^n}[Y^n] = A \right\} \\
&= \sup_{\sum_{n=1}^{N} a^n = A,\ Z^n \in M^{\phi_n},\ \mathbb{E}_{Q^n}[Z^n]=0\ \forall n} \left\{ \mathbb{E}\left[\sum_{n=1}^{N} u_n(X^n + a^n + Z^n)\right] \right\} \\
&= \sup_{\sum_{n=1}^{N} a^n = A} \sum_{n=1}^{N} \sup_{Y^n \in M^{\phi_n},\ \mathbb{E}_{Q^n}[Y^n]=a^n} \mathbb{E}[u_n(X^n + Y^n)] = \sup_{\sum_{n=1}^{N} a^n = A} \sum_{n=1}^{N} U_n(a^n),
\end{aligned}$$
(4.11)

which shows the first equality in (4.9). Then $\pi_A^{\mathbf{Q}}(\mathbf{X}) = \widetilde{\pi}_A^{\mathbf{Q}}(\mathbf{X}) = \widehat{\pi}_A^{\mathbf{Q}}(\mathbf{X})$ are consequences of (4.5) and the decompositions analogous to the one just obtained for $\pi_A^{\mathbf{Q}}(\mathbf{X})$ in (4.11). Set $A := \rho_B^{\mathbf{Q}}(\mathbf{X}) > -\infty$ then $B = \pi_A^{\mathbf{Q}}(\mathbf{X})$, by Proposition 4.2 item (a). Hence $B = \pi_A^{\mathbf{Q}}(\mathbf{X}) = \widetilde{\pi}_A^{\mathbf{Q}}(\mathbf{X}) = \widehat{\pi}_A^{\mathbf{Q}}(\mathbf{X})$ and, from Proposition 4.2 item (b), we obtain $A := \widetilde{\rho}_B^{\mathbf{Q}}(\mathbf{X}) = \widehat{\rho}_B^{\mathbf{Q}}(\mathbf{X})$ hence (4.10). $\square$

**Proposition 4.6.** *Let $A := \rho_B^{\mathbf{Q}}(\mathbf{X})$ and $\pi_A^{\mathbf{Q}}(\mathbf{X}) < +\infty$. There exists an optimal solution $a^* \in \mathbb{R}^N$ to problem (4.9), namely*

$$\pi_A^{\mathbf{Q}}(\mathbf{X}) = \max_{a \in \mathbb{R}^N\ s.t.\ \sum_{n=1}^{N} a^n = A,} \sum_{n=1}^{N} U_n(a^n) = \sum_{n=1}^{N} U_n(a_*^n) \quad \text{and} \quad \sum_{n=1}^{N} a_*^n = A. \quad (4.12)$$

*Proof.* Let $\mathbf{a}_m = (a_m^1, \cdots, a_m^N)_{m \in \mathbb{N}}$ be the approximating sequence of the supremum in (4.12). Then $\sum_{n=1}^{N} U_n(a_m^n) \geq \pi_A^{\mathbf{Q}}(\mathbf{X}) - \delta := C$ and $\sum_{n=1}^{N} a_m^n = A$ for each $m$. Then (4.12) is a consequence of the continuity of $U_n$ and of Lemma 4.7, which guarantees that $\mathbf{a}_m$ belongs to a compact set in $\mathbb{R}^N$. $\square$

**Lemma 4.7.** *For arbitrary constant $A, B \in \mathbb{R}$ set*

$$K := \left\{ \mathbf{a} \in \mathbb{R}^N \mid \sum_{n=1}^{N} a^n \leq A,\ \sum_{n=1}^{N} U_n(a^n) \geq B \right\}.$$

*Then $K$ is a bounded closed set in $\mathbb{R}^N$.*

*Proof.* See Appendix A.4. $\square$

We now turn our attention to the uniqueness of the optimal solution to the problem (3.2). The proof is in Appendix A.4 and employs the same arguments used in the proof of Proposition 2.4.

**Lemma 4.8.** *The penalty function can be written as*

$$\alpha_B(\mathbf{Q}) = \sup_{\mathbf{Z} \in M^\Phi} \left\{ \sum_{n=1}^N \mathbb{E}_{Q^n}[-Z^n] \mid \sum_{n=1}^N \mathbb{E}[u_n(Z^n)] = B \right\}$$

$$= \sup_{\mathbf{Z} \in L^1(\mathbb{P};\mathbf{Q})} \left\{ \sum_{n=1}^N \mathbb{E}_{Q^n}[-Z^n] \mid \mathbb{E}\left[\sum_{n=1}^N u_n(Z^n)\right] \geq B \right\}, \quad (4.13)$$

*and there exists at most one* $\mathbf{Z} \in L^1(\mathbb{P};\mathbf{Q})$ *satisfying*

$$\alpha_B(\mathbf{Q}) = \sum_{n=1}^N \mathbb{E}_{Q^n}[-Z^n] \quad \text{and} \quad \sum_{n=1}^N \mathbb{E}[u_n(Z^n)] \geq B. \quad (4.14)$$

*Remark* 4.9. From (4.10) and (3.5) we have

$$\widehat{\rho}_B^\mathbf{Q}(\mathbf{X}) = \widetilde{\rho}_B^\mathbf{Q}(\mathbf{X}) = \rho_B^\mathbf{Q}(\mathbf{X}) = -\sum_{n=1}^N \mathbb{E}_{Q^n}[X^n] - \alpha_B(\mathbf{Q}).$$

Hence with a proof similar to the one of Lemma 4.8, we may replace the inequality with an equality sign in the budget constraint in the definition of $\rho_B^\mathbf{Q}(\mathbf{X})$, $\widetilde{\rho}_B^\mathbf{Q}(\mathbf{X})$ and $\widehat{\rho}_B^\mathbf{Q}(\mathbf{X})$, and show the uniqueness of the optimizer $\mathbf{Y}$ in $\rho_B^\mathbf{Q}(\mathbf{X})$, $\widetilde{\rho}_B^\mathbf{Q}(\mathbf{X})$ and $\widehat{\rho}_B^\mathbf{Q}(\mathbf{X})$.

## 4.3 On the optimal solution of $\widehat{\rho}^\mathbf{Q}$ and comparison of optimal solutions

**Theorem 4.10.** *Suppose that* $\alpha_B(\mathbf{Q}) < +\infty$. *Then the random vector* $\widehat{\mathbf{Y}}_\mathbf{Q}$ *given by*

$$\widehat{Y}_\mathbf{Q}^n := -X^n - v'_n\left(\lambda^* \frac{dQ^n}{d\mathbb{P}}\right),$$

*where* $\lambda^*$ *is the unique solution to* (3.9), *satisfies* $\widehat{Y}_\mathbf{Q}^n \in L^1(Q^n)$, $u_n(X^n + \widehat{Y}_\mathbf{Q}^n) \in L^1(\mathbb{P})$, $\mathbb{E}\left[\sum_{n=1}^N u_n(X^n + \widehat{Y}_\mathbf{Q}^n)\right] = B$ *and*

$$\rho_B^\mathbf{Q}(\mathbf{X}) = \inf_{\mathbf{Y} \in M^\Phi} \left\{ \sum_{n=1}^N \mathbb{E}_{Q^n}[Y^n] \mid \mathbb{E}\left[\sum_{n=1}^N u_n(X^n + Y^n)\right] \geq B \right\} = \sum_{n=1}^N \mathbb{E}_{Q^n}\left[\widehat{Y}_\mathbf{Q}^n\right] \quad (4.15)$$

$$= \min_{\mathbf{Y} \in L^1(\mathbf{Q})} \left\{ \sum_{n=1}^N \mathbb{E}_{Q^n}[Y^n] \mid \mathbb{E}\left[\sum_{n=1}^N u_n(X^n + Y^n)\right] \geq B \right\} := \widehat{\rho}_B^\mathbf{Q}(\mathbf{X}), \quad (4.16)$$

*so that* $\widehat{\mathbf{Y}}_\mathbf{Q}$ *is the optimal solution for* $\widehat{\rho}_B^\mathbf{Q}(\mathbf{X})$.

*Proof.* Note that $\rho_B^\mathbf{Q}(\mathbf{X}) > -\infty$, as $\alpha_B(\mathbf{Q}) < +\infty$. The integrability conditions hold thanks to the results stated in Appendix A.3. From (3.5) and the expression (3.8) for the penalty, we compute

$$\rho_B^\mathbf{Q}(\mathbf{X}) = -\sum_{n=1}^N \mathbb{E}_{Q^n}[X^n] - \alpha_B(\mathbf{Q}) =$$

$$= \sum_{n=1}^N \mathbb{E}_{Q^n}\left[-X^n - v'_n\left(\lambda^* \frac{dQ^n}{d\mathbb{P}}\right)\right] = \sum_{n=1}^N \mathbb{E}_{Q^n}\left[\widehat{Y}_\mathbf{Q}^n\right].$$

We show that $\widehat{Y}_{\mathbf{Q}}^n$ satisfies the budget constraint:

$$\sum_{n=1}^{N} \mathbb{E}\left[u_n\left(X^n + \widehat{Y}_{\mathbf{Q}}^n\right)\right] = \sum_{n=1}^{N} \mathbb{E}\left[u_n\left(-v_n'\left(\lambda^* \frac{dQ^n}{d\mathbb{P}}\right)\right)\right]$$

$$= \sum_{n=1}^{N} \mathbb{E}\left[v_n\left(\lambda^* \frac{dQ^n}{d\mathbb{P}}\right)\right] - \lambda^* \sum_{n=1}^{N} \mathbb{E}_{Q^n}\left[v_n'\left(\lambda^* \frac{dQ^n}{d\mathbb{P}}\right)\right] = B$$

due to $u(-v'(y)) = v(y) - yv'(y)$ by Lemma A.5 and (3.9). Finally, $\rho_B^{\mathbf{Q}}(\mathbf{X}) = \widehat{\rho}_B^{\mathbf{Q}}(\mathbf{X})$ follows from (4.10), and the Remark 4.9 implies uniqueness. $\square$

When both solutions to the problems $\rho_B(\mathbf{X})$ and $\rho_B^{\mathbf{Q_X}}(\mathbf{X})$ exist, then they coincide.

**Proposition 4.11.** *Let $\mathbf{Y_X} \in \mathcal{C}_0 \cap M^\Phi$ be the optimal allocation for $\rho_B(\mathbf{X})$, $\mathbf{Q_X}$ be an optimal solution to the dual problem (3.1). Then $\mathbf{Y_X} = \widehat{\mathbf{Y}}_{\mathbf{Q_X}}$, i.e.,*

$$Y_{\mathbf{X}}^n = \widehat{Y}_{\mathbf{Q_X}}^n := -X^n - v_n'\left(\lambda^* \frac{dQ_{\mathbf{X}}^n}{d\mathbb{P}}\right).$$

*Proof.* Note that $\mathbf{Y_X}$ satisfies

$$\mathbb{E}\left[\sum_{n=1}^{N} u_n(X^n + Y_{\mathbf{X}}^n)\right] \geq B, \tag{4.17}$$

$$\sum_{n=1}^{N} Y_{\mathbf{X}}^n = \rho_B(\mathbf{X}), \tag{4.18}$$

$$\sum_{n=1}^{N} \mathbb{E}_{Q_{\mathbf{X}}^n}[Y_{\mathbf{X}}^n] \leq \sum_{n=1}^{N} Y_{\mathbf{X}}^n, \tag{4.19}$$

as $\mathbf{Y_X} \in \mathcal{C}$ and $\mathbf{Q_X} \in \mathcal{D}$. From (4.15), (3.5), (3.4) and (4.18) we deduce that

$$\sum_{n=1}^{N} \mathbb{E}_{Q_{\mathbf{X}}^n}[\widehat{Y}_{\mathbf{Q_X}}^n] = \rho_B^{\mathbf{Q_X}}(\mathbf{X}) = -\sum_{n=1}^{N} \mathbb{E}_{Q_{\mathbf{X}}^n}[X^n] - \alpha_B(\mathbf{Q_X}) = \rho_B(\mathbf{X}) = \sum_{n=1}^{N} Y_{\mathbf{X}}^n. \tag{4.20}$$

As $\mathbf{Y_X}$ satisfies (4.17), by definition of $\rho_B^{\mathbf{Q_X}}(\mathbf{X})$ we have

$$\sum_{n=1}^{N} Y_{\mathbf{X}}^n = \rho_B(\mathbf{X}) = \rho_B^{\mathbf{Q_X}}(\mathbf{X}) \leq \sum_{n=1}^{N} \mathbb{E}_{Q_{\mathbf{X}}^n}[Y_{\mathbf{X}}^n],$$

which shows, together with (4.19), that

$$\sum_{n=1}^{N} Y_{\mathbf{X}}^n = \sum_{n=1}^{N} \mathbb{E}_{Q_{\mathbf{X}}^n}[Y_{\mathbf{X}}^n]. \tag{4.21}$$

From (4.20) and (4.21) we then deduce

$$\alpha_B(\mathbf{Q_X}) = -\sum_{n=1}^{N} \mathbb{E}_{Q_{\mathbf{X}}^n}\left[X^n + \widehat{Y}_{\mathbf{Q_X}}^n\right], \quad \text{and}$$

$$\alpha_B(\mathbf{Q_X}) = -\sum_{n=1}^{N} \left(\mathbb{E}_{Q_{\mathbf{X}}^n}[X^n] + Y_{\mathbf{X}}^n\right) = -\sum_{n=1}^{N} \mathbb{E}_{Q_{\mathbf{X}}^n}[X^n + Y_{\mathbf{X}}^n].$$

As both $(\mathbf{X}+\mathbf{Y_X})$ and $(\mathbf{X}+\widehat{\mathbf{Y}}_{\mathbf{Q_X}})$ satisfy the budget constraints associated to $\alpha_B(\mathbf{Q_X})$ in equation (4.14), this implies that $\alpha_B(\mathbf{Q_X})$ is attained by both $(\mathbf{X}+\mathbf{Y_X})$ and $(\mathbf{X}+\widehat{\mathbf{Y}}_{\mathbf{Q_X}})$. The uniqueness shown in Lemma 4.8 allows us to conclude that $\mathbf{Y_X}=\widehat{\mathbf{Y}}_{\mathbf{Q_X}}$. □

*Remark* 4.12. (On $\mathbf{Y_X}=\widehat{\mathbf{Y}}_{\mathbf{Q_X}}$) In Theorem 4.19 we will prove the existence of $\widetilde{\mathbf{Y}}_{\mathbf{X}} \in \mathcal{C}_0 \cap L^1(\mathbb{P},\mathbf{Q_X})$ which satisfies (4.17), (4.18) and (4.19) with $\widetilde{\mathbf{Y}}_{\mathbf{X}}$ at the place of $\mathbf{Y_X}$. Then the above proof shows that $\widetilde{\mathbf{Y}}_{\mathbf{X}} = \widehat{\mathbf{Y}}_{\mathbf{Q_X}}$. Similarly, Corollary 4.13 also holds for such $\widetilde{\mathbf{Y}}_{\mathbf{X}} \in \mathcal{C}_0 \cap L^1(\mathbb{P},\mathbf{Q_X})$.

We now show that the maximizer of the dual representation is unique.

**Corollary 4.13.** *Suppose that there exists an optimal allocation $\mathbf{Y_X}$ to $\rho_B(\mathbf{X})$. Then the optimal solution $\mathbf{Q_X} = (Q_\mathbf{X}^1, \cdots, Q_\mathbf{X}^N)$ of the dual problem (3.1) is unique.*

*Proof.* Suppose that $\mathbf{Q}_1$ and $\mathbf{Q}_2$ are two optimizers of the dual problem (3.1). Then $\alpha_B(\mathbf{Q}_1) < +\infty$, $\alpha_B(\mathbf{Q}_2) < +\infty$ and, by Proposition 4.11 and Remark 4.12, we have for each $n$:

$$-X^n - v'_n\left(\lambda_1^* \frac{dQ_1^n}{d\mathbb{P}}\right) = \widehat{Y}_{\mathbf{Q}_1}^n = Y_{\mathbf{X}}^n = \widehat{Y}_{\mathbf{Q}_2}^n = -X^n - v'_n\left(\lambda_2^* \frac{dQ_2^n}{d\mathbb{P}}\right), \quad \mathbb{P} \text{ a.s.}$$

As $v'_n$ is invertible, we conclude that $\lambda_1^* \frac{dQ_1^n}{d\mathbb{P}} = \lambda_2^* \frac{dQ_2^n}{d\mathbb{P}}$, $\mathbb{P}$ a.s., which then implies $Q_1^n = Q_2^n$, as $\mathbb{E}\left[\frac{dQ_1^n}{d\mathbb{P}}\right] = \mathbb{E}\left[\frac{dQ_2^n}{d\mathbb{P}}\right] = 1$. □

### 4.4 On the existence of the optimal allocation for $\widetilde{\rho}_B$

#### 4.4.1 A first step

We first show that $\rho_B$ reaches its infimum for $\mathbf{Y} \in L^1(\mathbb{P};\mathbb{R}^N)$.

**Theorem 4.14.** *For $\mathcal{C} \subseteq \mathcal{C}_\mathbb{R} \cap M^\Phi$ and for any $\mathbf{X} \in M^\Phi$ there exists $\mathbf{Y} \in L^1(\mathbb{P};\mathbb{R}^N)$ such that*

$$\sum_{n=1}^N Y^n \in \mathbb{R}, \quad \mathbb{E}\left[\sum_{n=1}^N u_n(X^n+Y^n)\right] \geq B,$$

$$\rho_B(\mathbf{X}) := \inf\left\{\sum_{n=1}^N Z^n \mid \mathbf{Z} \in \mathcal{C}, \mathbb{E}\left[\sum_{n=1}^N u_n(X^n+Z^n)\right] \geq B\right\} = \sum_{n=1}^N Y^n,$$

*and a sequence $\{\mathbf{Y}_k\}_{k\in\mathbb{N}} \subset \mathcal{C}$ such that $\mathbb{E}\left[\sum_{n=1}^N u_n(X^n+Y_k^n)\right] \geq B$ and*

$$\mathbf{Y}_k \to \mathbf{Y} \; \mathbb{P}\text{-a.s.}$$

*Remark* 4.15. We note that the random vector $\mathbf{Y}$ in Theorem 4.14 satisfies all the conditions for being the optimal allocation for $\rho_B(\mathbf{X})$, except for the integrability condition $\mathbf{Y} \in M^\Phi$, which is replaced by $\mathbf{Y} \in L^1(\mathbb{P};\mathbb{R}^N)$. Furthermore $\mathbf{Y} = \lim_k \{\mathbf{Y}_k\}$, $\mathbb{P}$-a.s., for $\mathbf{Y_k} \in$ $\mathcal{C}_0 \cap M^\Phi$. If we assume that $\mathcal{C}_0$ is closed in $L^0(\mathbb{P})$, which is a reasonable assumption and holds true if $\mathcal{C} = \mathcal{C}^{(\mathbf{n})}$, in which case $\mathcal{C}_0^{(\mathbf{n})}$ is defined in (2.6), then $\mathbf{Y}$ would also belong to $\mathcal{C}_0$, but in general not to $\mathcal{C}$ (as $M^\Phi$ is in general not closed for $\mathbb{P}$-a.s. convergence). A special case is when the cardinality of $\Omega$ is finite and the set $\mathcal{C}$ is closed for the $\mathbb{P}$-a.s. convergence: under these assumptions $\mathbf{Y}$ belongs to $\mathcal{C}$ and $\mathbf{Y} = \mathbf{Y}_\mathbf{X} = \widehat{\mathbf{Y}}_{\mathbf{Q}_\mathbf{X}}$. In Section 4.4.2 we will show when $\mathbf{Y}$ also belongs to $\mathcal{C}_0 \cap L^1(\mathbf{Q}_\mathbf{X}; \mathbb{R}^N)$.

*Proof.* Take a sequence of vectors $(\mathbf{V}_k)_{k\in\mathbb{N}} \in \mathcal{C} \subseteq \mathcal{C}_\mathbb{R} \cap M^\Phi \subseteq L^1(\mathbb{P}; \mathbb{R}^N)$ such that $\mathbb{R} \ni c_k := \sum_{n=1}^N V_k^n \downarrow \rho_B(\mathbf{X})$ as $k \to +\infty$ and $\mathbb{E}\left[\sum_{n=1}^N u_n(X^n + V_k^n)\right] \geq B$. The sequence $(\mathbf{V}_k)_{k\in\mathbb{N}}$ is bounded for the $L^1(\mathbb{P}; \mathbb{R}^N)$ norm if and only if so is the sequence $(\mathbf{X} + \mathbf{V}_k)_{k\in\mathbb{N}}$. Given the following decomposition in positive and negative part

$$\sum_{n=1}^N \mathbb{E}[|X^n + V_k^n|] = \sum_{n=1}^N \mathbb{E}[(X^n + V_k^n)^+] + \sum_{n=1}^N \mathbb{E}[(X^n + V_k^n)^-], \quad (4.22)$$

we define the index sets:

$$N_\infty^+ = \left\{ n \in \{1,...,N\} \mid \limsup_{k \to +\infty} \mathbb{E}[(X^n + V_k^n)^+] = +\infty \right\},$$

$$N_b^+ = \left\{ n \in \{1,...,N\} \mid \limsup_{k \to +\infty} \mathbb{E}[(X^n + V_k^n)^+] < +\infty \right\},$$

and, similarly, $N_\infty^-$ and $N_b^-$ for the negative part. We can split the expression (4.22) as

$$\sum_{n \in N_\infty^+} E_\mathbb{P}[(X^n + V_k^n)^+] + \sum_{n \in N_b^+} E_\mathbb{P}[(X^n + V_k^n)^+] + \sum_{n \in N_\infty^-} E_\mathbb{P}[(X^n + V_k^n)^-] + \sum_{n \in N_b^-} E_\mathbb{P}[(X^n + V_k^n)^-].$$

If the sequence $(\mathbf{X} + \mathbf{V}_k)_{k\in\mathbb{N}}$ is not $L^1(\mathbb{P}; \mathbb{R}^N)$-bounded, then one of the sets $N_\infty^+$ or $N_\infty^-$ must be nonempty and therefore, because of the constraint $\sum_{n=1}^N V_k^n = c_k$, both $N_\infty^+$ and $N_\infty^-$ must be nonempty. From Lemma A.1 (a), by Jensen inequality and (4.2) we obtain

$$\begin{aligned}
B &\leq \sum_{n=1}^N \mathbb{E}[u_n(X^n + V_k^n)] \leq \sum_{n=1}^N u_n\left(\mathbb{E}[X^n + V_k^n]\right) \\
&= \sum_{n=1}^N u_n\left(\mathbb{E}[(X^n + V_k^n)^+]\right) + \sum_{n=1}^N u_n\left(-\mathbb{E}[(X^n + V_k^n)^-]\right) \\
&\leq b\left(\sum_{n \in N_\infty^+} \mathbb{E}[(X^n + V_k^n)^+] + \sum_{n \in N_b^+} \mathbb{E}[(X^n + V_k^n)^+]\right) \\
&\quad -2b\left(\sum_{n \in N_\infty^-} \mathbb{E}[(X^n + V_k^n)^-] + \sum_{n \in N_b^-} \mathbb{E}[(X^n + V_k^n)^-]\right) + const \\
&= b\left(c_k + \sum_{n=1}^N \mathbb{E}[X^n]\right) + const - b\left(\sum_{n \in N_\infty^-} \mathbb{E}[(X^n + V_k^n)^-] + \sum_{n \in N_b^-} \mathbb{E}[(X^n + V_k^n)^-]\right)
\end{aligned}$$

which is a contradiction, as the second term that multiplies $b$ in not bounded from above. Hence we exclude that our minimizing sequence $(\mathbf{V}_k)_{k\in\mathbb{N}}$ has unbounded $L^1(\mathbb{P};\mathbb{R}^N)$ norm and we may apply a Komlós compactness argument, as stated in Theorem 1.4 [22]. Applying this result to the sequence $(\mathbf{V}_k)_{k\in\mathbb{N}} \in \mathcal{C}$, we can find a sequence $\mathbf{Y}_k \in conv(\mathbf{V}_i, i \geq k) \in \mathcal{C}$, as $\mathcal{C}$ is convex, such that $\mathbf{Y}_k$ converges $\mathbb{P}$-a.s. to $\mathbf{Y} \in L^1(\mathbb{P};\mathbb{R}^N)$.

Observe that by construction $\sum_{n=1}^{N} Y_k^n$ is $\mathbb{P}$-a.s. a real number and, as a consequence, so is $\sum_{n=1}^{N} Y^n$. As $\mathbb{E}\left[\sum_{n=1}^{N} u_n(X^n+V_k^n)\right] \geq B$, also the $\mathbf{Y}_k$ satisfy such constraint and therefore $\rho_B(\mathbf{X}) \leq \sum_{n=1}^{N} Y_k^n$.

Let $\mathbf{Y}_k = \sum_{i \in J_k} \lambda_i^k \mathbf{V}_i \in conv(\mathbf{V}_i, i \geq k)$, for some finite convex combination $(\lambda_i^k)_{i \in J_k}$ such that $\lambda_i^k > 0$ and $\sum_{i \in J_k} \lambda_i^k = 1$, where $J_k$ is a finite subset of $\{k, k+1, ...\}$. For any fixed $k$ we compute

$$\sum_{n=1}^{N} Y_k^n = \sum_{n=1}^{N} \left(\sum_{i \in J_k} \lambda_i^k V_i^n\right)_j = \sum_{i \in J_k} \lambda_i^k \left(\sum_{n=1}^{N} V_i^n\right) = \sum_{i \in J_k} \lambda_i^k c_i \leq c_k \left(\sum_{i \in J_k} \lambda_i^k\right) = c_k \quad (4.23)$$

and from $\rho_B(\mathbf{X}) \leq \sum_{n=1}^{N} Y_k^n \leq c_k$, we then deduce that $\sum_{n=1}^{N} Y^n = \rho_B(\mathbf{X})$.

We now show that $\mathbf{Y}$ also satisfies the budget constraint. In case that all utility functions are bounded from above, this is an immediate consequence of Fatou Lemma, since

$$\sum_{n=1}^{N} \mathbb{E}[-u_n(X^n+Y^n)] = \sum_{n=1}^{N} \mathbb{E}[\varliminf_{k\to\infty}(-u_n(X^n+Y_k^n))]$$
$$\leq \varliminf_{k\to\infty} \sum_{n=1}^{N} \mathbb{E}[-u_n(X^n+Y_k^n)] \leq B.$$

In the general case, recall first that the sequence $\mathbf{V}_k$ is bounded in $L^1(\mathbb{P};\mathbb{R}^N)$, and the argument used in (4.23) shows that

$$\|\mathbf{X} + \mathbf{Y}_k\|_1 \leq \|\mathbf{X}\|_1 + \sup_k \|\mathbf{V}_k\|_1,$$

hence $\sup_k \|\mathbf{X} + \mathbf{Y}_k\|_1 < \infty$.

We now need to exploit the Inada condition at $+\infty$. Applying the Lemma A.1 (b) to the utility functions $u_n$, assumed null in 0, we get

$$-u_n(x) + \varepsilon x^+ + b(\varepsilon) \geq 0 \ \forall \, x \in \mathbb{R}.$$

Replacing $\mathbf{X} + \mathbf{Y}$ in the expression above, applying Fatou Lemma we have

$$\mathbb{E}\left[\sum_{n=1}^{N} -u_n(X^n+Y^n) + \varepsilon(X^n+Y^n)^+ + b(\varepsilon)\right]$$

$$= \mathbb{E}\left[\underline{\lim}_{k\to\infty}\left(\sum_{n=1}^{N} -u_n(X^n+Y_k^n) + \varepsilon(X^n+Y_k^n)^+ + b(\varepsilon)\right)\right]$$

$$\leq \underline{\lim}_{k\to\infty}\sum_{n=1}^{N}\mathbb{E}\left[-u_n(X^n+Y_k^n) + \varepsilon(X^n+Y_k^n)^+ + b(\varepsilon)\right]$$

$$\leq -B + \varepsilon\left(\sup_{k}\|\mathbf{X}+\mathbf{Y}_k\|_1\right) + b(\varepsilon).$$

As the term $b(\varepsilon)$ simplifies in the above inequality, we conclude that for all $\varepsilon > 0$

$$\mathbb{E}\left[\sum_{n=1}^{N} -u_n(X^n+Y^n)\right] \leq -B + \varepsilon\left(\sup_{k}\|\mathbf{X}+\mathbf{Y}_k\|_1 - \sum_{n=1}^{N}\mathbb{E}\left[(X^n+Y^n)^+\right]\right),$$

and since $\sup_k \|\mathbf{X}+\mathbf{Y}_k\|_1 < \infty$ we obtain

$$\mathbb{E}\left[\sum_{n=1}^{N} -u_n(X^n+Y^n)\right] \leq -B,$$

so that $\mathbf{Y}$ satisfies the constraint. $\square$

### 4.4.2 Second Step: The optimal allocation in $L^1(\mathbf{Q_X})$

We now prove further integrability properties of the random vector $\mathbf{Y}$ in Theorem 4.14.

**Lemma 4.16.** *The random vector $\mathbf{Y}$ in Theorem 4.14 satisfies $\mathbf{Y}^- \in L^1(\mathbf{Q_X})$.*

*Proof.* Applying (4.2) and $\phi_j(x) := -u_j(-|x|)$, note that for each fixed $1 \leq j \leq N$

$$\begin{aligned}
0 &\leq \mathbb{E}\left[\phi_j((X^j+Y^j)^-)\right] \leq \sum_{n=1}^{N}\mathbb{E}\left[\phi_n((X^n+Y^n)^-)\right] = \sum_{n=1}^{N}\mathbb{E}\left[-u_n(-(X^n+Y^n)^-)\right]\\
&= \sum_{n=1}^{N}\mathbb{E}\left[u_n(X^n+Y^n)^+\right] - \sum_{n=1}^{N}\mathbb{E}\left[u_n(X^n+Y^n)\right]\\
&\leq \sum_{n=1}^{N}u_n\left(\mathbb{E}\left[(X^n+Y^n)^+\right]\right) - B < \infty,
\end{aligned}$$

where we used Jensen inequality and $\mathbf{X}+\mathbf{Y} \in L^1(\mathbb{P};\mathbb{R}^N)$. This yields $(X^j+Y^j)^- \in L^{\phi_j} \subseteq L^1(Q_\mathbf{X}^j)$. From $Y^j = (X^j+Y^j)^+ - (X^j+Y^j)^- - X^j \geq -(X^j+Y^j)^- - X^j$ we get

$$0 \leq (Y^j)^- \leq (-(X^j+Y^j)^- - X^j)^- = ((X^j+Y^j)^- + X^j)^+.$$

Since, by assumption, $X^j \in M^{\phi_j} \subseteq L^1(Q_\mathbf{X}^j)$, then also $((X^j+Y^j)^- + X^j)^+ \in L^1(Q_\mathbf{X}^j)$ and so

$$(Y^j)^- \in L^1(Q_\mathbf{X}^j),\ 1 \leq j \leq N.$$

$\square$

**Lemma 4.17.** *The random vector* $\mathbf{Y}$ *in Theorem 4.14 satisfies* $\mathbf{Y}^+ \in L^1(\mathbf{Q_X})$.

*Proof.* We proved in Theorem 4.14 the existence of $\mathbf{Y}$ satisfying $\rho_B(\mathbf{X}) = \sum_{n=1}^N Y^n \in \mathbb{R}$ with $\mathbf{Y} \in L^1(\mathbb{P}; \mathbb{R}^N)$, $\mathbb{E}\left[\sum_{n=1}^N u_n(X^n+Y^n)\right] \geq B$ and $\mathbf{Y}$ is the $\mathbb{P}$-a.s. limit of a sequence $\{\mathbf{Y}_k\}_k$ in $\mathcal{C} \subseteq \mathcal{C}_\mathbb{R} \cap M^\Phi$ such that $\sum_{n=1}^N Y_k^n \to \rho_B(\mathbf{X})$, as $k \uparrow +\infty$, $\sum_{n=1}^N \mathbb{E}[u_n(X^n + Y_k^n)] \geq B$ and $\sum_{n=1}^N \mathbb{E}_{Q_\mathbf{X}^n}[Y_k^n] \leq \sum_{n=1}^N Y_k^n$. By passing to a subsequence, w.l.o.g we may assume $\sum_{n=1}^N Y_k^n \downarrow \rho_B(\mathbf{X})$. Let $j \in \{1, ..., N\}$. By Fatou's Lemma we get

$$\mathbb{E}_{Q_\mathbf{X}^j}[(Y^j)^+] \leq \varliminf_k \mathbb{E}_{Q_\mathbf{X}^j}[(Y_k^j)^+] \leq \sup_k \mathbb{E}_{Q_\mathbf{X}^j}[Y_k^j] + \sup_k \mathbb{E}_{Q_\mathbf{X}^j}[(Y_k^j)^-]. \tag{4.24}$$

First we show that $\sup_k \mathbb{E}_{Q_\mathbf{X}^j}[Y_k^j] < \infty$. Put $a_k^n = \mathbb{E}_{Q_\mathbf{X}^n}[Y_k^n]$. Then $\sum_{n=1}^N a_k^n \leq \widetilde{A} := \sum_{n=1}^N Y_k^n \leq \sum_{n=1}^N Y_1^n$ and $\sum_{n=1}^N U_n^{Q_\mathbf{X}^n}(a_k^n) \geq \sum_{n=1}^N \mathbb{E}[u_n(X^n + Y_k^n)] \geq B$ for all $k \in \mathbb{N}$. Thus by Lemma 4.7, $\{\mathbf{a}_k\}_{k \in \mathbb{N}}$ lies in a bounded set in $\mathbb{R}^N$ and thus

$$\sup_k \mathbb{E}_{Q_\mathbf{X}^j}[Y_k^j] < \infty. \tag{4.25}$$

Next we show $\sup_k \mathbb{E}_{Q_\mathbf{X}^j}[(Y_k^j)^-] < \infty$. For all $k \in \mathbb{N}$ it holds that

$$\begin{aligned}
0 &\leq \mathbb{E}\left[\phi_j((X^j + Y_k^j)^-)\right] \leq \sum_{n=1}^N \mathbb{E}\left[\phi_n((X^n + Y_k^n)^-)\right] = \sum_{n=1}^N \mathbb{E}\left[-u_n(-(X^n + Y_k^n)^-)\right] \\
&= \sum_{n=1}^N \mathbb{E}\left[u_n(X^n + Y_k^n)^+\right] - \sum_{n=1}^N \mathbb{E}[u_n(X^n + Y_k^n)] \leq \sum_{n=1}^N u_n\left(\mathbb{E}\left[(X^n + Y_k^n)^+\right]\right) - B,
\end{aligned}$$

where we used the fact that $\mathbf{Y}_k$ satisfies the constraint $\sum_{n=1}^N \mathbb{E}[u_n(X^n + Y_k^n)] > B$ and Jensen inequality. From the proof of Theorem 4.14 we know that $(X^n + Y_k^n)_{k \in \mathbb{N}}$ is $L^1(\mathbb{P})$-bounded for all $n = 1, ..., N$, and thus

$$0 \leq \sup_k \mathbb{E}\left[\phi_j((X^j + Y_k^j)^-)\right] \leq \sum_{n=1}^N u_n\left(\sup_k \mathbb{E}\left[(X^n + Y_k^n)^+\right]\right) - B < \infty.$$

By Remark 2.1 it then follows that $(X^j + Y_k^j)^-_{k \in \mathbb{N}}$ is $L^1(Q_\mathbf{X}^j)$-bounded. From $Y_k^j = (X^j + Y_k^j)^+ - (X^j + Y_k^j)^- - X^j \geq -(X^j + Y_k^j)^- - X^j$ we get

$$0 \leq (Y_k^j)^- \leq (-(X^j + Y_k^j)^- - X^j)^- = ((X^j + Y_k^j)^- + X^j)^+,$$

and thus

$$\sup_k \mathbb{E}_{Q_\mathbf{X}^j}[(Y_k^j)^-] \leq \sup_k \mathbb{E}_{Q_\mathbf{X}^j}[(X^j + Y_k^j)^-] + \mathbb{E}_{Q_\mathbf{X}^j}[|X^j|] < \infty, \tag{4.26}$$

where we recall that by assumption $X^j \in M^{\phi_j} \subseteq L^1(Q_\mathbf{X}^j)$. From (4.25) and (4.26) together with (4.24) the claim follows. □

### 4.4.3 The Final Step

For our final result on the existence we need one more assumption.

**Definition 4.18.** *We say that $\mathcal{C}_0$ is* closed under truncation *if for each $\mathbf{Y} \in \mathcal{C}_0$ there exists $m_Y \in \mathbb{N}$ and $\mathbf{c}_Y = (c_Y^1, ..., c_Y^N) \in \mathbb{R}^N$ such that $\sum_{n=1}^N c_Y^n = \sum_{n=1}^N Y^n := c_Y \in \mathbb{R}$ and for all $m \geq m_Y$*

$$\mathbf{Y}_m := \mathbf{Y} I_{\{\cap_{n=1}^N \{|Y^n|<m\}\}} + \mathbf{c}_Y I_{\{\cup_{n=1}^N \{|Y^n|\geq m\}\}} \in \mathcal{C}_0. \tag{4.27}$$

Note that in Definition 2.6, the set $\mathcal{C}_0^{(\mathbf{n})}$ is closed under truncation.

**Theorem 4.19.** *Let $\mathcal{C} = \mathcal{C}_0 \cap M^\Phi$ and suppose that $\mathcal{C}_0 \subseteq \mathcal{C}_\mathbb{R}$ is closed for the convergence in probability and closed under truncation. For any $\mathbf{X} \in M^\Phi$ there exists $\tilde{\mathbf{Y}}_\mathbf{X} \in \mathcal{C}_0 \cap L^1(\mathbb{P}; \mathbf{Q_X})$ such that*

$$\sum_{n=1}^N \tilde{Y}_\mathbf{X}^n \in \mathbb{R}, \quad \mathbb{E}\left[\sum_{n=1}^N u_n(X^n + \tilde{Y}_\mathbf{X}^n)\right] \geq B, \quad \sum_{n=1}^N \left(\mathbb{E}_{Q_X^n}\left[\tilde{Y}_\mathbf{X}^n\right] - \tilde{Y}_\mathbf{X}^n\right) = 0,$$

*and*

$$\rho_B(\mathbf{X}) = \inf\left\{\sum_{n=1}^N Y^n \mid \mathbf{Y} \in \mathcal{C}_0 \cap M^\Phi, \mathbb{E}\left[\sum_{n=1}^N u_n(X^n + Y^n)\right] \geq B\right\} = \sum_{n=1}^N \tilde{Y}_\mathbf{X}^n$$

$$= \min\left\{\sum_{n=1}^N Y^n \mid \mathbf{Y} \in \mathcal{C}_0 \cap L^1(\mathbb{P}; \mathbf{Q_X}), \mathbb{E}\left[\sum_{n=1}^N u_n(X^n + Y^n)\right] \geq B\right\} := \widetilde{\rho}_B(\mathbf{X}),$$
$$\tag{4.28}$$

*so that $\tilde{\mathbf{Y}}_\mathbf{X}$ is the optimal solution to the extended problem $\widetilde{\rho}_B(\mathbf{X})$.*

*Proof.* The optimal solution $\tilde{\mathbf{Y}}_\mathbf{X}$ coincides with the vector $\mathbf{Y}$ in Theorem 4.14, which belongs to $L^1(\mathbb{P}; \mathbf{Q_X})$, by Theorem 4.14, Lemma 4.16, Lemma 4.17, and to $\mathcal{C}_0$, as $\mathcal{C}_0$ is closed for the convergence in probability and $\mathbf{Y} = \lim_m \mathbf{Y}_m$ $\mathbb{P}$-a.s. and $(\mathbf{Y}_m) \subset \mathcal{C}_0$. Comparing Theorem 4.19 with Theorem 4.14 we see that it remains to prove $\rho_B = \widetilde{\rho}_B$ and $\sum_{n=1}^N \left(\mathbb{E}_{Q_X^n}\left[\tilde{Y}_\mathbf{X}^n\right] - \tilde{Y}_\mathbf{X}^n\right) \leq 0$: this is done in Proposition 4.22 and Proposition 4.20 and requires the truncation assumption on $\mathcal{C}_0$. The opposite inequality

$$\sum_{n=1}^N \tilde{Y}_\mathbf{X}^n = \rho_B(\mathbf{X}) = \rho_B^{\mathbf{Q_X}}(\mathbf{X}) \leq \sum_{n=1}^N \mathbb{E}_{Q_X^n}\left[\tilde{Y}_\mathbf{X}^n\right]$$

holds as $\tilde{\mathbf{Y}}_\mathbf{X}$ fulfills the budget constraints of $\rho_B^{\mathbf{Q_X}}(\mathbf{X})$. □

**Proposition 4.20.** *Suppose that $\mathcal{C}_0$ is closed under truncation. Then*

$$\sum_{n=1}^N \mathbb{E}_{Q_\mathbf{X}^n}[Y^n] \leq \sum_{n=1}^N Y^n, \text{ for all } \mathbf{Y} \in \mathcal{C}_0 \cap L^1(\mathbf{Q_X}; \mathbb{R}^N).$$

*Proof.* Let $\mathbf{Y} \in \mathcal{C}_0 \cap L^1(\mathbf{Q_X}; \mathbb{R}^N)$ and consider $\mathbf{Y}_m$ for $m \in \mathbb{N}$ as defined in (4.27), where w.l.o.g. we assume $m_Y = 1$. Note that $\sum_{n=1}^N Y_m^n = c_Y (= \sum_{n=1}^N Y^n)$ for all $m \in \mathbb{N}$. By boundedness of $\mathbf{Y}_m$ and (4.27), we have $\mathbf{Y}_m \in \mathcal{C}_0 \cap M^\Phi$ for all $m \in \mathbb{N}$. Further, $\mathbf{Y}_m \to \mathbf{Y}$ $\mathbf{Q_X}$-a.s. for $m \to \infty$, and thus, since $|\mathbf{Y}_m| \le \max\{|\mathbf{Y}|, |\mathbf{c}_Y|\} \in L^1(\mathbf{Q_X}; \mathbb{R}^N)$ for all $m \in \mathbb{N}$, also $\mathbf{Y}_m \to \mathbf{Y}$ in $L^1(\mathbf{Q_X}; \mathbb{R}^N)$ for $m \to \infty$ by dominated convergence. We then obtain

$$\sum_{n=1}^N \mathbb{E}_{Q_{\mathbf{X}}^n}[Y^n] = \lim_{m \to \infty} \sum_{n=1}^N \mathbb{E}_{Q_{\mathbf{X}}^n}[Y_m^n] \le \lim_{m \to \infty} \sum_{n=1}^N Y_m^n = c_Y = \sum_{n=1}^N Y^n.$$

□

The map $\widetilde{\rho}_B$ is defined on $M^\Phi$ but the admissible claims $\mathbf{Y}$ belongs to the set $\mathcal{C}_0 \cap L^1(\mathbb{P}; \mathbf{Q_X})$, not included in $M^\Phi$. As $L^1(\mathbb{P}; \mathbf{Q_X}) \subseteq L^1(\mathbb{P}; \mathbb{R}^N)$ with the same argument used in the proof of Proposition 2.4, we can show that $\widetilde{\rho}_B(\mathbf{X}) > -\infty$ for all $\mathbf{X} \in M^\Phi$. By the same argument in the proof of Proposition 2.4 and by (2.3) we also deduce that $\widetilde{\rho}_B(\mathbf{X}) < +\infty$ for all $\mathbf{X} \in M^\Phi$, so that

$$\widetilde{\rho}_B : M^\Phi \to \mathbb{R}$$

is convex and monotone decreasing on its domain $dom(\widetilde{\rho}) = M^\Phi$. From Theorem A.2, we then know that the penalty functions of $\rho_B$ and $\widetilde{\rho}_B$ are defined as:

$$\alpha_B(\mathbf{Q}) := \sup\left\{\sum_{n=1}^N \mathbb{E}_{Q^n}[-X^n] - \rho_B(\mathbf{X}) \mid \mathbf{X} \in M^\Phi\right\},$$

$$\widetilde{\alpha}_B(\mathbf{Q}) := \sup\left\{\sum_{n=1}^N \mathbb{E}_{Q^n}[-X^n] - \widetilde{\rho}_B(\mathbf{X}) \mid \mathbf{X} \in M^\Phi\right\}.$$

**Lemma 4.21.** *If $\mathcal{C}_0$ is closed under truncation, then $\widetilde{\alpha}_B(\mathbf{Q_X}) = \alpha_B(\mathbf{Q_X})$.*

*Proof.* In the proof, we will suppress the label $B$ from the penalty functions. Set $\mathbb{E}[\Lambda(\mathbf{X} + \mathbf{Z})] = \mathbb{E}\left[\sum_{n=1}^N u_n(X^n + Z^n)\right]$. We then have that

$$\widetilde{\alpha}(\mathbf{Q_X}) = \sup\left\{\sum_{n=1}^N \mathbb{E}_{Q_{\mathbf{X}}^n}[-X^n] - \widetilde{\rho}_B(\mathbf{X}) \mid \mathbf{X} \in M^\Phi\right\}$$

$$= \sup_{\mathbf{X} \in M^\Phi}\left\{\sum_{n=1}^N \mathbb{E}_{Q_{\mathbf{X}}^n}[-X^n] + \sup\left\{-\sum_{n=1}^N Z^n \mid \mathbf{Z} \in \mathcal{C}_0 \cap L^1(\mathbb{P}; \mathbf{Q_X}), \mathbb{E}[\Lambda(\mathbf{X} + \mathbf{Z})] \ge B\right\}\right\}$$

$$= \sup\left\{\sum_{n=1}^N \mathbb{E}_{Q_{\mathbf{X}}^n}[-X^n] - \sum_{n=1}^N Z^n \mid \mathbf{Z} \in \mathcal{C}_0 \cap L^1(\mathbb{P}; \mathbf{Q_X}), \mathbf{X} \in M^\Phi, \mathbb{E}[\Lambda(\mathbf{X} + \mathbf{Z})] \ge B\right\}$$

$$\le \sup\left\{\sum_{n=1}^N \mathbb{E}_{Q_{\mathbf{X}}^n}[-X^n] - \sum_{n=1}^N Z^n \mid \mathbf{Z} \in \mathcal{C}_0 \cap L^1(\mathbb{P}; \mathbf{Q_X}), \mathbf{X} \in L^1(\mathbb{P}; \mathbf{Q_X}), \mathbb{E}[\Lambda(\mathbf{X} + \mathbf{Z})] \ge B\right\}$$

$$= \sup\left\{\sum_{n=1}^N \mathbb{E}_{Q_{\mathbf{X}}^n}[-W^n] + \sum_{n=1}^N \mathbb{E}_{Q_{\mathbf{X}}^n}[Z^n] - \sum_{n=1}^N Z^n \mid \mathbf{Z} \in \mathcal{C}_0 \cap L^1(\mathbb{P}; \mathbf{Q_X}), \mathbf{W} \in L^1(\mathbb{P}; \mathbf{Q_X}), \mathbb{E}[\Lambda(\mathbf{W})] \ge B\right\}$$

$$= \sup_{\mathbf{W} \in L^1(\mathbb{P}; \mathbf{Q_X})} \left\{ \sum_{n=1}^N \mathbb{E}_{Q_{\mathbf{X}}^n} [-W^n] \mid \mathbb{E}[\Lambda(\mathbf{W})] \geq B \right\} + \sup \left\{ \sum_{n=1}^N \left( \mathbb{E}_{Q_{\mathbf{X}}^n}[Z^n] - Z^n \right) \mid \mathbf{Z} \in \mathcal{C}_0 \cap L^1(\mathbb{P}; \mathbf{Q_X}) \right\}$$

$$\leq \sup_{\mathbf{W} \in L^1(\mathbb{P}; \mathbf{Q_X})} \left\{ \sum_{n=1}^N \mathbb{E}_{Q_{\mathbf{X}}^n}[-W^n] \mid \mathbb{E}[\Lambda(\mathbf{W})] \geq B \right\} = \alpha(\mathbf{Q_X}),$$

because $\sum_{n=1}^N \left( \mathbb{E}_{Q_{\mathbf{X}}^n}[Z^n] - Z^n \right) \leq 0$ for all $\mathbf{Z} \in \mathcal{C}_0 \cap L^1(\mathbb{P}; \mathbf{Q_X}) \subseteq \mathcal{C}_0 \cap L^1(\mathbf{Q_X}, \mathbb{R}^N)$, as shown in Proposition 4.20. The last equality follows from (4.13).

The opposite inequality is trivial, as $\widetilde{\rho}_B \leq \rho_B$ implies

$$\widetilde{\alpha}(\mathbf{Q_X}) = \sup \left\{ \sum_{n=1}^N \mathbb{E}_{Q_{\mathbf{X}}^n}[-X^n] - \widetilde{\rho}_B(\mathbf{X}) \mid \mathbf{X} \in M^\Phi \right\}$$

$$\geq \sup \left\{ \sum_{n=1}^N \mathbb{E}_{Q_{\mathbf{X}}^n}[-X^n] - \rho_B(\mathbf{X}) \mid \mathbf{X} \in M^\Phi \right\} = \alpha(\mathbf{Q_X}).$$

$\square$

**Proposition 4.22.** *If $\mathcal{C}_0$ is closed under truncation, then*

$$\rho_B(\mathbf{X}) = \widetilde{\rho}_B(\mathbf{X}) = \inf_{\mathbf{Y} \in L^1(\mathbb{P}; \mathbf{Q_X})} \left\{ \sum_{n=1}^N Y^n \mid \mathbf{Y} \in \mathcal{C}_0, \mathbb{E}\left[ \sum_{n=1}^N u_n(X^n + Y^n) \right] \geq B \right\}. \quad (4.29)$$

*Proof.* We know that $\widetilde{\rho}_B : M^\Phi \to \mathbb{R}$ is convex and monotone decreasing. By definition, $\widetilde{\rho}_B \leq \rho_B$. Under the truncation assumption, Lemma 4.21 shows that $\widetilde{\alpha}_B(\mathbf{Q_X}) = \alpha_B(\mathbf{Q_X})$. Then, by Theorem A.2,

$$\widetilde{\rho}_B(\mathbf{X}) = \sup \left\{ \sum_{n=1}^N \mathbb{E}_{Q^n}[-X^n] - \widetilde{\alpha}_B(\mathbf{Q}) \mid \frac{d\mathbf{Q}}{dP} \in L^{\Phi^*} \right\} \geq \sum_{n=1}^N \mathbb{E}_{Q_{\mathbf{X}}^n}[-X^n] - \widetilde{\alpha}_B(\mathbf{Q_X})$$

$$= \sum_{n=1}^N \mathbb{E}_{Q_{\mathbf{X}}^n}[-X^n] - \alpha_B(\mathbf{Q_X}) = \rho_B(\mathbf{X}).$$

$\square$

**Corollary 4.23.** *Under the same assumptions of Theorem 4.19, the following holds true:*

$$\rho_B(\mathbf{X}) = \rho_B^{\mathbf{Q_X}}(\mathbf{X}) = \widetilde{\rho}_B^{\mathbf{Q_X}}(\mathbf{X}) = \widehat{\rho}_B^{\mathbf{Q_X}}(\mathbf{X}) = \widetilde{\rho}_B(\mathbf{X}), \quad (4.30)$$

$$\pi_A(\mathbf{X}) = \pi_A^{\mathbf{Q_X}}(\mathbf{X}) = \widetilde{\pi}_A^{\mathbf{Q_X}}(\mathbf{X}) = \widehat{\pi}_A^{\mathbf{Q_X}}(\mathbf{X}), \quad (4.31)$$

*for $A := \rho_B(\mathbf{X})$, and the unique optimal solutions to the extended problems $\widehat{\rho}_B^{\mathbf{Q_X}}(\mathbf{X})$, $\widehat{\rho}_B^{\mathbf{Q_X}}(\mathbf{X})$, $\widetilde{\rho}_B(\mathbf{X})$, and $\widehat{\pi}_A^{\mathbf{Q_X}}(\mathbf{X})$, $\widetilde{\pi}_A^{\mathbf{Q_X}}(\mathbf{X})$ exist and coincide with*

$$\widetilde{\mathbf{Y}}_{\mathbf{X}} = \widehat{\mathbf{Y}}_{\mathbf{Q_X}} = \left( -X^n - v_n' \left( \lambda^* \frac{dQ_{\mathbf{X}}^n}{d\mathbb{P}} \right) \right)_n \in \mathcal{C}_0 \cap L^1(\mathbb{P}; \mathbf{Q_X}),$$

*and $\mathbf{Q_X}$ is the unique optimal solution to the dual problem (3.1).*

*Proof.* From (4.29), (4.10), (4.9), (3.6) and Corollary 4.3 we already know that (4.30) and (4.31) hold true, when $A := \rho_B(\mathbf{X})$. By Theorem 4.19 there exists an optimal solution $\widetilde{\mathbf{Y}}_\mathbf{X} \in \mathcal{C}_0 \cap L^1(\mathbb{P}; \mathbf{Q_X})$ to $\widetilde{\rho}_B(\mathbf{X})$ and by Proposition 4.11 and Remark 4.12 it coincides with the unique optimal solution $\widehat{\mathbf{Y}}_{\mathbf{Q_x}}$ for $\widehat{\rho}_B^{\mathbf{Q_x}}(\mathbf{X})$. By 4.16, $\widetilde{\rho}_B^{\mathbf{Q_x}}(\mathbf{X}) = \widehat{\rho}_B^{\mathbf{Q_x}}(\mathbf{X}) = \sum_{n=1}^N \mathbb{E}_{Q_X^n}\left[\widehat{Y}_{\mathbf{Q_x}}^n\right]$ and then $\widehat{\mathbf{Y}}_{\mathbf{Q_x}} = \widetilde{\mathbf{Y}}_\mathbf{X} \in \mathcal{C}_0 \cap L^1(\mathbb{P}; \mathbf{Q_X})$ proves that it is also the optimal solution for $\widetilde{\rho}_B^{\mathbf{Q_x}}(\mathbf{X})$. From (4.30) and (4.31), we know that $B = \widetilde{\pi}_A^{\mathbf{Q_x}}(\mathbf{X}) = \widehat{\pi}_A^{\mathbf{Q_x}}(\mathbf{X})$ and $A = \widehat{\rho}_B^{\mathbf{Q_x}}(\mathbf{X}) = \widehat{\rho}_B^{\mathbf{Q_x}}(\mathbf{X})$. Therefore, Proposition 4.2 (d) shows that $\widetilde{\mathbf{Y}}_\mathbf{X}$ is the unique optimal solution to $\widetilde{\pi}_A^{\mathbf{Q_x}}(\mathbf{X})$ and $\widehat{\pi}_A^{\mathbf{Q_x}}(\mathbf{X})$. □

## 5 Additional Properties of $\mathbf{Q_X}$ and Fair Risk Allocation

In this section we provide additional properties for the systemic risk measure $\rho(\mathbf{X})$ given by (1.4) and for the systemic risk allocations $\rho^n(\mathbf{X}) = \mathbb{E}_{Q_\mathbf{X}^n}[Y_\mathbf{X}^n], n = 1, ..., N$, introduced in (1.7). We argue that the choice of $\mathbf{Q_X}$ as the systemic vector of probability measures is fair both from the point of view of the system and from the point of view of the individual banks.

### 5.1 Cash additivity and marginal risk contribution

In this section we provide a sensitivity analysis of $\rho(\mathbf{X})$ with respect to changes in the positions $\mathbf{X}$, which also shows the relevance of the dual optimizer $\mathbf{Q_X}$. We first show that $\rho(\mathbf{X})$ is *cash additive*.

**Lemma 5.1.** *Define $\mathcal{W_C} := \{\mathbf{Z} \in \mathcal{C}_\mathbb{R} \mid \mathbf{Y} \in \mathcal{C} \iff \mathbf{Y} - \mathbf{Z} \in \mathcal{C}\} \cap M^\Phi$. Then, the risk measure $\rho$ is cash additive on $\mathcal{W_C}$, i.e.,*

$$\rho(\mathbf{X} + \mathbf{Z}) = \rho(\mathbf{X}) - \sum_{n=1}^N Z^n \text{ for all } \mathbf{Z} \in \mathcal{W_C} \text{ and all } \mathbf{X} \in M^\Phi, \tag{5.1}$$

*and it satisfies*

$$\frac{d}{d\varepsilon}\rho(\mathbf{X}+\varepsilon\mathbf{V})|_{\varepsilon=0} = -\sum_{n=1}^N V^n, \tag{5.2}$$

*for all $\mathbf{V}$ such that $\varepsilon\mathbf{V} \in \mathcal{W_C}$ for all $\varepsilon \in (0, 1]$.*

*Proof.* Let $\mathbf{Z} \in \mathcal{W_C}$. Then, $\mathbf{W} := \mathbf{Z} + \mathbf{Y} \in \mathcal{C} \subseteq \mathcal{C}_\mathbb{R}$ for any $\mathbf{Y} \in \mathcal{C}$. For any $\mathbf{X} \in M^\Phi$ we have

$$\rho(\mathbf{X} + \mathbf{Z}) = \inf\{\sum_{n=1}^N Y^n \mid \mathbf{Y} \in \mathcal{C}, \, \mathbb{E}[\Lambda(\mathbf{X} + \mathbf{Z} + \mathbf{Y})] \geq B\}$$

$$= \inf\{\sum_{n=1}^N W^n - \sum_{n=1}^N Z^n \mid \mathbf{W} \in \mathcal{C}, \, \mathbb{E}[\Lambda(\mathbf{X} + \mathbf{W})] \geq B\} = \rho(\mathbf{X}) - \sum_{n=1}^N Z^n.$$

In particular, $\rho(\mathbf{X}+\varepsilon\mathbf{V}) = \rho(\mathbf{X}) - \varepsilon\sum_{n=1}^N V^n$ for $\varepsilon\mathbf{V} \in \mathcal{W_C}$ and (5.2) follows. □

*Example* 5.2. In case of the set $\mathcal{C}^{(\mathbf{n})}$ in Example 2.5, $\rho$ is cash additive on $\mathcal{W}_{\mathcal{C}^{(\mathbf{n})}} = \mathcal{C}^{(\mathbf{n})}$ This equality holds since we are assuming no restrictions on the vector $d = (d, \cdots, d_m) \in \mathbb{R}^m$, which determines the grouping.

*Remark* 5.3. Under Assumption 2.2 we have $\mathbb{R}^N \subseteq \mathcal{W}_{\mathcal{C}}$ and (5.2) holds for all $\mathbf{V} \in \mathbb{R}^N$.

The *marginal risk contribution* $\frac{d}{d\varepsilon}\rho(\mathbf{X}+\varepsilon\mathbf{V})|_{\varepsilon=0}$ was also considered in [13] and [3] and is an important quantity which describes the sensitivity of the risk of $\mathbf{X}$ with respect to the impact $\mathbf{V} \in L^0(\mathbb{R}^N)$. The property (5.2) cannot be immediately generalized to the case of random vectors $\mathbf{V}$ as $\sum_{n=1}^N V^n \notin \mathbb{R}$ in general. In the following, we obtain the general local version of cash additivity, which extends (5.2) to a random setting.

**Proposition 5.4.** *Let* $\mathbf{X}$ *and* $\mathbf{V} \in M^\Phi$. *Let* $\mathbf{Q_X}$ *be the optimal solution to the dual problem (3.1) associated to* $\rho(\mathbf{X})$ *and assume that* $\rho(\mathbf{X}+\varepsilon\mathbf{V})$ *is differentiable with respect to* $\varepsilon$ *at* $\varepsilon = 0$, *and* $\frac{d\mathbf{Q_{X+\varepsilon V}}}{d\mathbb{P}} \to \frac{d\mathbf{Q_X}}{d\mathbb{P}}$ *in* $\sigma^*(L^{\Phi^*}, M^\Phi)$, *as* $\varepsilon \to 0$. *Then,*

$$\frac{d}{d\varepsilon}\rho(\mathbf{X}+\varepsilon\mathbf{V})|_{\varepsilon=0} = -\sum_{n=1}^N \mathbb{E}_{Q_\mathbf{X}^n}[V^n]. \tag{5.3}$$

*Proof.* As the penalty function $\alpha_B$ does not depend on $\mathbf{X}$, by (3.4) we deduce

$$\frac{d}{d\varepsilon}\rho(\mathbf{X}+\varepsilon\mathbf{V})|_{\varepsilon=0} = \frac{d}{d\varepsilon}\left\{\sum_{n=1}^N \mathbb{E}_{Q_{\mathbf{X}+\varepsilon\mathbf{V}}^n}[-X^n-\varepsilon V^n] - \alpha_B(\mathbf{Q_{X+\varepsilon V}})\right\}|_{\varepsilon=0}$$

$$= \frac{d}{d\varepsilon}\left\{\sum_{n=1}^N \mathbb{E}_{Q_{\mathbf{X}+\varepsilon\mathbf{V}}^n}[-X^n] - \alpha_B(\mathbf{Q_{X+\varepsilon V}})\right\}|_{\varepsilon=0}$$

$$+ \sum_{n=1}^N \frac{d}{d\varepsilon}\left(\varepsilon\mathbb{E}_{Q_{\mathbf{X}+\varepsilon\mathbf{V}}^n}[-V^n]\right)|_{\varepsilon=0} \tag{5.4}$$

$$= 0 + \sum_{n=1}^N \lim_{\varepsilon\to 0}\mathbb{E}_{Q_{\mathbf{X}+\varepsilon\mathbf{V}}^n}[-V^n] = \sum_{n=1}^N \mathbb{E}_{Q_\mathbf{X}^n}[-V^n], \tag{5.5}$$

where the equality between (5.4) and (5.5) is justified by the optimality of $\mathbf{Q_X}$ and the differentiability of $\rho(\mathbf{X}+\varepsilon\mathbf{V})$, while the last equality is guaranteed by the convergence of $\frac{d\mathbf{Q_{X+\varepsilon V}}}{d\mathbb{P}}$. □

*Remark* 5.5. We emphasize that the generalization (5.3) of (5.2) holds because we are computing the expectation with respect to the vector $\mathbf{Q_X}$ and hints at the implementation of the maximizer $\mathbf{Q_X}$ of the dual problem as explained in the following Sections. The assumptions of Proposition 5.4 are satisfied in the case of the exponential utility functions, which is considered in Section 6.

## 5.2 Interpretation and implementation of $\rho(\mathbf{X})$

Going back to the definition (1.4) of $\rho(\mathbf{X})$, once an aggregation function $\Lambda$, an acceptance set $\mathbb{A}$, and a class $\mathcal{C} \subseteq \mathcal{C}_\mathbb{R}$ have been chosen, then $\rho(\mathbf{X})$ represents the minimal total cash amount needed to make the system acceptable at time $T$. For the sake of notation's simplicity, in the sequel we simply write $\mathbf{Y_X}$ for the solution of $\rho_B(\mathbf{X})$, i.e., we do not specify if we work with $\mathbf{Y_X}$ or $\tilde{\mathbf{Y}}_\mathbf{X}$. As already mentioned in the Introduction and as a result of Proposition 4.1, one relevant economic justification for $\rho$ is that the optimal allocation $\mathbf{Y_X}$ of $\rho(\mathbf{X})$ maximizes the expected system utility among all random allocations of cost less or equal to $\rho(\mathbf{X})$.

We notice also that the class $\mathcal{C}$ may determine the level of *risk sharing* (as explained below in (b)) between the banks, ranging from no risk sharing in the case of deterministic allocations $\mathcal{C} = \mathbb{R}^N$, to the case of full risk sharing $\mathcal{C} = \mathcal{C}_\mathbb{R}$, and other constraints in between as in the grouping Example 2.5. We now discuss two features of our systemic risk measure.

*Implementation of the scenario-dependent allocation:*

(a) In practice, the mechanism can be described as a *default fund* as in the case of a CCP (see [3]). The amount $\rho(\mathbf{X})$ would be collected at time 0 according to some systemic risk allocation $\rho^n(\mathbf{X}), n = 1, \cdots, N$, satisfying $\sum_{n=1}^{N} \rho^n(\mathbf{X}) = \rho(\mathbf{X})$. Then, at time $T$, this exact same amount would be redistributed among the banks according to the optimal scenario-dependent allocations $Y_\mathbf{X}^n$'s satisfying $\sum_{n=1}^{N} Y_\mathbf{X}^n = \rho(\mathbf{X})$, so that the fund acts as a clearing house.

(b) An alternative interpretation and implementation more in the spirit of monetary risk measures is in terms of *capital requirements* together with a *risk sharing mechanism*. Consider again a given systemic risk allocation $\rho^n(\mathbf{X}), n = 1, \cdots, N$, At time 0, a capital requirement $\rho^n(\mathbf{X})$ is imposed on each bank $n$, $n = 1, ..., N$. Then, at time $T$, a risk sharing mechanism takes place: each bank provides (if negative) or collects (if positive) the amount $Y_\mathbf{X}^n - \rho^n(\mathbf{X})$. Note that in sum the financial position of bank $n$ at time T is $X^n + \rho^n(\mathbf{X}) + (Y_\mathbf{X}^n - \rho^n(\mathbf{X})) = X^n + Y_\mathbf{X}^n$ as required. Further, this risk sharing mechanism is made possible because of the *clearing property* $\sum_{n=1}^{N} (Y_\mathbf{X}^n - \rho^n(\mathbf{X})) = 0$ which follows from $\sum_{n=1}^{N} Y_\mathbf{X}^n = \rho(\mathbf{X})$ and the full risk allocation requirement $\sum_{n=1}^{N} \rho^n(\mathbf{X}) = \rho(\mathbf{X})$. The incentive for a singe bank to enter in such a mechanism will be made clear below after we introduce the choice of a fair risk allocation in Section 5.3.

*Total risk reduction and dependence structure of* $\mathbf{X}$

From a system-wide point of view, considering the optimal random allocation $\mathbf{Y_X}$ implies a reduction of the total amount needed to secure the system (compared with the optimal deterministic allocation). This reduction is also a consequence of our framework of scenario-dependent allocations that allows for taking into account the dependence structure of $\mathbf{X}$. An example showing these features can be found in Example 7.1 [7]. In the case of the aggregation function $\Lambda$ being a sum of utility functions as considered in this paper, one can see directly that the dependence structure of $\mathbf{X}$ is taken into account from the constraint $\mathbb{E}\left[\sum_{n=1}^{N} u_n(X^n + Y^n)\right] \geq B$ in (1.4), which depends only on the marginal distributions of $\mathbf{X}$ in the case of deterministic $Y^n$'s.

## 5.3 Fair systemic risk allocation $\rho^n(\mathbf{X})$

We now address the problem of choosing a systemic risk allocations $(\rho^n(\mathbf{X}))_n \in \mathbb{R}^N$ (or individual contributions at time zero) as introduced in Definition 1.2. Note that in our setting, besides providing a *ranking* of the institutions in terms of their systemic riskiness, a risk allocation $\rho^n(\mathbf{X})$ can be interpreted as a capital contribution/requirement for institution $n$ in order to secure the system.

From (5.3) we see that $\mathbb{E}_{\mathbf{Q_X}}[\cdot]$ defined by $\mathbb{E}_{\mathbf{Q_X}}[\mathbf{Y}] = \sum_{n=1}^{N} \mathbb{E}_{Q_{\mathbf{X}}^n}[Y^n]$ already appeared as a multivariate valuation operator and, on the other hand, we have obtained in (4.21) that the minimizer $\mathbf{Y_X}$ and the maximizer $\mathbf{Q_X}$ of the dual problem satisfy

$$\rho(\mathbf{X}) = \sum_{n=1}^{N} Y_{\mathbf{X}}^n = \sum_{n=1}^{N} \mathbb{E}_{Q_{\mathbf{X}}^n}[Y_{\mathbf{X}}^n],$$

which shows that $\left(\rho^n(\mathbf{X})\right)_n = \left(\mathbb{E}_{Q_{\mathbf{X}}^n}[Y_{\mathbf{X}}^n]\right)_n$ provides a systemic risk allocation.

Any vector of probability measures $\mathbf{Q} = (Q^n)_n$ gives rise to a valuation operator $\mathbb{E}_{\mathbf{Q}}[\cdot]$ and to the systemic risk measure $\rho^{\mathbf{Q}}$ given by (1.9). Note, however, that in (1.9) the clearing condition $\sum_{n=1}^{N} Y^n = \rho(\mathbf{X})$ is not guaranteed since the optimization is there performed over all $\mathbf{Y} \in M^{\Phi}$. Now, using the valuation $\mathbb{E}_{\mathbf{Q_X}}[\cdot]$ given by the dual optimizer we know by Proposition 4.11 that the optimal allocation in (1.9) fulfills the clearing condition $\mathbf{Y_X} \in \mathcal{C}_{\mathbb{R}}$, and is in fact the same as the optimal allocation for the original systemic risk measure in (1.4). From (4.20) and (4.21) we obtain

$$\sum_{n=1}^{N} Y_{\mathbf{X}}^n = \rho(\mathbf{X}) = \rho^{\mathbf{Q_X}}(\mathbf{X}) = \sum_{n=1}^{N} \mathbb{E}_{Q_{\mathbf{X}}^n}[Y_{\mathbf{X}}^n], \tag{5.6}$$

which shows that the valuation with $\mathbb{E}_{\mathbf{Q_X}}[\cdot]$ is in line with the systemic risk measure $\rho(\mathbf{X})$. This supports the introduction of $\mathbb{E}_{\mathbf{Q_X}}[\cdot]$ as a suitable systemic valuation operator.

The essential question for a financial institution is now whether its allocated share of the total systemic risk determined by the risk allocation $(\mathbb{E}_{Q_{\mathbf{X}}^1}[Y_{\mathbf{X}}^1], \cdots, (\mathbb{E}_{Q_{\mathbf{X}}^N}[Y_{\mathbf{X}}^N])$ is fair.

With the choice $\mathbf{Q} = \mathbf{Q_X}$, from Corollary 4.3, Lemma 4.5 and equation (4.12) we have

$$\pi_A(\mathbf{X}) = \pi_A^{\mathbf{Q_X}}(\mathbf{X}) = \max_{\sum_{n=1}^N a^n = A,} \sum_{n=1}^N \sup_{\mathbb{E}_{Q_\mathbf{X}^n}[Y^n] = a^n} \mathbb{E}\left[u_n(X^n + Y^n)\right]. \tag{5.7}$$

Choosing $A = \rho_B(\mathbf{X})$, we obtain by Proposition 4.2 and the fact that, then, $\mathbf{Y_X}$ is the optimal solution of $\pi_A^{\mathbf{Q_X}}(\mathbf{X})$, that $\mathbb{E}_{Q_{\mathbf{X}^n}}[Y_\mathbf{X}^n] = a_*^n$, $\sum_{n=1}^N \mathbb{E}_{Q_{\mathbf{X}^n}}[Y_\mathbf{X}^n] = A$, and (5.7) can be rewritten as

$$\pi_A(\mathbf{X}) = \pi_A^{\mathbf{Q_X}}(\mathbf{X}) = \sum_{n=1}^N \sup_{\mathbb{E}_{Q_\mathbf{X}^n}[Y^n] = \mathbb{E}_{Q_\mathbf{X}^n}[Y_\mathbf{X}^n]} \mathbb{E}\left[u_n(X^n + Y^n)\right].$$

This means that by using $\mathbf{Q_X}$ for valuation, the system utility maximization in (1.8) reduces to individual utility maximization problems for the banks without the "systemic" constraint $\mathbf{Y} \in \mathcal{C}$:

$$\forall n, \quad \sup_{Y^n} \left\{ \mathbb{E}\left[u_n(X^n + Y^n)\right] \mid \mathbb{E}_{Q_\mathbf{X}^n}[Y^n] = \mathbb{E}_{Q_\mathbf{X}^n}[Y_\mathbf{X}^n] \right\}.$$

The optimal allocation $Y_\mathbf{X}^n$ and its value $\mathbb{E}_{Q_\mathbf{X}^n}[Y_\mathbf{X}^n]$ can thus be considered fair by the $n^{th}$ bank, as $Y_\mathbf{X}^n$ maximizes the individual expected utility of bank $n$ among all random allocations (not constrained to be in $\mathcal{C}_\mathbb{R}$) with value $\mathbb{E}_{Q_\mathbf{X}^n}[Y_\mathbf{X}^n]$. In particular, it is clear then that for individual banks it is more advantageous to use random allocations than the deterministic $\rho_n$ as the supreme will be larger, as previoulsy stated in Section 5.2 (a) and (b). This finally argues for the fairness of the risk allocation $(\mathbb{E}_{Q_\mathbf{X}^1}[Y_\mathbf{X}^1], \cdots, \mathbb{E}_{Q_\mathbf{X}^N}[Y_\mathbf{X}^N])$ as fair valuation of the optimal scenario-dependent allocation $(Y_\mathbf{X}^1, \cdots, Y_\mathbf{X}^N)$.

## 6 The exponential case

In this section, we focus on a relevant case under Assumption 2.2, i.e., we set $\mathcal{C} = \mathcal{C}^{(\mathbf{n})}$, see Examples 2.5 and 3.5, and we choose $u_n(x) = -e^{-\alpha_n x}/\alpha_n$, $\alpha_n > 0$, $n = 1, \cdots, N$, as in Example 3.6. Then $v_n(y) = \frac{1}{\alpha_n}[y \ln(y) - y]$, $v_n'(y) = \frac{1}{\alpha_n} \ln(y)$. We select $B < \sum_{n=1}^N u_n(+\infty) = 0$. Under these assumptions, $\phi_n(x) := -u_n(-|x|) + u_n(0) = \frac{1}{\alpha_n}(e^{\alpha_n |x|} - 1)$,

$$M^{\phi_n} = M^{\exp} := \left\{ X \in L^0(\mathbb{R}) \mid \mathbb{E}[e^{c|X|}] < +\infty \text{ for all } c > 0 \right\},$$

the Orlicz Hearts $M^{\phi_n}$, $n = 1, \cdots, N$, coincide with the single Orlicz Heart $M^{\exp}$ associated to the exponential Young function $x \to e^{|x|} - 1$ and the random variable $\overline{X} := \sum_n X^n \in M^{\exp}$ is well defined. The systemic risk measure (2.5) $\rho : (M^{\exp})^N \to \mathbb{R}$ becomes

$$\rho(\mathbf{X}) = \inf \left\{ \sum_{n=1}^N Y^n \mid \mathbf{Y} \in \mathcal{C}^{(\mathbf{n})}, \mathbb{E}\left[-\sum_{n=1}^N \frac{1}{\alpha_n} \exp\left[-\alpha_n(X^n + Y^n)\right]\right] = B \right\}. \tag{6.1}$$

Recall that each set $\mathcal{C}^{(\mathbf{n})}$ is closed in probability and closed by truncation. From Proposition 2.4 and Corollary 4.23 we deduce:

**Proposition 6.1.** *The map $\rho$ in (6.1) is finitely valued, monotone decreasing, convex, continuous and subdifferentiable on the Orlicz Heart $M^\Phi = (M^{\exp})^N$, and the problem $\widetilde{\rho}(\mathbf{X})$ admits the unique optimal solution $\widetilde{\mathbf{Y}}_{\mathbf{X}}$ given in Corollary 4.23.*

For a given partition $\mathbf{n}$ and allocations $\mathcal{C}^{(\mathbf{n})}$ we can explicitly compute the optimal value $\rho(\mathbf{X})$, the unique optimal allocation of (6.1) and the unique optimizer $\mathbf{Q}_{\mathbf{X}}$ of the corresponding dual problem (3.10). Notice that in this exponential case $\widetilde{\mathbf{Y}}_{\mathbf{X}} = \mathbf{Y}_{\mathbf{X}} \in (M^{\exp})^N$ is the optimal solution for $\rho(\mathbf{X})$ and for $\widetilde{\rho}(\mathbf{X})$.

**Theorem 6.2.** *For $m = 1, \cdots, h$, and for $k \in I_m$ we have:*

$$d_m = \beta_m \log\left(-\frac{\beta}{B}\mathbb{E}\left[\exp\left(-\frac{\overline{X}_m}{\beta_m}\right)\right]\right), \tag{6.2}$$

$$Y_m^k = -X^k + \frac{1}{\beta_m \alpha_k}\overline{X}_m + \frac{1}{\beta_m \alpha_k}d_m \in M^{\exp}, \tag{6.3}$$

*where $\overline{X}_m = \sum_{k \in I_m} X^k$, $\beta_m = \sum_{k \in I_m} \frac{1}{\alpha_k}$, $\beta = \sum_{i=1}^N \frac{1}{\alpha_i}$, and*

$$\rho(\mathbf{X}) = \sum_{i=1}^N Y^i = \sum_{m=1}^h d_m.$$

*The vector $\mathbf{Q}_{\mathbf{X}}$ of probability measures with densities*

$$\frac{dQ_{\mathbf{X}}^m}{d\mathbb{P}} := \frac{e^{-\frac{1}{\beta_m}\overline{X}_m}}{\mathbb{E}\left[e^{-\frac{1}{\beta_m}\overline{X}_m}\right]} \quad m = 1, \cdots, h. \tag{6.4}$$

*is the optimal solution of the dual problem (3.10), i.e.,*

$$\rho(\mathbf{X}) = \sum_{m=1}^h \mathbb{E}_{Q_{\mathbf{X}}^m}[-\overline{X}_m] - \alpha_B(\mathbf{Q}_{\mathbf{X}}), \tag{6.5}$$

*and $\mathbb{E}_{Q_{\mathbf{X}}^m}[Y_{\mathbf{X}}^n]$, $m = 1, \cdots, h$, $n \in I_m$, is a systemic risk allocation, as in Definition 1.1.*

*Proof.* By (3.11) we note that $\mathbf{Q}_{\mathbf{X}}$, defined in (6.4), belongs to $\mathcal{D}$. Using $\mathbf{Q}_{\mathbf{X}}$ and selecting, from Example 3.6, $\lambda^* = -\frac{B}{\beta}$, it is easy to verify that the random variable $Y_{\mathbf{X}}^n := -X^n - v_n'\left(\lambda^* \frac{dQ_{\mathbf{X}}^n}{d\mathbb{P}}\right)$ assigned in Corollary 4.23 coincides with the expression in (6.3) and $\sum_{n \in I_m} Y_{\mathbf{X}}^n = d_m$. We prove below that $\sum_{m=1}^h d_m = \sum_{m=1}^h \mathbb{E}_{Q_{\mathbf{X}}^m}\left[-\overline{X}_m\right] - \alpha_B(\mathbf{Q}_{\mathbf{X}})$. A priori these equations would not be sufficient to prove that $(\mathbf{Y}_{\mathbf{X}}, \mathbf{Q}_{\mathbf{X}})$ were indeed the optimal solutions to the primal and dual problems, as one needs to know that one of the two is indeed an optimizer of the corresponding problem.

The proofs that $\mathbf{Y}_{\mathbf{X}}$ defined in (6.3) is the optimizer of $\rho(\mathbf{X})$, uses the Lagrange's method and several estimates of lengthy computations and is omitted[2].

---
[2]The proof can be obtained upon request from the authors.

We now prove (6.5). First notice that:

$$H(Q_{\mathbf{X}}^m, \mathbb{P}) = \mathbb{E}_{Q_{\mathbf{X}}^m}\left[\ln\left(\frac{dQ_{\mathbf{X}}^m}{d\mathbb{P}}\right)\right] = \frac{1}{\beta_m}\mathbb{E}_{Q_{\mathbf{X}}^m}\left[-\overline{X}_m\right] - \ln\mathbb{E}\left[e^{-\frac{1}{\beta_m}\overline{X}_m}\right]. \quad (6.6)$$

By (3.13), $\alpha_B(\mathbf{Q_X})$ can be rewritten as

$$\begin{aligned}
\alpha_B(\mathbf{Q_X}) &= \sum_{m=1}^{h}\sum_{i\in I_m}\left\{\frac{1}{\alpha_i}H(Q_{\mathbf{X}}^m, \mathbb{P}) + \frac{1}{\alpha_i}\ln\left(-\frac{B}{\beta}\right)\right\} \\
&= \sum_{m=1}^{h}\left(\mathbb{E}_{Q_{\mathbf{X}}^m}\left[-\overline{X}_m\right] - \beta_m\ln\left(-\frac{\beta}{B}\mathbb{E}\left[e^{-\frac{1}{\beta_m}\overline{X}_m}\right]\right)\right) \\
&= \sum_{m=1}^{h}\left(\mathbb{E}_{Q_{\mathbf{X}}^m}\left[-\overline{X}_m\right] - d_m\right) = \sum_{m=1}^{h}\mathbb{E}_{Q_{\mathbf{X}}^m}\left[-\overline{X}_m\right] - \rho(\mathbf{X}),
\end{aligned}$$

as $\rho(\mathbf{X}) = \sum_{i=1}^{N} Y^i = \sum_{m=1}^{h} d_m$. Then (3.12) concludes the proof. □

*Remark* 6.3. Note that if we arbitrarily change the components of the vector $\mathbf{X}$, but keep fixed the components in one given subgroup, say $I_{m_0}$, then the risk measure $\rho(\mathbf{X})$ will of course change, but $d_{m_0}$ and $Y_{m_0}^k$ for $k \in I_{m_0}$ remain the same.

### 6.1 Sensitivity analysis

Let $\mathbf{X} \in (M^{\exp})^N$, $\mathbf{V} \in (M^{\exp})^N$ and set $\overline{V}_m := \sum_{k\in I_m} V_k$, for $m = 1, \cdots, h$. We consider a perturbation $\varepsilon\mathbf{V}$, $\varepsilon \in \mathbb{R}$, and perform a sensitivity analysis in the exponential case. Consider the optimal allocations $Y^i_{\mathbf{X}+\varepsilon\mathbf{V}}$ and the optimal solution $\mathbf{Q}_{\mathbf{X}+\varepsilon\mathbf{V}}$ of the dual problem associated to $\rho(\mathbf{X} + \varepsilon\mathbf{V})$, see (6.4). By (6.3) and (6.2) we have

$$Y^n_{\mathbf{X}+\varepsilon\mathbf{V}} = -X^n - \varepsilon V^n + \frac{1}{\beta_m\alpha_n}\left(\overline{X}_m + \varepsilon\overline{V}_m\right) + \frac{1}{\beta_m\alpha_n}d_m(\mathbf{X} + \varepsilon\mathbf{V}), \quad (6.7)$$

where

$$d_m(\mathbf{X} + \varepsilon\mathbf{V}) = \beta_m\log\left(-\frac{\beta}{B}\mathbb{E}\left[\exp\left(-\frac{\overline{X}_m + \varepsilon\overline{V}_m}{\beta_m}\right)\right]\right). \quad (6.8)$$

**Proposition 6.4.** *Let $\rho$ be the systemic risk measure defined in (6.1). Then*

1. *Marginal risk contribution of group $m$:*

$$\left.\frac{d}{d\epsilon}d_m(\mathbf{X} + \varepsilon\mathbf{V})\right|_{\varepsilon=0} = \mathbb{E}_{Q_{\mathbf{X}}^m}[-\overline{V}_m], \quad m = 1, ..., h.$$

2. *Local causal responsibility:*

$$\left.\frac{d}{d\varepsilon}\mathbb{E}_{Q_{\mathbf{X}}^m}[Y^n_{\mathbf{X}+\varepsilon\mathbf{V}}]\right|_{\varepsilon=0} = \mathbb{E}_{Q_{\mathbf{X}}^m}[-V^n], \quad n \in I_m.$$

3. $\left.\frac{d}{d\varepsilon}\mathbb{E}_{Q_{\mathbf{X}+\varepsilon\mathbf{V}}^m}[Z]\right|_{\varepsilon=0} = -\frac{1}{\beta_m}COV_{Q_{\mathbf{X}}^m}[\overline{V}_m, Z]$, *for any $Z \in M^{\exp}$,*

4. Marginal risk allocation of institution $n \in I_m$:

$$\frac{d}{d\varepsilon}\mathbb{E}_{Q^m_{\mathbf{X}+\varepsilon\mathbf{V}}}[Y^n_{\mathbf{X}+\varepsilon\mathbf{V}}]\bigg|_{\varepsilon=0} = \mathbb{E}_{Q^m_{\mathbf{X}}}[-V^n] - \frac{1}{\beta_m}COV_{Q^m_{\mathbf{X}}}[\overline{V}_m, Y^n_{\mathbf{X}}] \quad (6.9)$$

$$= \mathbb{E}_{Q^m_{\mathbf{X}}}[-V^n] + \frac{1}{\beta_m}COV_{Q^m_{\mathbf{X}}}[\overline{V}_m, X^n] - \frac{1}{\alpha_n}\frac{1}{\beta_m^2}COV_{Q^m_{\mathbf{X}}}[\overline{V}_m, \overline{X}_m], \quad (6.10)$$

5. Sensitivity of the penalty function:

$$\frac{d}{d\varepsilon}\alpha_B(\mathbf{Q}_{\mathbf{X}+\varepsilon\mathbf{V}})\bigg|_{\varepsilon=0} = \sum_{m=1}^{h} \frac{1}{\beta_m}COV_{Q^m_{\mathbf{X}}}[\overline{V}_m, \overline{X}_m],$$

6. Systemic marginal risk contribution:

$$\frac{d}{d\varepsilon}\rho(\mathbf{X}+\varepsilon\mathbf{V})\bigg|_{\varepsilon=0} = \sum_{m=1}^{h}\sum_{i\in I_m} \mathbb{E}_{Q^m_{\mathbf{X}}}[-V^i] = \sum_{m=1}^{h} \mathbb{E}_{Q^m_{\mathbf{X}}}[-\overline{V}_m].$$

*Proof.* The proof is the results of lengthy computations and is omitted[3]. □

The interpretation of these formulas is not simple because we are *dealing with the systemic probability measure $Q^m_{\mathbf{X}}$ and not with the "physical" measure $\mathbb{P}$*. Think of the difference between the physical measure $\mathbb{P}$ and a martingale measure. If we replace $Q^m_{\mathbf{X}}$ with $\mathbb{P}$, none of the results of Proposition 6.4 will hold in general.

The first term $\mathbb{E}_{Q^m_{\mathbf{X}}}[-V^n]$ in (6.9) or (6.10) is easy to interpret: it is not a systemic contribution, as it only involves the increment $V^n$ in the (same) bank $n$. If we sum over all $n$ in the same group, we obtain from (6.9) or (6.10)

$$\sum_{n\in I_m} \frac{d}{d\varepsilon}\mathbb{E}_{Q^m_{\mathbf{X}+\varepsilon\mathbf{V}}}[Y^n_{\mathbf{X}+\varepsilon\mathbf{V}}]\bigg|_{\varepsilon=0} = \mathbb{E}_{Q^m_{\mathbf{X}}}[-\overline{V}_m] = \frac{d}{d\varepsilon}d_m(\mathbf{X}+\varepsilon V)\bigg|_{\varepsilon=0}. \quad (6.11)$$

So, this first term $\mathbb{E}_{Q^m_{\mathbf{X}}}[-V^n]$ is the contribution to the marginal risk allocation of bank $n$ regardless of any systemic influence. Equation (6.11) is the Local Casual Responsibility for the whole group, but not for the single bank inside each group. The sign of the increment $V^n$ in the first term of (6.9) is here relevant: an increment (positive) corresponds to a risk reduction, regardless of the dependence structure. If $\mathbf{V}$ is deterministic, the marginal risk allocation to bank $n$ is exactly $\mathbb{E}_{Q^m_{\mathbf{X}}}[-V^n] = -V^n$ and no other correction terms are present.

To understand the other terms in (6.9) or (6.10), take $\mathbf{V} = V^j\mathbf{e}_j$ with $j \neq n$. In this way, the first term in (6.9) disappears ($V^n = 0$) and we obtain

$$\frac{d}{d\varepsilon}\mathbb{E}_{Q^m_{\mathbf{X}+\varepsilon V^j\mathbf{e}_j}}[Y^n_{\mathbf{X}+\varepsilon V^j\mathbf{e}_j}]\bigg|_{\varepsilon=0} = \frac{1}{\beta_m}COV_{Q^m_{\mathbf{X}}}[V^j, X^n] - \frac{1}{\alpha_n}\frac{1}{\beta_m^2}COV_{Q^m_{\mathbf{X}}}[V^j, \overline{X}_m].$$

---
[3]The proof can be obtained upon request from the authors.

To fix the ideas, suppose that $COV_{Q_{\mathbf{X}}^m}[V^j, X^n] < 0$, and examine for the moment only the contribution of $\frac{1}{\beta_m}COV_{Q_{\mathbf{X}}^m}[V^j, X^n]$. This component does not depend on the specific $\alpha_n$, but it depends on the dependence structure between $(V^j, X^n)$. If the systemic risk probability $Q_{\mathbf{X}}^m$ attributes negative correlation to $(V^j, X^n)$, then, from the systemic perspective this is good (independently of the sign of $V^j$): a decrement in bank $j$ is balanced by bank $n$, and viceversa. If bank $n$ is negatively correlated (as seen by $Q_{\mathbf{X}}^m$) with the increment of bank $j$, then the risk allocation of bank $n$ should decrease. Therefore, bank $n$ takes advantage of this, as its risk allocation is reduced ($\frac{1}{\beta_m}COV_{Q_{\mathbf{X}}^m}[V^j, X^n] < 0$). Since the overall marginal risk allocation of the group $m$ is fixed (equal to $\mathbb{E}_{Q_{\mathbf{X}}^m}[-\overline{V}_m] = \mathbb{E}_{Q_{\mathbf{X}}^m}[-V^j]$, from (6.11)), someone else has to pay for such advantage to bank $n$. This is the last term in (6.10), discussed next.

For the third component in (6.10), we distinguish between the systemic component $-\frac{1}{\beta_m^2}COV_{Q_{\mathbf{X}}^m}[V^j, \overline{X}_m]$, which only depends on the aggregate group $\overline{X}_m$, and the systemic relevance $\frac{1}{\alpha_n}$ of bank $n$. The systemic quantity is therefore distributed among the various banks according to $\frac{1}{\alpha_n}$. In addition, this term must compensate for the possible risk reduction term (the second term in (6.10)), as the overall risk allocation to group $m$ is determined by $\mathbb{E}_{Q_{\mathbf{X}}^m}[-\overline{V}_m] = \mathbb{E}_{Q_{\mathbf{X}}^m}[-V^j]$.

Finally Items 1 and 6 express the same property (which holds in general, as shown in Proposition 5.4) respectively for one group or for the entire system.

## 6.2 Monotonicity

Another desirable fairness property is *monotonicity*. It is clear that if $\mathcal{C}_1 \subseteq \mathcal{C}_2 \subseteq \mathcal{C}_{\mathbb{R}}$, then $\rho_1(\mathbf{X}) \geq \rho_2(\mathbf{X})$ for the corresponding systemic risk measures

$$\rho_i(\mathbf{X}) := \inf\left\{\sum_{n=1}^N Y^n \mid \mathbf{Y} \in \mathcal{C}_i, \mathbb{E}\left[\sum_{n=1}^N u_n(X^n + Y^n)\right] \geq B\right\}, \quad i = 1, 2.$$

The two extreme cases occur for $\mathcal{C}_1 := \mathbb{R}^N$ (the deterministic case) and $\mathcal{C}_2 := \mathcal{C}_{\mathbb{R}}$ (the unconstraint scenario dependent case). Hence, we know that when going from deterministic to scenario-dependent allocations the total systemic risk decreases. It is then desirable that each institution profits from this decrease in total systemic risk in the sense that also its individual risk allocation decreases:

$$\rho_1^n(\mathbf{X}) \geq \rho_2^n(\mathbf{X}) \text{ for each } n = 1, ..., N. \quad (6.12)$$

The opposite would clearly be perceived as unfair. In the next proposition (see in particular equation (6.15)) we prove that (6.12) holds true, in the context of the grouping Example 2.5, when we compute the risk allocation $\rho^n(\mathbf{X}) = \mathbb{E}_{Q_{\mathbf{X}}^n}[Y_{\mathbf{X}}^n]$ using $\mathbf{Q}_{\mathbf{X}}$. If we would select a vector of probability measures $\mathbf{R}$ different from $\mathbf{Q_X}$ to compute the risk allocation with the formula $\mathbb{E}_{R^n}[Y_{\mathbf{X}}^n]$, the property (6.12) would be in general lost.

For a given partition $\mathbf{n}$ and $\mathcal{C} = \mathcal{C}^{(\mathbf{n})}$, let $Y_r^k$, $k \in I_r$, $r = 1, \cdots, h$, be the corresponding optimal allocations of the primal problem (6.1) and $Q_{\mathbf{X}}^r$, $r = 1, \cdots, h$, be the optimal solutions of the corresponding dual problem (3.10) (in this section we suppress the label $\mathbf{X}$ from the optimal allocation $\mathbf{Y_X}$ to $\rho(\mathbf{X})$).

Consider for some $m \in \{1, \cdots, h\}$ a non empty subgroup $I_m'$ of the group $I_m$. Set $I_m'' := I_m \setminus I_m'$. Then the $(h+1)$ groups $I_1, I_2, \cdots, I_m', I_m'', I_{m+1}, \cdots, I_h$ corresponds to a new partition $\mathbf{n}'$. The optimal allocations of the primal problem (6.1) with $\mathcal{C} = \mathcal{C}^{(\mathbf{n}')}$ coincide with $Y_r^k$, $k \in I_r$, for $r \neq m$. For $r = m$, $i \in I_m'$, we have the following.

**Proposition 6.5.** *Define with $Y_{m'}^i$, $i \in I_m'$, the optimal allocation to the primal problem with $\mathcal{C} = \mathcal{C}^{(\mathbf{n}')}$. Then*

$$\mathbb{E}_{Q_{\mathbf{X}}^m}\left[\sum_{i \in I_m'} Y_m^i\right] \leq \sum_{i \in I_m'} Y_{m'}^i := d_m'. \tag{6.13}$$

*In particular, if the group $I_m'$ consists of only one single element $\{i\}$, then $Y_{m'}^i$ is deterministic and*

$$\mathbb{E}_{Q_{\mathbf{X}}^m}[Y_m^i] \leq Y_{m'}^i \quad \text{for each } i \in I_m. \tag{6.14}$$

*If we compare the deterministic optimal allocation $\mathbf{Y}^*$ (corresponding to $\mathcal{C} = \mathbb{R}^N$) with the (random) optimal allocations $\mathbf{Y}$ associated to one single group ($\mathcal{C} = \mathcal{C}_\mathbb{R} \cap (M^{\exp})^N$), we conclude*

$$\mathbb{E}_{Q_{\mathbf{X}}}[Y^n] \leq (Y^*)^n \text{ for each } n = 1, \cdots, d, \tag{6.15}$$

*where $Q_{\mathbf{X}}$ is the unique optimal solution of the dual problem associated to $\mathcal{C} = \mathcal{C}_\mathbb{R} \cap (M^{\exp})^N$.*

*Proof.* Given the subgroup $I_m'$, define $\beta_m' := \sum_{k \in I_m'} \frac{1}{\alpha_k}$. Then the optimal value with respect to $\mathcal{C}^{(\mathbf{n}')}$ is given by

$$d_m' = \beta_m' \ln\left\{-\frac{\beta}{B}\mathbb{E}\left[\exp\left(-\frac{1}{\beta_m'}\sum_{k \in I_m'} X^k\right)\right]\right\}.$$

Summing the components of the solutions relative to $\mathcal{C}^{(\mathbf{n})}$ over $k \in I_m'$, we get

$$\sum_{k \in I_m'} \mathbf{Y}_m^k = \sum_{k \in I_m'}\left(\frac{1}{\beta_m \alpha_k}\overline{X}_m - X^k\right) + \sum_{k \in I_m'}\frac{1}{\beta_m \alpha_k}d_m$$

$$= \left(\frac{\beta_m'}{\beta_m}\overline{X}_m - \sum_{k \in I_m'} X^k\right) + \frac{\beta_m'}{\beta_m}d_m.$$

Using Jensen inequality we obtain

$$\mathbb{E}_{Q_{\mathbf{x}}^m}\left[\sum_{k\in I'_m}\mathbf{Y}_m^k\right]$$

$$=\beta'_m\ln\left\{\exp\left(\frac{1}{\beta'_m}\mathbb{E}_{Q_{\mathbf{x}}^m}\left[\left(\frac{\beta'_m}{\beta_m}\overline{X}_m-\sum_{k\in I'_m}X^k\right)\right]\right)\right\}+\frac{\beta'_m}{\beta_m}\beta_m\ln\left(-\frac{\beta}{B}\mathbb{E}\left[\exp\left(-\frac{\overline{X}_m}{\beta_m}\right)\right]\right)$$

$$\leq\beta'_m\ln\left\{\mathbb{E}_{Q_{\mathbf{x}}^m}\left[\exp\left(\frac{1}{\beta_m}\overline{X}_m-\frac{1}{\beta'_m}\sum_{k\in I'_m}X^k\right)\right]\right\}+\beta'_m\ln\left(-\frac{\beta}{B}\mathbb{E}\left[\exp\left(-\frac{\overline{X}_m}{\beta_m}\right)\right]\right)$$

$$=\beta'_m\ln\left\{\mathbb{E}\left[\frac{\exp\left(-\frac{\overline{X}_m}{\beta_m}\right)\exp\left(\frac{1}{\beta_m}\overline{X}_m\right)\exp\left(-\frac{1}{\beta'_m}\sum_{k\in I'_m}X^k\right)}{\mathbb{E}\left[e^{-\frac{1}{\beta_m}\overline{X}_m}\right]}\right]\right\}$$

$$+\beta'_m\ln\left(-\frac{\beta}{B}\mathbb{E}\left[\exp\left(-\frac{\overline{X}_m}{\beta_m}\right)\right]\right)$$

$$=\beta'_m\ln\left\{-\frac{\beta}{B}\mathbb{E}\left[\exp\left(-\frac{1}{\beta'_m}\sum_{k\in I'_m}X^k\right)\right]\right\}=d'_m.$$

We have that (6.14) and (6.15) directly follow by (6.13). $\square$

# A   Appendix

## A.1   Properties

**Lemma A.1.** *Under Assumption 2.2, if* $\lim_{x\to-\infty}\left(\frac{u_n(x)}{x}\right)=+\infty$ *and if* $\lim_{x\to+\infty}\frac{u_n(x)}{x}=0$, *then*

(a) *there exists* $c\in\mathbb{R}$ *and* $b\in\mathbb{R}_+$ *such that*

  (i) $u_n(x)\leq bx+c$ *for all* $x\geq 0$ *and all* $n$;

  (ii) $u_n(x)\leq 2bx+c$ *for all* $x\leq 0$ *and all* $n$.

(b) $\forall\varepsilon>0$ *there exists* $b=b(\varepsilon)>0$ *such that* $u_n(x)\leq\varepsilon x+b$ *for* $x\geq 0$ *and all* $n$.

*Proof.* Note that $dom(u_n)=\mathbb{R}$ for each $n$. Hereafter the left derivatives of the concave increasing functions $u_n$ are denoted by $u'_n$ and satisfy $u'_n(x)\geq 0$ for all $x\in\mathbb{R}$.

(a) We have the following.

  (i) The concavity of each $u_n$ implies that $u_n(x)\leq u'_n(0)x+c_n$ for all $x\in\mathbb{R}$ (for some $c_n$) and therefore, setting $b:=\max_n u'_n(0)\geq 0$ and $c\geq\max_n c_n$, $u_n(x)\leq bx+c$ for all $x\geq 0$.

(ii) We prove that for every $M > 0$ there exists a constant $d > 0$ with $u_n(x) \leq Mx + d$ for all $n$ and $x \leq 0$. By taking $M = 2b$ we obtain (ii). The assumption $\lim_{x \to -\infty} \left(\frac{u_n(x)}{x}\right) = +\infty$, implies that there exists $K > 0$ (which depends on $M$) such that for all $n$ $u_n(x) \leq Mx$ for $x \leq -K$. Hence $Mx - u_n(x) \geq 0$ for $x \in (-\infty, -K)$. It is clear now that since the function $Mx - u_n(x)$ is continuous on $[-K, 0]$ we may add a properly chosen $d > 0$ so that $Mx + d - u_n(x) \geq 0$ for all $x \in (-\infty, 0]$ and all $n$.

(b) The assumption $\lim_{x \to +\infty} \frac{u_n(x)}{x} = 0$ guarantees the existence of a constant $K > 0$, which depends on $\varepsilon$, such that $u_n(x) \leq \varepsilon x$ for $x \geq K$ and all $n$. Hence

$$u_n(x) \leq \varepsilon x + K\varepsilon + \sup_n \left(\sup_{[0,K]} u_n(s)\right) \quad \forall x \geq 0.$$

$\square$

*Proof.* [of Proposition 2.4] To show $\rho > -\infty$, we suppose by contradiction that $\rho(\mathbf{X}) = -\infty$, for some $\mathbf{X} \in M^\Phi \subseteq L^1(\mathbb{P}, \mathbb{R}^N)$. Let $\mathbf{Y}_m \in \mathcal{C}$ satisfy $\sum_{n=1}^N Y_m^n \downarrow -\infty$, as $m \to +\infty$ and $\Lambda(\mathbf{X} + \mathbf{Y}_m) \in \mathbb{A}$ for each $m$. The condition $\sum_{n=1}^N Y_m^n \downarrow -\infty$, as $m \to +\infty$ implies $\sum_{n=1}^N \mathbb{E}[Y_m^n] \downarrow -\infty$, as $m \to +\infty$. Note also that, by Jensen inequality,

$$B \leq \mathbb{E}[\Lambda(\mathbf{X} + \mathbf{Y}_m)] \leq \Lambda(\mathbb{E}[\mathbf{X} + \mathbf{Y}_m]) = \sum_{n=1}^N u_n(\mathbb{E}[X^n] + \mathbb{E}[Y_m^n]). \tag{A.1}$$

We now prove that $\sum_{n=1}^N u_n(\mathbb{E}[X^n] + \mathbb{E}[Y_m^n]) \downarrow -\infty$, as $m \to +\infty$, which is in contradiction with (A.1). Set $\mathbf{x}_m := (x_m^n)_{n=1}^N$ where $x_m^n := \mathbb{E}[Y_m^n]$. Since $\sum_{n=1}^N x_m^n \downarrow -\infty$, there must exist $n_0 \in \{1, \cdots, N\}$ and a subsequence $\mathbf{x}_{h_m}$ such that $x_{h_m}^{n_0} \downarrow -\infty$ as $m \to +\infty$. With an abuse of notation, denote again such subsequence $\mathbf{x}_{h_m}$ with $\mathbf{x}_m$. Then we have $x_m^{n_0} \downarrow -\infty$. If there exists another coordinate $n_1 \in \{1, \cdots, N\} \setminus n_0$ such that $\underline{\lim}_{m \to \infty} x_m^{n_1} = -\infty$, take the subsequence $\mathbf{x}_{k_m}$ such that $x_{k_m}^{n_1} \downarrow -\infty$. By diagonal procedure, we obtain one single sequence denoted again by $\mathbf{x}_m$ such that $x_m^{n_0} \downarrow -\infty$ and $x_m^{n_1} \downarrow -\infty$, as $m \to +\infty$. We may adopt this procedure (at most $N$ times) also in the case $\limsup_{m \to \infty} x_m^{n_2} = +\infty$ for some coordinate $n_2$. At the end, we will obtain one single sequence $\mathbf{x}_m$ and three disjoint sets of coordinate indices $N_-$, $N_+$, $N^*$ such that

$$\begin{aligned} x_m^n \downarrow -\infty & \quad \text{if } n \in N_- \subseteq \{1, \cdots, N\}, \\ x_m^n \uparrow +\infty & \quad \text{if } n \in N_+ \subseteq \{1, \cdots, N\}, \\ |x_m^n| \leq K & \quad \text{for all } m \text{ and all } n \in N^* = \{1, \cdots, N\} \setminus (N_- \cup N_+), \end{aligned}$$

where $K$ is a constant independent of $m$. We know that $N_- \neq \emptyset$, since $n_0 \in N_-$ (but the other two sets $N_+$ and $N^*$ may be empty). Since $\sum_{n=1}^N x_m^n \downarrow -\infty$, we deduce that, for large $m$, $\sum_{n=1}^{N} x_m^n \leq 0$ so that

$$\sum_{n \in N_+} x_m^n \leq -\sum_{n \in N_-} x_m^n - \sum_{n \in N^*} x_m^n \leq -\sum_{n \in N_-} x_m^n + NK, \text{ for each fixed (large) } m. \quad (A.2)$$

Now we use the inequalities of Lemma A.1 (a) and in (A.2). We obtain, for each fixed large $m$, that

$$\sum_{n=1}^{N} u_n(\mathbb{E}[X^n] + \mathbb{E}[Y_m^n]) = \sum_{n \in N_+} u_n(\mathbb{E}[X^n] + x_m^n) + \sum_{n \in N_-} u_n(\mathbb{E}[X^n] + x_m^n) + \sum_{n \in N^*} u_n(\mathbb{E}[X^n] + x_m^n)$$

$$\leq C_1 + \sum_{n \in N_+} b x_m^n + \sum_{n \in N_-} 2b x_m^n + \sum_{n \in N^*} u_n(K)$$

$$\leq C_2 - \sum_{n \in N_-} b x_m^n + bNK + \sum_{n \in N_-} 2b x_m^n = C_3 + b \sum_{n \in N_-} x_m^n$$

with the constants $C_1, C_2, C_3$ all independent from $m$. Then $b \sum_{n \in N_-} x_m^n \downarrow -\infty$, as $m \to +\infty$, since $x_m^n \downarrow -\infty$ for each $n \in N_-$. This shows $\rho(\mathbf{X}) > -\infty$ for all $\mathbf{X} \in M^\Phi$.

Let $\mathbf{X} \in M^\Phi$. Then $\mathbb{E}[\Lambda(\mathbf{X})] > -\infty$ and $\mathbf{X} + m\mathbf{1} \uparrow +\infty$ $\mathbb{P}$-a.s. if $m \to +\infty$, $m \in \mathbb{R}$, where $\mathbf{1} = (1, \cdots, 1)$. As $\mathbb{E}[\Lambda(\mathbf{X})] > -\infty$, we have that $\mathbb{E}[\Lambda(\mathbf{X} + m\mathbf{1})] > -\infty$ for $m > 0$, and by monotone convergence it follows that $\mathbb{E}[\Lambda(\mathbf{X} + m\mathbf{1})] \uparrow \Lambda(+\infty) > B$. Since $\mathbb{R}^N \subseteq \mathcal{C}$, then $m\mathbf{1} \in \mathcal{C}$ and $\{\mathbf{Y} \in \mathcal{C}, \Lambda(\mathbf{X} + \mathbf{Y}) \in \mathbb{A}\} \neq \emptyset$, so that $\rho(\mathbf{X}) < +\infty$.

Hence $\rho : M^\Phi \to \mathbb{R}$ and then convexity and monotonicity are straightforward. The remaining properties in (a) are a consequence of Theorem A.2 and the fact that $M^\Phi$ is a Banach space.

We now prove (b). We claim that if $\mathbb{E}[\Lambda(\mathbf{X} + \mathbf{Y})] > B$ then $\mathbf{Y} \in \mathcal{C}$ can not be optimal:

$$\mathbf{Y} \in \mathcal{C} \text{ and } \mathbb{E}[\Lambda(\mathbf{X} + \mathbf{Y})] > B \implies \sum_{n=1}^{N} Y^n > \rho^=(\mathbf{X}). \quad (A.3)$$

Indeed, the continuity of $u_n$ and $\mathbb{E}[u_n(Z^n)] > -\infty$ for all $\mathbf{Z} \in M^\Phi$ imply the existence of $\delta \in \mathbb{R}_+^N$, $\delta \neq \mathbf{0}$, such that $\mathbb{E}[\Lambda(\mathbf{X} + \mathbf{Y} - \delta)] = B$ and so, as $\mathbf{Y} - \delta \in \mathcal{C}$, $\rho^=(\mathbf{X}) \leq \sum_{n=1}^{N}(Y^n - \delta^n) < \sum_{n=1}^{N} Y^n$. This implies $\rho(\mathbf{X}) = \rho^=(\mathbf{X})$, otherwise if $\rho(\mathbf{X}) < \rho^=(\mathbf{X})$, then by definition of $\rho(\mathbf{X})$, there would exist $\varepsilon > 0$ and $\mathbf{Y} \in \mathcal{C}$ satisfying $\mathbb{E}[\Lambda(\mathbf{X} + \mathbf{Y})] > B$ and $\sum_{n=1}^{N} Y^n \leq \rho(\mathbf{X}) + \varepsilon < \rho^=(\mathbf{X})$, which contradicts (A.3).

We now show uniqueness by contradiction. Suppose that $\rho(\mathbf{X})$ is attained by two distinct $\mathbf{Y}_1 \in \mathcal{C}$ and $\mathbf{Y}_2 \in \mathcal{C}$, so that $\mathbb{P}(Y_1^j \neq Y_2^j) > 0$ for some $j$. Then we have

$$\rho(\mathbf{X}) = \sum_{n=1}^{N} Y_1^n = \sum_{n=1}^{N} Y_2^n \quad \text{and} \quad \mathbb{E}\left[\sum_{n=1}^{N} u_n(X^n + Y_k^n)\right] = B \text{ for } k = 1, 2.$$

For $\lambda \in [0, 1]$ set $\mathbf{Y}_\lambda := \lambda \mathbf{Y}_1 + (1 - \lambda) \mathbf{Y}_2$. Then $\mathbf{Y}_\lambda \in \mathcal{C}$, as $\mathcal{C}$ is convex. This implies

$$\sum_{n=1}^{N} Y_\lambda^n = \lambda \sum_{n=1}^{N} Y_1^n + (1 - \lambda) \sum_{n=1}^{N} Y_2^n = \rho(\mathbf{X}), \forall \lambda \in [0, 1],$$

and for $\lambda \in (0,1)$

$$B = \lambda \mathbb{E}\left[\sum_{n=1}^{N} u_n(X^n + Y_1^n)\right] + (1-\lambda)\mathbb{E}\left[\sum_{n=1}^{N} u_n(X^n + Y_2^n)\right] <$$
$$< \mathbb{E}\left[\sum_{n=1}^{N} u_n(\lambda X^n + \lambda Y_1^n + (1-\lambda)X^n + (1-\lambda)Y_2^n)\right] = \mathbb{E}\left[\sum_{n=1}^{N} u_n(X^n + Y_\lambda^n)\right],$$

where we used that $u_j$ is strictly concave and $\mathbb{P}(Y_1^j \neq Y_2^j) > 0$. This is a contradiction with $\rho(\mathbf{X}) = \rho^=(\mathbf{X})$ and (A.3). $\square$

## A.2 Orlicz setting

We now recall an important result for the characterization of systemic risk measures of the form (2.5) on the Orlicz Heart.

**Theorem A.2** (Theorem 1, [9]). *Suppose that $\mathcal{L}$ is a Fréchet lattice and $\rho : \mathcal{L} \to \mathbb{R} \cup \{+\infty\}$ is convex and monotone decreasing. Then*

1. *$\rho$ is continuous in the interior of $dom(\rho)$, with respect to the topology of $\mathcal{L}$,*

2. *$\rho$ is subdifferentiable in the interior of $dom(\rho)$,*

3. *for all $\mathbf{X} \in int(dom(\rho))$*

$$\rho(\mathbf{X}) = \max_{Q \in \mathcal{L}_+^*} \{Q(-X) - \alpha(Q)\},$$

*where $\mathcal{L}^*$ is the dual of $\mathcal{L}$ (for the topology for which $\mathcal{L}$ is a Frechet lattice), $\mathcal{L}_+^* = \{Q \in \mathcal{L}^* \mid Q \text{ is positive}\}$ and $\alpha : \mathcal{L}^* \to \mathbb{R} \cup \{+\infty\}$, defined by*

$$\alpha(Q) = \sup_{\mathbf{X} \in \mathcal{L}} \{Q(-\mathbf{X}) - \rho(\mathbf{X})\},$$

*is $\sigma(\mathcal{L}^*, \mathcal{L})$-lsc and convex.*

### A.2.1 Dual representation in the Orlicz setting

*Proof.* **[of Proposition 3.4]**

Consider the convex functional $\Theta_n : M^{\phi_n}(\mathbb{R}) \to \mathbb{R}$ defined by $\Theta_n(Z) := \mathbb{E}[-u_n(Z)]$ and let $\Theta_n^*$ be its convex conjugate. We have: $\Theta_n(Z^n) > -\infty$, as $M^{\phi_n}(\mathbb{R}) \subseteq L^1(\mathbb{P})$ and $\mathbb{E}[u_n(Z^n)] \leq u_n(\mathbb{E}[Z^n]) < +\infty$; $\Theta_n(Z^n) < +\infty$, as $Z^n \in M^{\phi_n}(\mathbb{R})$ implies $\mathbb{E}[u_n(Z^n)] > -\infty$. Then we have $\Theta_n^*(\xi) = \mathbb{E}[v_n(-\xi)]$, for $\xi \in L^{\phi_n^*}(\mathbb{R})$ by [9], Section 5.2. Let $f : M^\Phi \to \mathbb{R}$ be defined by $f(\mathbf{Z}) := \sum_{n=1}^{N} \mathbb{E}[-u_n(Z^n)] + B = \sum_{n=1}^{N} \Theta_n(Z^n) + B$, and observe that

$$\mathcal{A} := \left\{\mathbf{Z} \in M^\Phi \mid \sum_{n=1}^{N} \mathbb{E}[u_n(Z^n)] \geq B\right\} = \{\mathbf{Z} \in M^\Phi \mid f(\mathbf{Z}) \leq 0\}.$$

We have that $f$ is convex and decreasing with respect to the componentwise order relation. Let $f^*(\xi)$ be its convex conjugate, for $\xi \in L^{\Phi^*}$. We assume that $\xi \neq \mathbf{0}$. By the Fenchel inequality $\mathbb{E}[\mathbf{Z}\xi] \leq f(\mathbf{Z}) + f^*(\xi)$, we obtain for all $\mathbf{Z} \in \mathcal{A}$ and $\lambda > 0$,

$$\mathbb{E}[-\mathbf{Z}\xi] = \lambda \mathbb{E}[\mathbf{Z}(-\tfrac{1}{\lambda}\xi)] \leq \lambda[f(\mathbf{Z}) + f^*(-\tfrac{1}{\lambda}\xi)] \leq \lambda f^*(-\tfrac{1}{\lambda}\xi), \ \mathbb{P}\text{-a.s.}$$

Hence

$$\alpha_B(\xi) := \sup_{\mathbf{Z} \in \mathcal{A}} \{\mathbb{E}[-\mathbf{Z}\xi]\} \leq \inf_{\lambda > 0} \lambda f^*(-\tfrac{1}{\lambda}\xi). \tag{A.4}$$

By definition of the convex Fenchel conjugate and the fact that $M^\Phi$ is a product space, we have

$$\begin{aligned} f^*(\xi) &:= \sup_{\mathbf{Z} \in M^\Phi} \{\mathbb{E}[\xi \mathbf{Z}] - f(\mathbf{Z})\} = -B + \sup_{\mathbf{Z} \in M^\Phi} \left\{ \sum_{n=1}^N \mathbb{E}[\xi_n Z^n] - \sum_{n=1}^N \Theta_n(Z^n) \right\} \\ &= -B + \sum_{n=1}^N \left( \sup_{Z \in M^\Phi(\mathbb{R})} \{\mathbb{E}[\xi_n Z] - \Theta_n(Z)\} \right) = -B + \sum_{n=1}^N \Theta_n^*(\xi_n), \end{aligned}$$

where we have used (2.4), and therefore

$$\inf_{\lambda > 0} \lambda f^*(-\tfrac{1}{\lambda}\xi) = \inf_{\lambda > 0} \left( -B\lambda + \lambda \sum_{n=1}^N \Theta_n^*(-\tfrac{1}{\lambda}\xi_n) \right) = \inf_{\lambda > 0} \left( -B\lambda + \lambda \sum_{n=1}^N \mathbb{E}\left[ v_n\left(\tfrac{1}{\lambda}\xi_n\right) \right] \right).$$

To prove (3.7) we need only to show that there is no duality gap in (A.4), i.e., if $\alpha_B(\xi) < +\infty$ then

$$\alpha_B(\xi) = \inf_{\lambda > 0} \lambda f^*(-\tfrac{1}{\lambda}\xi). \tag{A.5}$$

Observe that, by the definition of $f^*$, we have for each $\lambda > 0$

$$\lambda f^*(-\tfrac{1}{\lambda}\xi) := \sup_{\mathbf{Z} \in M^\Phi} \{\mathbb{E}[-\xi\mathbf{Z}] - \lambda f(\mathbf{Z})\}.$$

As $\xi$ is not identically equal to $\mathbf{0}$ and $M^\Phi$ is a linear space, we have $\sup_{\mathbf{Z} \in M^\Phi} \{\mathbb{E}[-\xi\mathbf{Z}]\} = +\infty$ and therefore

$$\inf_{\lambda > 0} \lambda f^*(-\tfrac{1}{\lambda}\xi) = \inf_{\lambda > 0} \sup_{\mathbf{Z} \in M^\Phi} \{\mathbb{E}[-\xi\mathbf{Z}] - \lambda f(\mathbf{Z})\} = \inf_{\lambda \geq 0} \sup_{\mathbf{Z} \in M^\Phi} \{\mathbb{E}[-\xi\mathbf{Z}] - \lambda f(\mathbf{Z})\}.$$

We claim that

$$\inf_{\lambda \geq 0} \sup_{\mathbf{Z} \in M^\Phi} \{\mathbb{E}[-\xi\mathbf{Z}] - \lambda f(\mathbf{Z})\} = \sup_{\mathbf{Z} \in M^\Phi} \inf_{\lambda \geq 0} \{\mathbb{E}[-\xi\mathbf{Z}] - \lambda f(\mathbf{Z})\}. \tag{A.6}$$

Assuming (A.6), we may immediately conclude that

$$\begin{aligned} \inf_{\lambda > 0} \lambda f^*(-\tfrac{1}{\lambda}\xi) &= \sup_{\mathbf{Z} \in M^\Phi} \inf_{\lambda \geq 0} \{\mathbb{E}[-\xi\mathbf{Z}] - \lambda f(\mathbf{Z})\} = \sup_{\mathbf{Z} \in M^\Phi} \left\{ \mathbb{E}[-\xi\mathbf{Z}] - \sup_{\lambda \geq 0} \lambda f(\mathbf{Z}) \right\} \\ &= \sup_{\mathbf{Z} \in \mathcal{A}} \{\mathbb{E}[-\xi\mathbf{Z}]\} := \alpha_B(\xi). \end{aligned}$$

We now prove (A.6) by showing the equivalent condition:

$$\sup_{\lambda \geq 0} \inf_{\mathbf{Z} \in M^{\Phi}} \{\mathbb{E}[\xi \mathbf{Z}] + \lambda f(\mathbf{Z})\} = \inf_{\mathbf{Z} \in M^{\Phi}} \sup_{\lambda \geq 0} \{\mathbb{E}[\xi \mathbf{Z}] + \lambda f(\mathbf{Z})\}. \tag{A.7}$$

In order to make an easy comparison with the results in [43], let $f_0(\mathbf{Z}) := \mathbb{E}[\xi \mathbf{Z}]$. Consider the function $F : M^{\Phi} \times \mathbb{R} \to \mathbb{R} \cup \{+\infty\}$, defined by

$$F(\mathbf{Z}, u) = \begin{cases} f_0(\mathbf{Z}) & \text{if } \mathbf{Z} \in M^{\Phi} \text{ and } f(\mathbf{Z}) \leq u, \\ +\infty & \text{otherwise}, \end{cases}$$

see (2.8) in [43], and the associated Lagrangian $K(\mathbf{Z}, \lambda)$, see (4.4) in [43]. Then (A.7) can be rewritten as

$$\sup_{\lambda \geq 0} \inf_{\mathbf{Z} \in M^{\Phi}} K(\mathbf{Z}, \lambda) = \inf_{\mathbf{Z} \in M^{\Phi}} \sup_{\lambda \geq 0} K(\mathbf{Z}, \lambda). \tag{A.8}$$

As $f : M^{\Phi} \to \mathbb{R}$ is convex decreasing and finite valued, Theorem A.2 guarantees that it is continuous on $M^{\Phi}$ (for the $M^{\Phi}$-norm). Therefore, see Example 1 on pages 7 and 22 in [43], the function $F$ is closed convex in $(\mathbf{Z}, u)$.

Then the absence of duality gap in (A.5) is now expressed by (A.8) and it follows from Theorems 17 and 18 of [43], provided that the (convex) optimal value function, defined in (4.7) [43],

$$\varphi(u) := \inf_{\mathbf{Z} \in M^{\Phi}} F(\mathbf{Z}, u), \ u \in \mathbb{R},$$

is bounded from above in a neighborhood of 0. This is easily verified by showing the existence of an element $\mathbf{Z}_0 \in M^{\Phi}$ such that $u \to F(\mathbf{Z}_0, u)$ is bounded from above in a neighborhood of 0. This concludes the proof of (3.7).

To prove (3.8) we set $\xi_n := \frac{dQ^n}{d\mathbb{P}} \geq 0$ a.s.. Recall from Lemma A.5 that $v_n$ is strictly convex with $v_n(+\infty) = +\infty$, $v_n(0^+) = u_n(+\infty)$, $\lim_{z \to +\infty} \frac{v_n(z)}{z} = +\infty$ because of Assumption 2.2, item 2, and $v_n$ is continuously differentiable. As $u'_n(+\infty) = 0$ and $u'_n(-\infty) = +\infty$, we get $v'_n(0) = -\infty$ and $v'_n(+\infty) = +\infty$.

Set $\eta = \frac{1}{\lambda} \in (0, +\infty)$ and consider the differentiable function $F : (0, +\infty) \to \mathbb{R}$ defined by

$$F(\eta) := -B\eta + \eta \sum_{n=1}^{N} \mathbb{E}\left[v_n\left(\frac{1}{\eta}\xi_n\right)\right].$$

Then $\alpha_B(\xi) = \inf_{\eta > 0} F(\eta)$ and (3.9) can be rewritten as

$$F'(\eta) = 0 \tag{A.9}$$

with

$$F'(\eta) = -B + \sum_{n=1}^{N} \mathbb{E}\left[v_n\left(\frac{1}{\eta}\xi_n\right)\right] - \frac{1}{\eta}\sum_{n=1}^{N} \mathbb{E}\left[\xi_n v'_n\left(\frac{1}{\eta}\xi_n\right)\right].$$

Note that if $\eta^* > 0$ is the solution to (A.9), then by replacing such $\eta^*$ into $F(\eta)$ we immediately obtain (3.8).

Next, thanks to the integrability conditions provided by Lemma A.4, we show the existence of the solution $\eta^* > 0$ of (A.9). First we consider $\eta \to +\infty$. Since $\sum_{n=1}^{N} v_n(0^+) = \sum_{n=1}^{N} u_n(+\infty) > B$ by Assumption 2.2, we have that

$$\underline{\lim}_{\eta \to +\infty} - B + \sum_{n=1}^{N} \mathbb{E}\left[v_n\left(\frac{1}{\eta}\xi_n\right)\right] > 0.$$

Moreover, $v_n'(0) = -\infty$ shows that

$$\underline{\lim}_{\eta \to +\infty} - \frac{1}{\eta}\sum_{n=1}^{N} \mathbb{E}\left[\xi_n v_n'\left(\frac{1}{\eta}\xi_n\right)\right] \geq 0.$$

Hence $\underline{\lim}_{\eta \to +\infty} F'(\eta) > 0$. We now look at $\eta \to 0$:

$$\begin{aligned}
\lim_{\eta \to 0} F'(\eta) &= -B + \lim_{\eta \to 0}\sum_{n=1}^{N} \mathbb{E}\left[v_n\left(\frac{1}{\eta}\xi_n\right)\right] - \frac{1}{\eta}\sum_{n=1}^{N}\mathbb{E}\left[\xi_n v_n'\left(\frac{1}{\eta}\xi_n\right)\right] \\
&= -B + \lim_{t \to +\infty}\sum_{n=1}^{N}\mathbb{E}\left[v_n(t\xi_n)\right] - t\sum_{n=1}^{N}\mathbb{E}\left[\xi_n v_n'(t\xi_n)\right] \\
&= -B + \sum_{n=1}^{N}\lim_{t \to +\infty}\mathbb{E}\left[v_n(t\xi_n) - t\xi_n v_n'(t\xi_n)\right].
\end{aligned}$$

The convexity of $v_n$ implies that for any fixed $z_0 > 0$ and $z > z_0$

$$v_n(z) - v_n(z_0) \leq v_n'(z)(z - z_0).$$

From $\lim_{z \to +\infty} \frac{v(z)}{z} = +\infty$, $v_n'(z) \to +\infty$ as $z \to +\infty$ and

$$v_n(z) - z v_n'(z) \leq v_n(z_0) - z_0 v_n'(z) \downarrow -\infty \text{ as } z \to +\infty,$$

we have by monotone convergence

$$\lim_{t \to +\infty}\mathbb{E}\left[v_n(t\xi_n) - t\xi_n v_n'(t\xi_n)\right] = -\infty,$$

so that $\underline{\lim}_{\eta \to 0} F'(\eta) = -\infty$. By the continuity of $F'$ we obtain the existence of the solution $\eta^* > 0$ for (A.9). Uniqueness follows from the strict convexity of $F$. $\square$

*Remark* A.3. In [28], (A.5) is deduced, by different means, in a $L^\infty(\mathbb{R})$ setting and in the one-dimensional case. In [3], (A.5) is obtained, by different means, in the multi-dimensional deterministic case, i.e. in $\mathbb{R}^N$.

## A.3 Auxiliary results for existence

The following auxiliary result is standard and can be found in many articles on utility maximization, see for example Lemma 18, [8]. Recall that we are working under Assumption 2.2, Item 4.

**Lemma A.4.** *Let $v : \mathbb{R}_+ \to \mathbb{R}$ be a strictly convex differentiable function with $v'(0^+) = -\infty$, $v'(+\infty) = +\infty$ and let $Q \ll \mathbb{P}$. Then*

(a) $v'(\lambda \frac{dQ}{d\mathbb{P}}) \in L^1(Q)\ \forall \lambda > 0$;

(b) $F(\lambda) \triangleq \mathbb{E}[\frac{dQ}{d\mathbb{P}} v'(\lambda \frac{dQ}{d\mathbb{P}})]$ *defines a bijection between $(0, +\infty)$ and $(-\infty, +\infty)$.*

By applying the classical convex duality real valued theory from [42] we get:

**Lemma A.5.** *The convex conjugate function $v : \mathbb{R} \to (-\infty, +\infty]$ of $u$, given by $v(y) = \sup_{x \in \mathbb{R}} \{u(x) - xy\}$, is a proper lsc convex function, equal to $+\infty$ on $(-\infty, 0)$, bounded from below on $\mathbb{R}$, finite valued strictly convex, continuously differentiable on $(0, +\infty)$ and satisfying*

$$v(+\infty) = +\infty,\ v(0^+) = u(+\infty),\ v'(0^+) = -\infty,\ v'(+\infty) = +\infty,$$

$$u'(x) = (v')^{-1}(-x), u(-v'(y)) = -yv'(y) + v(y), \quad \forall y \geq 0,$$

*where the usual rule $0 \cdot \infty = 0$ is applied.*

**Proposition A.6** (Proposition 3.6, [11]). *Let $Q \ll \mathbb{P}$. For all $c \in \mathbb{R}$ the optimizer $\lambda(c; Q)$ of*

$$\min_{\lambda > 0} \left\{ \mathbb{E}\left[v\left(\lambda \frac{dQ}{d\mathbb{P}}\right)\right] + \lambda c \right\}$$

*is the unique positive solution of the first order condition*

$$\mathbb{E}_Q\left[v'\left(\lambda \frac{dQ}{d\mathbb{P}}\right)\right] + c = 0. \tag{A.10}$$

*If $\sup\{\mathbb{E}[u(g)] \mid g \in L^1(Q) \text{ and } \mathbb{E}_Q[g] \leq c\} < u(+\infty)$, the random variable $\widehat{g} := -v'(\lambda(c;Q)\frac{dQ}{d\mathbb{P}})$ belongs to the set $\{g \in L^1(Q) \mid \mathbb{E}_Q[g] = c\}$, satisfies $u(\widehat{g}) \in L^1(\mathbb{P})$, and*

$$\min_{\lambda>0}\left\{\mathbb{E}\left[v\left(\lambda\frac{dQ}{d\mathbb{P}}\right)\right] + \lambda c\right\} = \sup\{\mathbb{E}[u(g)] \mid g \in L^1(Q) \text{ and } \mathbb{E}_Q[g] \leq c\} = \mathbb{E}[u(\widehat{g})] < u(+\infty).$$

## A.4 Proofs for Section 4.2

*Proof.* **[of Proposition 4.4]** From $M^{\phi_n} \subseteq L^1(\mathbb{P}, Q^n) \subseteq L^1(Q^n)$ we clearly have $U_n(a^n) \leq \widetilde{U}_n(a^n) \leq \widehat{U}_n(a^n) \leq u_n(+\infty)$, so that

$$\text{if } U_n(a^n) = u(+\infty) \text{ then } U_n(a^n) = \widetilde{U}_n(a^n) = \widehat{U}_n(a^n) = u_n(+\infty). \tag{A.11}$$

By the Fenchel inequality we get

$$\mathbb{E}\left[u_n(X^n + W)\right] \leq \lambda \left(\mathbb{E}_{Q^n}[X^n] + \mathbb{E}_{Q^n}[W]\right) + \mathbb{E}\left[v_n\left(\lambda \frac{dQ^n}{dP}\right)\right],$$

and hence

$$U_n(a^n) \leq \widetilde{U}_n(a^n) \leq \widehat{U}_n(a^n) \leq \inf_{\lambda>0}\left\{\lambda\left(\mathbb{E}_{Q^n}[X^n]+a^n\right)+\mathbb{E}\left[v_n\left(\lambda\frac{dQ^n}{dP}\right)\right]\right\}<+\infty, \quad (A.12)$$

as $\mathbb{E}\left[v_n\left(\lambda\frac{dQ^n}{dP}\right)\right]<+\infty$. Therefore (4.5) is a consequence of (A.11) and (4.7). To show (4.7), consider the integral functional $I:M^{\phi_n}\to\mathbb{R}$ defined by $I(X^n)=\mathbb{E}\left[u_n(X^n)\right]$. It is finite valued, monotone increasing and concave on $M^{\phi_n}$ (as $\mathbb{E}[u_n(X^n)]\leq u_n(\mathbb{E}[X^n])<+\infty$), and therefore, by the Theorem A.2, it is norm-continuous on $M^{\phi_n}$. We can then follow the well known duality approach (see for example [11]). Consider the convex cone $D^0:=\left\{W\in M^{\phi_n}\mid\mathbb{E}_{Q^n}[W]\leq 0\right\}$ which is the polar cone of the one dimensional cone $D:=\left\{\lambda\frac{dQ^n}{dP}\mid\lambda\geq 0\right\}$, so that the bipolar $D^{00}$ coincide with $D$. Let $\delta_{D^0}:M^{\phi_n}\to\mathbb{R}\cup\{+\infty\}$ be the support functional of $D^0$. By [39], or directly by hand, the concave conjugate $I^*:L^{\phi_n^*}\to\mathbb{R}\cup\{-\infty\}$ is given by $I^*(\xi^n)=\mathbb{E}\left[-v_n(\xi^n)\right]$ and so, by Fenchel duality Theorem,

$$\begin{aligned}
U_n(a^n) &= \sup_{W\in D^0}\mathbb{E}\left[u_n(X^n+a^n+W)\right] = \sup_{Z\in D^0+X^n+a^n}\mathbb{E}\left[u_n(Z)\right] \\
&= \sup_{Z\in M^{\phi_n}}\left\{\mathbb{E}\left[u_n(Z)\right]-\delta_{D^0+X^n+a^n}(Z)\right\} = \min_{\xi^n\in L^{\phi_n^*}}\left\{\delta^*_{D^0+X^n+a^n}(\xi^n)-\mathbb{E}\left[-v_n(\xi^n)\right]\right\} \\
&= \min_{\xi^n\in L^{\phi_n^*}}\left\{\mathbb{E}[\xi^n(X^n+a^n)]+\delta_{D^{00}}(\xi^n)+\mathbb{E}\left[v_n(\xi^n)\right]\right\} \\
&= \min_{\xi^n\in D^{00}}\left\{\mathbb{E}[\xi^n(X^n+a^n)]+\mathbb{E}\left[v_n(\xi^n)\right]\right\} = \min_{\lambda>0}\left\{\lambda\left(\mathbb{E}_{Q^n}[X^n]+a^n\right)+\mathbb{E}\left[v_n\left(\lambda\frac{dQ^n}{dP}\right)\right]\right\},
\end{aligned}$$

where we used $\delta^*_{D^0}=\delta_{D^{00}}$, $D^{00}=D$ and the fact that the minimizer is obtained at $\lambda>0$, otherwise if $\lambda=0$ then $U_n(a^n)=\mathbb{E}\left[v_n(0)\right]=u_n(+\infty)$. We conclude the proof by proving (4.6). From the inequality (A.12), it is clear that $U_n(-\infty)=-\infty$. Define

$$V_n(\lambda):=\mathbb{E}\left[v_n\left(\lambda\frac{dQ_n}{d\mathbb{P}}\right)\right]+\lambda\mathbb{E}_{Q_n}[X^n].$$

When $U_n(a^n)<u_n(+\infty)$, from (4.7) we have that

$$U_n(a^n)=\inf_{\lambda>0}\left\{V_n(\lambda)+\lambda a^n\right\},$$

which shows that $U_n$ and $V_n$ are conjugate of each other, i.e., $V_n(\lambda)=\sup_{a^n>0}\left\{U_n(a^n)-\lambda a^n\right\}$. From Lemmas A.4 and A.5 we know that the convex function $V_n$ is differentiable on $(0,+\infty)$ and therefore $U_n$ is differentiable on $(-\infty,+\infty)$ and

$$U'_n(a)=(V'_n)^{-1}(-a)>0.$$

We only need to show the last two conditions. As $v_n(0^+)=u_n(+\infty)=+\infty$ then $V_n(0^+)=+\infty$. Since $v'_n(0^+)=-\infty$ we get $V'_n(0^+)=-\infty$ and $U'_n(+\infty)=0$. Moreover

$$\begin{aligned}
V'_n(+\infty) &= \lim_{\lambda\to+\infty}\frac{V_n(\lambda)}{\lambda} = \lim_{\lambda\to+\infty}\frac{1}{\lambda}\mathbb{E}\left[v_n\left(\lambda\frac{dQ_n}{d\mathbb{P}}\right)\right]+\mathbb{E}_{Q_n}[X^n] \\
&\overset{Jensen}{\geq} \lim_{\lambda\to+\infty}\frac{1}{\lambda}v_n(\lambda)+\mathbb{E}_{Q_n}[X^n]=v'_n(\infty)+\mathbb{E}_{Q_n}[X^n]=+\infty
\end{aligned}$$

which implies $U'_n(-\infty)=+\infty$. $\square$

*Proof.* **[of Lemma 4.7]** The set $K$ is clearly closed. We show that is bounded. For $N = 1$ it is true. Let $N > 1$. First we prove that, for all $j = 1, ..., N$,

$$U_j(a)\left\{1 + \frac{\sum_{n \neq j} U_n(A - (N-1)a)}{U_j(a)}\right\} \to -\infty \text{ as } a \downarrow -\infty. \tag{A.13}$$

Recall that $U_n(-\infty) = -\infty$ and $U_n(+\infty) \leq u_n(+\infty)$ for all $n$. Suppose that for some $k \in \{1, ..., N\}$, $u_k(+\infty) < +\infty$. Then $U_k(+\infty) < +\infty$ and for all $j = 1, ..., N$

$$\lim_{a \to -\infty}\left\{\frac{U_k(A - (N-1)a)}{U_j(a)}\right\} = 0. \tag{A.14}$$

Now suppose that for some $k \in \{1, ..., N\}$, $u_k(+\infty) = +\infty$. Then Proposition 4.4 shows that $U_k(a^k) < +\infty = u_k(+\infty), U_k' > 0, U_k'(-\infty) = +\infty, U_k'(+\infty) = 0$. By l'Hopital's rule, for all $j = 1, ..., N$ we obtain again

$$\lim_{a \to -\infty}\left\{\frac{U_k(A - (N-1)a)}{U_j(a)}\right\} = \lim_{a \to -\infty}\frac{-(N-1)U_k'(A - (N-1)a)}{U_j'(a)} = 0. \tag{A.15}$$

From (A.14) and (A.15) we deduce that (A.13) holds true.

We conclude that for any constant $B$ there exists a constant $R$ such that for all $j = 1, ..., N$ and $a < R$

$$U_j(a)\left\{1 + \frac{\sum_{n \neq j} U_n(A - (N-1)a)}{U_j(a)}\right\} < B.$$

Let $\mathbf{a} \in K$ and let $i$ be such that $a^i = \min\{a^1, ..., a^N\}$. Note that $a^j \leq A - (N-1)a^i$ for all $j = 1, ..., N$ because $\sum_{n=1}^N a^n \leq A$ holds. Assume that $a^i < R$. Then

$$B \leq \sum_{n=1}^N U_n(a^n) \leq U_i(a^i)\left\{1 + \frac{\sum_{n \neq i} U_n(A - (N-1)a^i)}{U_i(a^i)}\right\}, \tag{A.16}$$

which is a contradiction. Thus $a^j \geq R$ for all $j = 1, ..., N$, and then also $a^j \leq A - (N-1)R$ for all $j = 1, ..., N$ because $\sum_{n=1}^N a^n \leq A$ holds. This proves the claim. $\square$

Let $\mathbf{X} \in M^\Phi$ and consider the function $F(\delta) := \mathbb{E}\left[\sum_{n=1}^N u_n(X^n + Y^n - \delta)\right]$, $\delta \in \mathbb{R}$. If $\mathbf{Y} \in M^\Phi$, then $F$ is finite valued and concave on $\mathbb{R}$, hence continuous on $\mathbb{R}$ (see the discussion at the beginning of Section 4.2). However, when $\mathbf{Y} \in L^1(\mathbf{Q})$ satisfies $\mathbb{E}\left[\sum_{n=1}^N u_n(X^n + Y^n)\right] > B$ (with the understanding that $u_n(X^n + Y^n) \in L^1(\mathbb{P})$ for each $n$), it is not any more evident if $F$ is continuous on $\mathbb{R}$, as one has to guarantee that $\mathbb{E}\left[\sum_{n=1}^N u_n(X^n + Y^n - \delta)\right] > -\infty$, for $\delta > 0$.

**Lemma A.7.** *If $\mathbf{X} \in M^\Phi$ and $\mathbf{Z} \in L^1(\mathbf{Q})$ satisfy $\mathbb{E}\left[\sum_{n=1}^N u_n(X^n + Z^n)\right] > B$ then there exists $\widetilde{\mathbf{Z}} \in L^1(\mathbf{Q})$ satisfying $\sum_{n=1}^N \mathbb{E}_{Q^n}[\widetilde{Z}^n] < \sum_{n=1}^N \mathbb{E}_{Q^n}[Z^n]$ and $\mathbb{E}\left[\sum_{n=1}^N u_n(X^n + \widetilde{Z}^n)\right] = B$.*

*Proof.* Set $A_n := \{X^n + Z^n > k_n\}$ and let $k_n \in \mathbb{R}$ satisfy $\mathbb{P}(A_n) > 0$ and $Q^n(A_n) > 0$. For any $\delta > 0$, set $\widetilde{Z}^n := (Z^n - \delta 1_{A_n})_n \in L^1(\mathbf{Q})$ and $G(\delta) := \mathbb{E}\left[\sum_{n=1}^N u_n(X^n + Z^n - \delta 1_{A_n})\right]$. Then

$$G(\delta) = \mathbb{E}\left[\sum_{n=1}^N u_n(X^n + Z^n) 1_{A_n^C}\right] + \mathbb{E}\left[\sum_{n=1}^N u_n(X^n + Z^n - \delta) 1_{A_n}\right]$$

$$\geq \mathbb{E}\left[\sum_{n=1}^N u_n(X^n + Z^n) 1_{A_n^C}\right] + \mathbb{E}\left[\sum_{n=1}^N u_n(k_n - \delta) 1_{A_n}\right] > -\infty,$$

which implies that $G$ is continuous on $\mathbb{R}_+$ and the thesis follows. $\square$

*Proof.* **[of Lemma 4.8]** From (3.5) and $\rho_B^{\mathbf{Q}}(\mathbf{X}) = \widetilde{\rho}_B^{\mathbf{Q}}(\mathbf{X})$, note that the penalty function can also be written as

$$\alpha_B(\mathbf{Q}) = -\sum_{n=1}^N \mathbb{E}_{Q^n}[X^n] - \rho_B^{\mathbf{Q}}(\mathbf{X}) = -\sum_{n=1}^N \mathbb{E}_{Q^n}[X^n] - \widetilde{\rho}_B^{\mathbf{Q}}(\mathbf{X}) =$$

$$= \sup\left\{\sum_{n=1}^N \mathbb{E}_{Q^n}[-Z^n] \mid \mathbf{Z} \in L^1(\mathbb{P}; \mathbf{Q}), \mathbb{E}[\Lambda(\mathbf{Z})] \geq B\right\}.$$

Set

$$c^=(\mathbf{Q}) := \inf\left\{\sum_{n=1}^N \mathbb{E}_{Q^n}[Z^n] \mid \mathbf{Z} \in L^1(\mathbb{P}; \mathbf{Q}), \mathbb{E}[\Lambda(\mathbf{Z})] = B\right\}.$$

Similarly to the proof of (A.3), we show that:

$$\mathbf{Z} \in L^1(\mathbb{P}; \mathbf{Q}) \text{ and } \mathbb{E}[\Lambda(\mathbf{Z})] > B \Longrightarrow \sum_{n=1}^N \mathbb{E}_{Q^n}[Z^n] > c^=(\mathbf{Q}), \tag{A.17}$$

Indeed, Lemma A.7 implies the existence $\widetilde{\mathbf{Z}} \in L^1(\mathbb{P}; \mathbf{Q})$ satisfying $\sum_{n=1}^N \mathbb{E}_{Q^n}[\widetilde{Z}^n] < \sum_{n=1}^N \mathbb{E}_{Q^n}[Z^n]$ and $\mathbb{E}[\Lambda(\widetilde{\mathbf{Z}})] = B$ and therefore $c^=(\mathbf{Q}) \leq \sum_{n=1}^N \mathbb{E}_{Q^n}[\widetilde{Z}^n] < \sum_{n=1}^N \mathbb{E}_{Q^n}[Z^n]$. It follows that

$$c(\mathbf{Q}) := -\alpha_B(\mathbf{Q}) = \inf\left\{\sum_{n=1}^N \mathbb{E}_{Q^n}[Z^n] \mid \mathbf{Z} \in L^1(\mathbb{P}; \mathbf{Q}), \mathbb{E}[\Lambda(\mathbf{Z})] \geq B\right\} = c^=(\mathbf{Q}).$$

Indeed, $-\infty < c(\mathbf{Q}) \leq c^=(\mathbf{Q})$ and assume by contradiction that $c(\mathbf{Q}) < c^=(\mathbf{Q})$. By definition of $c(\mathbf{Q})$, there exist $\varepsilon > 0$ and $\mathbf{Z} \in L^1(\mathbb{P}; \mathbf{Q})$ such that $\mathbb{E}[\Lambda(\mathbf{Z})] > B$ and $\sum_{n=1}^N \mathbb{E}_{Q^n}[Z^n] \leq c(\mathbf{Q}) + \varepsilon < c^=(\mathbf{Q})$, which contradicts (A.17).

Uniqueness then follows from an argument similar to the one applied at the end of the proof of Proposition 2.4, replacing $\sum_{n=1}^N Y^n$ with $\sum_{n=1}^N \mathbb{E}_{Q^n}[Y^n]$. $\square$